\documentclass[12pt]{article}

\usepackage[dvips]{color}

\newcommand{\bftau}{\mbox{\boldmath $\tau$}}

\newcommand{\bfsigma}{\mbox{\boldmath $\sigma$}}
\newcommand{\fr}[2]{\frac{{\displaystyle #1}}{{\displaystyle #2}}}

\def\lsim{\mathrel{\rlap{\lower4pt\hbox{\hskip1pt$\sim$}}
    \raise1pt\hbox{$<$}}}         

\def\gsim{\mathrel{\rlap{\lower4pt\hbox{\hskip1pt$\sim$}}
    \raise1pt\hbox{$>$}}}         

\renewcommand{\theequation}{\arabic{section}.\arabic{equation}}

\begin{document}
\pagenumbering{roman}
\title{ Extended Theory of Finite Fermi Systems:\\
 {\Large Collective Vibrations in Closed Shell Nuclei}}

\author {S.Kamerdzhiev$^{a,b}$,  J.Speth$^{a,c}$, and G.Tertychny$^{a,b}$\\
{\it $^a$Institut f\"ur Kernphysik, Forschungszentrum J\"ulich,}\\
{\it 52425 J\"ulich, Germany}\\
{\it $^b$Institute of Physics and Power Engineering,}\\
{\it  249020 Obninsk, Russia}\\
{\it $^c$Special Research Center for the Structure of Matter,}\\
{\it University of Adelaide,}\\
{\it South Australia 5005, Australia}}

\date{}
\maketitle \tableofcontents



\newpage
\pagenumbering{arabic}
\begin{abstract}
We review an extension of Migdal's Theory of Finite Fermi Systems
which has been developed and applied to collective vibrations in
closed shell nuclei in the past ten years. This microscopic
approach is based on a consistent use of the Green function
method. Here one considers in a consistent way more complex
1p1h$\otimes$phonon
configurations beyond the RPA correlations. Moreover, these   
configurations are not only included in the excited states but
also explicitly
in the ground states of nuclei. The method has been applied to   
the calculation of the strength distribution and transition
densities of giant electric and magnetic resonances in stable and
unstable magic nuclei. Using these microscopic transition
densities, cross sections for inelastic electron and alpha
scattering have been calculated                                     
and compared with the available experimental data. The method also       
allows one to extract in a consistent way the magnitude of the      
strength of the various multipoles in the energy regions in which   
several multipoles overlap. We compare the microscopic transition
densities,  the strength distributions and the various multipole
strengths with their values extracted phenomenologically.          

\end{abstract}


Transition densities.
\newpage
\setcounter{equation}{0}
\section{Introduction}
\subsection{Aims of the review}

The atomic nucleus is a complicated quantum mechanical many-body
system with a rich excitation spectrum. The experimental
investigations of nuclei during the past 50 years have provided us
with an overwhelming amount of excellent data which have to be
understood in the framework of a quantum mechanical many-body
theory. Moreover, we expect new data from the coming radioactive
beam facilities for atomic nuclei far from the stability line,      
which will give us information about, for example, nuclei with a    
large neutron-to-proton ratio. Some of the features of the nuclei,
such as their level fluctuations, can be described in terms of
random matrix theory, which is now interpreted as the
manifestation of chaotic motion. This stochastic nature of the
levels is, however, only one aspect of  the spectrum of the atomic
nucleus which shows, on the other hand, many "regular" features,
such as single-particle structure and collective modes. This
regular behavior is a consequence of the Pauli principle and the
special structure of the nucleon-nucleon interaction that leads to 
the concept of a mean field. As a consequence of the mean field,
one is able to define quasiparticles in the sense of Landau and to
apply Landau's theory of Fermi liquids \cite{Landau} to the
nuclear many body problem, as has been done by Migdal in his
theory of finite Fermi systems (TFFS) \cite{AB}.                   

In this article we focus  on the question of how nuclei respond to  
a weak external field. This corresponds in lowest order to linear   
response and  Migdal's  equations look  formally very similar to    
it. As we will show in the next section, however, the range of      
validity is much larger than one would expect from the
conventional derivation of the linear response equations. Migdal's  
equations are derived within the  many body Green functions (GF)
theory. Using this powerful formulation of the many-body problem    
one is able to obtain equations which in principle are ``exact''.
This is due to a renormalization procedure similar to the one used
in quantum field theory. In the original version of the TFFS one
considers explicitly only the propagation of particle-hole pairs    
in the nuclear medium. All the other configurations, such as the two   
particle-two holes ones, are renormalized into an effective
two-body interaction and effective operators. The final equations
have the same form as the corresponding equations of the
conventional linear response theory, or random phase approximation  
(RPA).

The important observation by Migdal was that the effective          
interaction and effective operators so defined depend  only weakly  
on the mass number and the energy. Therefore, as in Landau's         
original theory, these effective quantities can be parameterized    
and the corresponding parameters should be the same for all nuclei
except for the lightest ones. Moreover, these parameters have been
chosen to be density--dependent, as the interaction inside and
outside of nuclei may be quite different. These assumptions have been shown to be  
correct in numerous applications and they were and still are the most   
important features of TFFS.
One can hope that the same parameters are also applicable to        
nuclei far from stability where we shall next apply the theory.        


It is well known that pairing correlations are important in
non-magic nuclei. In this case on has to define quasiparticles     
analogously to the BCS theory, which include the pairing gap, and   
in the linear response equation one has to consider as well the     
change of the pairing gaps \cite{AB,LM63} in a consistent way. The
equations of the extended theory look formally identical to the
quasiparticle RPA (QRPA) \cite{Baranger60}.                         


The standard TFFS is not a self-consistent theory in the sense      
that one starts with an effective interaction from which one        
obtains simultaneously the mean-field quantities such as            
single-particle energies, single-particle wave functions and the
particle-hole interaction for the linear response. Instead, one     
has independent parameters for the single-particle model and the
particle-hole interaction.
Of the various extensions of Migdal's original theory that have     
been made over the years, the first were directed at taking         
self-consistency effects into account \cite{sfkh78,khs82,AB2}.      

It has been shown
that the TFFS can also be applied to strongly deformed nuclei
\cite{zps78}. In that case the appropriate quasiparticles can be
deduced from the deformed shell model of Nilsson.  The extension    
of TFFS to a second order response theory has been developed
\cite{Speth70} and the formalism for the calculation of electric
and magnetic moments of excited states in even-even nuclei,
transitions between excited states, as well as isomer shifts of
rotational states in deformed nuclei, can be found in
refs.~\cite{Ring and Speth,Meyer and Speth}.

In addition, the $\Delta$-isobar has been included in the space of  
quasiparticles \cite{ost91} and, in the effective interaction,     
the effect of the one pion exchange, which turned out to be         
crucial for the existence of pion condensation and the properties   
of spin modes \cite{AB2,sk80,bstf81}.

As mentioned earlier, both the linear response theory and           
(effectively) the TFFS are  based on the RPA, which is restricted   
to a configuration space which includes one particle--one hole
(1p1h) configurations only. If applied to collective modes in
nuclei it gives values of, for
example, the centroid energies and total strengths of giant         
resonances in good agreement with the data.  The widths of the      
resonances and their fine structure, however, are not.
These can only be obtained if one                                   
includes in the conventional RPA and its related approaches the
coupling of the 1p1h excitations to more complex configurations.
Therefore, if one wants to apply TFFS to a realistic theoretical
interpretation of giant resonance experiments one has to extend     
the formalism to include explicitly in the theory more complex      
configurations than 1p1h pairs.                                    

The main aim of this article is therefore to discuss such           
extensions of Migdal's theory and their application to giant        
resonances in nuclei. The essential feature---that which            
determines the necessity and the title of our approach---is         
of course the explicit taking of complex configurations into        
account. After a short review of the standard TFFS we discuss in a
qualitative way possible extensions of the theory. Here we will
give physical arguments for why the coupling of the most
collective low-energy phonons to the single-particle and
single-hole propagators is the most important and which higher      
configurations are necessary for a quantitative description  of
the widths of giant resonances.

In section~2 we derive the basic equations of the extended TFFS     
(ETFFS) for closed shell nuclei. As in the original TFFS, the GF
method is used, and for convenience we restrict ourselves to
doubly closed shell nuclei, where pairing correlations can be
neglected. As we shall see, there exist  several stages of
sophistication of the
extension, and we will discuss these in some detail. As a first     
approximation we include the
complex configurations in the excited states only. This already
gives rise to the fragmentation of the multipole strength, but the
total strength remains unchanged. In the next step we include also  
the more complex configurations in the  ground state of the         
nuclei. This gives a further fragmentation but, in addition, gives  
changes in the magnitudes of the electric and magnetic transition   
strengths, which turns out to be important for a quantitative
comparison of the theoretical results with the data. The final
formulas of the extended theory are given in {\bf r}-space, which   
is especially appropriate for including                             
effects of the single-particle continuum.                           

In section~3 we apply the various versions of our theory to giant
multipole resonances (GMR) in medium and heavy mass nuclei. The     
comparison with experiment demonstrates the power of our new        
theoretical framework and, simultaneously,  the importance of the
different steps of our approach.

In section~4 the calculations of transition densities within our
approach are reviewed. With this information one is able to derive
cross sections that can be compared directly with electron and      
alpha scattering experiments. One major difference between the      
microscopically calculated transition densities and the
phenomenological ones is an energy dependence of the microscopic
ones. In this connection we discuss also the problems that arise
in the analysis of giant resonances in nuclei in which the
resonances are very broad and where the various multipolarities
overlap.

Finally, we summarize our review and discuss possible extensions
that may be important for a theoretical understanding of nuclei     
far from the stability line.


\subsection{Brief review of the standard theory}                     

Migdal \cite{AB} has applied Landau's theory of interacting Fermi
systems to atomic nuclei. Here one has first to deal with two        
kinds of fermions---protons and neutrons---and second
with a relatively small number of particles. In the past 30 years    
Migdal's theory has been applied successfully by many groups to
various nuclear structure problems. (See, for example, the reviews    
\cite{sfkh78,khs82,ost91,bstf81,SWW77,Kaev79,kns88} and the book
\cite{AB2}.)                                                         


Landau's theory deals with infinite  systems of  interacting
fermions, such as liquid $^3$He or nuclear matter. If there were
no interaction, the system would simply be a collection of
independent particles, each characterized by its spin and a wave
number {\bf k}. Landau's basic assumption is that the interacting
system can be obtained from the non-interacting one by an            
adiabatic switching on of the interaction. In particular, there
should be a one-to-one correspondence between the single-particle
states of non-interacting systems and the so-called quasiparticle
states in interacting systems. These quasiparticles  behave in the
correlated  system like  real particles in a non-interacting
system. They obey Fermi-Dirac statistics and occupy, like the        
non-interacting particles, corresponding quasiparticle states up
to the Fermi energy. In order to define such quasiparticles Landau
considers the total energy of an interacting system as a
functional of the occupation function $ n({\bf k})$ of the
quasiparticle states {\bf k}. The quasiparticle energies are then
given as  first functional derivatives of the total energy with
respect to $n({\bf  k})$ and the interaction between the
quasiparticles is defined as its second functional derivative.
Using this approach, one can calculate properties of the          
excited system of a Fermi liquid, such as the zero sound mode in
$^3$He. In infinite systems a quasiparticle differs from a real
particle essentially by its mass, because both can be described by
a plane wave. The
effective mass of the quasiparticle is, in general,
momentum--dependent and is deduced from experiments. The
renormalized quasiparticle interaction depends on spin and momenta
and is expanded at the Fermi surface in terms of Legendre
polynomials with free parameters---the well-known Landau
parameters---that
are also determined from experiments.                                

Migdal extended  these ideas to finite Fermi systems and applied     
his TFFS to  atomic nuclei. Here the quasiparticles are the
single-particle states of the nuclear shell model, which can be
obtained experimentally from the neighboring odd-mass nuclei of      
closed shell nuclei. The quasiparticle interaction here is defined
in the same way but it is
isospin dependent. As
in the infinite system, one expands the interaction at the Fermi     
surface in terms of Legendre polynomials and the parameters of the
expansion, the famous Landau-Migdal parameters, are considered
universal. They are also determined from experiment.

The TFFS is for that reason a semi-phenomenological microscopic
theory. It is a microscopic theory because all its fundamental        
equations are derived rigorously from first principles, however it
also contains phenomenological aspects such as the above-mentioned
quasiparticle energies and the quasiparticle interaction. All       
the parameters are well-defined microscopically and in principle
could be calculated starting from the bare nucleon-nucleon
interaction. Such calculations, however, are very involved and in actual       
calculations one has to make severe approximations so that we can
not expect to obtain full quantitative agreement with the
phenomenological parameters. Nevertheless the calculated              
interaction parameters are in surprisingly  good agreement with
the phenomenological ones \cite{kns88}.

\subsubsection{The Landau-Migdal interaction}

As mentioned above, the energy $E$ of an interacting system may be    
considered to be a functional of the occupation functions $n({\bf
k)}$ of the quasiparticles $E=E(n({\bf k}))$. If one excites the
system, one basically changes the occupation functions by an
amount $\delta n({\bf k})$. The corresponding change of the energy
is
\begin{equation}\begin{array}{c}
\delta E = \sum_{{\bf k}}\epsilon^{0}({\bf k})\delta n({\bf k}) +
\sum_{{\bf k,k'}}f({\bf k,k'})\delta n({\bf k})\delta n({\bf k'}) =\\[4mm]
\sum_{{\bf k}}[\epsilon^{0}({\bf k}) + \sum_{{\bf k'}}f({\bf
k,k'})\delta n({\bf k'})] \delta n({\bf k}) = \sum_{{\bf
k}}\epsilon({\bf k})\delta n({\bf k}),
\end{array}\end{equation}
where the $\epsilon^{0}({\bf k})$ are the equilibrium energies of     
the quasiparticles.

The quasiparticle energies $\epsilon({\bf k})$ and the interaction
between a quasiparticle and a quasi-hole  $f({\bf k,
k'})$ defined in this way are the first and second derivatives of    
the energy functional with respect to the occupation functions:
\begin{equation}
 \epsilon({\bf k})=\frac {\delta E}{\delta n({\bf k})}, \quad f({\bf k,k'}) = \frac{\delta^{2}E}{\delta n({\bf k})\delta n({\bf
 k'})}.
\end{equation}                                                       

In nuclear matter we have an explicit spin and isospin dependence:   
\begin{equation}
f({\bf k,k'}) = F({\bf k,k'}) + F'({\bf k,k'}) {\bf {\bftau}}
\cdot {\bf {\bftau}'} + [G({\bf k,k'}) + G'({\bf k,k'}) {\bf
{\bftau}} \cdot {\bf {\bftau}'} ] {\bf {\bfsigma}} \cdot {\bf
{\bfsigma}'}.
\end{equation}

In addition to the central interaction one has in principle also     
to consider tensor and spin-orbit forces which, however, have been
neglected in most of the actual calculations. If one denotes the     
momenta before the collision by ${\bf k}_{1}={\bf k}$ and ${\bf
k}_{2}= {\bf k}'$ then, by translational invariance, the momenta
after the collision are given by ${\bf k}_{3}={\bf k}+{\bf q}$ and
${\bf k}_4={\bf k'}-{\bf q}$, where ${\bf q}$ is the momentum
transfer.
In the third article of ref.~\cite{Landau} Landau showed that the     
main contribution (singularity) to the full scattering amplitude
should come from small ${\bf q}$. He also renormalized the            
integral equation for this amplitude in such a way that it
contained a microscopic analog of the function $f({\bf k,k'})$ and
an integration  over a small region near the Fermi surface. In
this case the interaction depends only on the angle $\theta$
between ${\bf k}$ and ${\bf k}'$. This suggests an expansion in
Legendre polynomials:
 \begin{equation}
F({\bf k,k'}) = \sum_{l}F_{l}P_{l}(\cos \theta).
\end{equation}

The constants $F_{l}$ are the famous Landau-Migdal parameters. One
introduces dimensionless parameters  by defining
\begin{equation}
 F_{l}= C_{0}f_{l},
\end{equation}
where $C_{0}$ is the inverse density of states at the     
Fermi surface:
 \begin{equation}
C_{0} = (\frac{dn}{d\epsilon})^{-1}|_{\epsilon = \epsilon_{F}}.
\end{equation}
Likewise, one may also expand the other components of the
interaction (1.3) and define the parameters $f'_{l},\; g_{l},\;
g'_{l}$ for the various terms.

Some of the Landau parameters can be related to bulk properties of
the nucleus, such as the compression modulus $K$,                    
\begin{equation}
K = k_{F}^{2}\frac{{d\/}^{2}(E/A)}{{d\/} k_{F}^{2}} = 6               
\frac{\hbar^{2}k_{F}^{2}}{2m^{*}}(1+2f_{0}),
\end{equation}
the symmetry energy $\beta$,

\begin{equation}
\beta =\frac{1}{3}\:\frac{\hbar^2k_F^2}{2m^*}(1+2f'_0)
\end{equation}
and the effective mass $m$*                                     
\begin{equation}
 {m*}/m = 1+\frac{2}{3}f_{1}.
\end{equation}

The various parameters have to be determined from experiments and
can then be used to predict other experimental quantities. In
particular, these parameters enter into the equation for $\delta
n({\bf k,\omega})$, the solution of which determines small          
amplitude excitation of Bose type in  Fermi systems. In nuclei
these excitations correspond to the giant multipole resonances
that we are going to investigate.                                    

If one restricts the expansion in Eq.~(1.4) to the lowest order,      
$l=0$, then the interaction Eq.~(1.3) is a constant in momentum
space, which corresponds in ${\bf r}$-space to a delta function in
${(\bf r-r')}$. The next order, $l=1$, is the derivative of a
delta function, which introduces a momentum dependence into the
particle-hole  (ph)
interaction. In the application of TFFS to nuclei nearly all
calculations have been performed with only the $l=0$ component of
the interaction, which corresponds in ${\bf r}$-space to the
following form of the Landau-Migdal interaction:
\begin{eqnarray}                                                    
F({\bf r},{\bf r'}) = C_{0}[f(r) + f'( r) {\bf {\bftau}_{1}}
\cdot{\bf {\bftau}_{2}} + (g + g' {\bf {\bftau}_{1}} \cdot {\bf
{\bftau}_{2}} ) {\bf {\bfsigma}_{1}} \cdot {\bf {\bfsigma}_{2}}
] {\delta} ({\bf r} - {\bf r'}),
\end{eqnarray}

In {\it finite}  nuclei,  one has to introduce density-dependent     
parameters because it is obvious that the interaction inside a
nucleus is  different from the interaction in the outer region of
the nucleus. The most often used ansatz is                            
\begin{equation}
f(r) = f_{ex} + (f_{in} - f_{ex})\rho_{0}(r)/\rho_{0}(0)
\end{equation}
and similarly for the other parameters. Here $\rho_{0}(r)$ is the
density distribution  in the ground state of the nucleus under
consideration and $f_{in}$ and $f_{ex}$ are the parameters inside
and outside of the nucleus. In actual calculations it turned out
that $g$ and $g'$ depend only weakly on the density, so that one     
uses the same parameters inside and outside. This density
dependence of the Landau-Migdal interaction is the basic reason
for its success and universal applicability.

\subsubsection{General theoretical background}
Most of the results discussed so far were obtained within the        
phenomenological version of the theory of  Fermi liquids. In the
third article of ref.~\cite{Landau} Landau gave a microscopic         
justification of his original phenomenological theory using the GF
technique. There he derived in a fully microscopic way the basic
equations of his theory. He showed, in particular, that the
scattering amplitude of two quasiparticles is connected with the
response function $R$ (and two-particle GF) in the                  
ph channel and that the quantity $f({\bf k,k'})$ is connected
with the forward scattering amplitude of two quasiparticles at the
Fermi surface.

The starting point of the TFFS are the equations for the energies
and transition amplitudes of excited states in even-even nuclei
and the equations for moments and transitions in odd mass nuclei.
As in Landau's theory, these equations can be also derived using    
the GF technique.

The one-particle GF is defined by
\begin{equation}
G_{12}(t_{1},t_{2}) = -\imath\langle
A0|T{a_{1}(t_{1})a^{+}_{2}(t_{2})}|A0\rangle
\end{equation}
and the two-particle GF by
\begin{equation}
K_{1234}(t_{1},t_{2};t_{3},t_{4}) =
\langle
A0|Ta_{1}(t_{1})a_{2}(t_{2})a^{+}_{3}(t_{3})a^{+}_{4}(t_{4})|A0\rangle.
\end{equation}
Here $a$ and $a^{+}$ denote time-dependent annihilation and creation  
operators, $|A0\rangle$ is the exact eigenfunction of the ground
state of an $A$-particle system and $T$ is the time-ordering
operator. In the following we use the single-particle states
$\varphi_{\lambda}({\bf r},s)$ of a nuclear shell model as basis
states; therefore the subscripts 1...4  stand for a set of          
single-particle quantum numbers.

We also need the response function $R$, which is defined as         
\begin{equation}
R(12,34) = K(23,14) - G(21)G(34),
\end{equation}
where the indices denote space and time coordinates.

The calculation of the excitation spectra of even-even nuclei and   
their transition probabilities can often be reduced to the
calculation of the  strength function, which describes the
distribution of the transition strength in a  nucleus induced by
an external field $V^{0}(\omega)$

\begin{equation}
S(\omega) = \sum_{n \neq 0} |\langle
An|V^{0}|A0\rangle|^{2}\delta(\omega - \omega_{n}),
\end{equation}
where $\omega_{n} = E_{n} - E_{0}$ is the excitation energy, while
$|An\rangle$  and $|A0\rangle$ refer to the excited and ground
states of a nucleus with mass number $A$, respectively. As the     
response function  $R(\omega)$ in the energy representation has
the spectral expansion

\begin{equation}
 R_{12,34}(\omega) = \int^{\infty}_{-\infty} \frac{d\epsilon}{2\pi\imath}R_{12,34}(\omega,\epsilon)
                  = \sum_{n} \left(\frac{\phi^{n0*}_{21}\phi^{n0}_{43}}{\omega + \omega_{n} - \imath\delta}
                   - \frac{\phi^{n0}_{12}\phi^{n0*}_{34}}{\omega - \omega_{n} +
                   \imath\delta}\right),
\end{equation}
where

\begin{equation}
\phi^{n0}_{12}= \langle An | a_{1}^+ a_{2} | A0 \rangle
\end{equation}
is the transition amplitude between the ground state and the
excited state $n$, the strength function $S(\omega)$ is completely
determined by the response function $R(\omega$):

\begin{equation}
S(\omega) = \frac{1}{\pi} \lim_{\Delta\rightarrow+0} \; {\rm Im}
\; \sum_{1234} V^{0\ast}_{21}R_{12,34}
       (\omega+\imath\Delta)V^{0}_{43}.
\end{equation}

This response function is defined by the Bethe--Salpeter equation
in the ph channel \cite{AB}:
\begin{eqnarray}
R_{12,34}(\omega,\epsilon) = -G_{31}(\epsilon + \omega)G_{24}(\epsilon)   + \hspace{2.0in}\nonumber\\
    \sum_{5678} G_{51}(\epsilon + \omega)G_{26}(\epsilon)
            \int^{\infty}_{-\infty}\frac{d\epsilon'}{2\pi\imath} U_{56,78}(\omega,\epsilon,\epsilon')
             R_{78,34}(\omega,\epsilon'),
\label{eq:1}
\end{eqnarray}
which is shown graphically in Fig.~1.1.                              

In Eq.~(1.19) $G$ is the exact one-particle GF, and $U$ is the
irreducible amplitude in the ph channel, which is the
unrenormalized ph interaction. (For details see
ref.~\cite{KTs86}.) The
one-particle GF  $G$ and the response $R$ are related by a system    
of nonlinear equations  \cite{AB,agd}. In particular, this system
of equations includes the relation
\begin{equation}
 U(12,34) = \imath\frac{\delta\Sigma(3,4)}{\delta G(1,2)},
\end{equation}
where $\Sigma$ is the so-called  mass operator
related to the one-particle GF by the Dyson equation
\begin{equation}
G_{12}(\epsilon) = G^{0}_{12}(\epsilon) + \sum_{34}
G^{0}_{13}(\epsilon) \Sigma_{34}(\epsilon)
                   G_{42}(\epsilon)
\end{equation}
and G$^{0}_{12}(\epsilon) = ((\epsilon - {\bf
p}^{2}/2m)^{-1})_{12}$ is the one-particle GF of a free particle.

\subsubsection[Microscopic derivation of the basic eqns of the TFFS]{Microscopic derivation of the basic equations of the TFFS}
Here we will briefly describe the derivation of the basic             
equations of the standard TFFS in such a way as to clarify the
explicit inclusion of complex configurations to be discussed in
section~2.  For a more detailed derivation see                      
refs.~\cite{AB,SWW77}.

As mentioned before, the GF $G$ and the response function $R$ are     
determined self-consistently by a system of non-linear equations.
This is in principle an exact formulation of the
(non-relativistic) $A$-particle problem, but is of little use for     
practical applications.

In order to arrive at solvable equations, one applies Landau's
quasiparticle concept and his renormalization procedure, which he
developed for the microscopic theory of Fermi liquids
\cite{Landau}. For  the nuclear many-body problem, it was done by     
Migdal \cite{AB} in his TFFS.

As a first step, one splits the one-particle GF into a               
quasiparticle pole part which is diagonal in the shell model basis
and a remainder,
\begin{equation}
G_{12}(\epsilon) = \delta_{12}a_{1}(\frac{1-n_{1}}{\epsilon -
\epsilon_{1} + \imath\delta}
                     + \frac{n_{1}}{\epsilon - \epsilon_{1} - \imath\delta}) +
                      G^{r}_{12}(\epsilon).
\end{equation}
Here $a_{1}$ denotes the single--particle strength of the shell
model pole, $\epsilon_{1}$ is the single-particle energy, $n$ is
the quasiparticle occupation number ($1$ or $0$) and
$G^{r}(\epsilon)$ is that part of the exact one-particle GF that    
remains when the shell model pole has been removed. It is assumed
in Migdal's TFFS that $G^{r}(\epsilon)$ is a smooth function of
$\epsilon$ in the vicinity of $\epsilon_{F}$. The dominance of       
the first term in Eq.~(1.22) with                                     
respect to the $\epsilon$--dependence of the one-particle GF is
one of the basic assumptions on which the TFFS rests. The validity
of this assumption is crucial for the reliability of the results
of all TFFS calculations. As we will see in the following, there
might be additional pole terms in the expansion of Eq.~(1.22) that    
originate from more complex configurations and which have to be
considered in addition to the simple shell model poles. This
extension of the theory will be the main topic of this review.

The main goal in the following is to obtain an equation for the      
response function that can be solved in practice because
$R(\omega)$ contains all the information we need. In Eq.~(1.19)
the product of two one-particle Green functions enters. Here it is
important to realize that with the quasiparticle ansatz in
Eq.~(1.22) this product can be separated at low transferred energies  
into a part $A$, which depends strongly on the energy $\omega$
and has a $\delta$--function maximum with respect to $\epsilon$       
for $\epsilon_{1}$ approaching $\epsilon_{2}$ (that is, near          
$\epsilon_{F} $)
 and a weakly energy--dependent part $B$,                             
\begin{equation}
 - G(\epsilon+\omega)G(\epsilon)=
 A(\epsilon,\omega)+B(\epsilon,\omega).
 \end{equation}
 $A$ is given by
\begin{equation}
A_{12,34}(\epsilon,\omega)=2\pi\imath                                 
 a_{1}a_{2}\delta_{13}\delta_{24}\frac{n_{2}-n_{1}}                   
{\epsilon_{1}-\epsilon_{2}-\omega}
\delta(\epsilon-
\frac{\epsilon_{1}+\epsilon_{2}}{2})                                 
\end{equation}
whereas $B$, which contains all the rest, does not give rise to a    
pronounced $\omega$--dependence.

We now insert Eq.~(1.24) into the equation for the response
function and integrate over $\epsilon$ to obtain in compact
notation                                                            
\begin{equation}
R(\omega) = (A + B) - (A + B)UR(\omega),                            
\end{equation}
where $A$ is the shell-model ph-propagator:                         
\begin{equation}
A_{12,34}(\omega)=
 \int^{\infty}_{-\infty}\frac{d\epsilon}{2\pi\imath} A_{12,34}(\epsilon,\omega) =
a_{1}a_{2}\delta_{13}\delta_{24}\frac{n_{2}-n_{1}}{\epsilon_{1}-\epsilon_{2}-\omega}.
\end{equation}

With the help of Landau's renormalization procedure one can        
rewrite Eq.~(1.25) in such a way that only the known $A$ appears
explicitly in the equation, whereas the unknown $B$ changes $U$
into the renormalized ph interaction $F$ and gives rise to
effective charges for the external fields.  We introduce a
renormalized  response function $\tilde{R}(\omega)$, which is
connected with the original response function $R(\omega)$ by the
relation
\begin{equation}
R = e_{q}\tilde{R}e_{q} + Be_{q},
\end{equation}
where
\begin{equation}
e_{q} = 1 - FB
\end{equation}
 and the renormalized ph interaction $F$ satisfies the integral equation
\begin{equation}
F = U - UBF.                                                              
\end{equation}

A detailed investigation of these equations shows that $F$ depends
smoothly on the energy variables. For that reason the energy         
dependence is neglected when $F$ is parameterized. The equation
for $\tilde{R}(\omega)$ is given by
\begin{equation}
\tilde{R} = A -AF\tilde{R}
\end{equation}
and the explicit form is
\begin{equation}
\tilde{R}_{12,34}(\omega) =
\frac{n_{2}-n_{1}}{\epsilon_{1}-\epsilon_{2}-\omega}\delta_{13}\delta_{24}-            
\frac{n_{2}-n_{1}}{\epsilon_{1}-\epsilon_{2}-\omega}\sum_{56}F_{12,56}\tilde{R}_{56,34}.
\end{equation}
This is the basic equation of the TFFS.

Due to the renormalization procedure only the 1p1h configurations   
appear explicitly in the final equation, whereas all the more
complex configurations give rise to renormalized quantities: the
ph-interaction $F$ and the effective charges $e_{q}$. These
quantities are not calculated within the theoretical framework but
are parameterized in the way that was discussed before for the
ph-interaction.                                                        
Due to conservation laws the electric proton and neutron operators
have $e_{q}^{p}=1$, $e_{q}^{n}= 0$, respectively, and only
spin-dependent operators have to be renormalized. In all practical
cases the second term on the right side of Eq.~(1.27) does not
contribute, so it is sufficient to solve Eq.~(1.31). Here all         
quantities are known. The single-particle energies are given by
the nuclear shell model or taken, as far as possible, from
experiment and the parameters of the interaction and the effective
operators have been deduced from experimental data.

Equation (1.31) has exactly the same form as the conventional RPA    
equation that, however, has been derived using approximations from
the outset. In the present derivation, Eqs. (1.27) and (1.28) are
still exact. For that reason one is able to obtain relations (the
Ward identities) between the effective operators and the effective
interaction. In cases where conservation laws exist these
relations determine the effective operators completely. In
addition, the present derivation shows more clearly the range of
validity of the theory, which naturally also applies to the
conventional RPA equation. Moreover, the GF formalism provides a
natural basis for an extension of the theory, as we will see
in the next section.                                                 

For practical reasons one solves not Eq.~(1.31), but the related       
equation for the change of the density matrix $\rho_{12}(\omega)$     
 in the external
field $V^{0}(\omega)$, which is defined as

\begin{equation}
 \rho_{12} (\omega) = -\sum_{34} \tilde{R}_{12,34}(\omega)e_{q}V^{0}_{43}(\omega).
\end{equation}
The equation for $\rho_{12}$ follows from Eq.~(1.31) and has the
form
\begin{equation}
\rho_{12}(\omega) = -\sum_{34} A_{12,34} e_{q} V^{0}_{43} -
\sum_{3456} A_{12,34} F_{34,56} \rho_{56}(\omega).
\end{equation}
 The expression for the strength function is then given by
\begin{equation}
S(\omega,\Delta) = -\frac{1}{\pi}
 \; {\rm Im} \;\sum_{12}e_{q}V^{0*}_{21}\rho_{12}(\omega+\imath\Delta),
\end{equation}
where $\Delta$ is a (finite) smearing parameter which simulates a finite       
experimental resolution and at the same time phenomenologically can include    
configurations  not dealt with explicitly in the approach under consideration. 

Equations (1.33) and (1.34) are the  main equations that are used   
in the calculations within the TFFS.

\subsubsection{Coordinate representation}

Most of the calculation that we will present here have been          
performed in ${\bf r}$-space and not in the configuration space of
a shell model basis. For the cases of RPA and TFFS the method  was
suggested in refs.~\cite{sb75,sfk76} and was included by us in the
ETFFS \cite{kstt93}. The main reason for this choice is that the        
${\bf r}$-space representation is much more appropriate for the
treatment of the single-particle continuum, as first pointed out
in refs.~\cite{sb75,sfk76}. Therefore we give here some relevant      
equations in the coordinate representation.

Equation(1.33) has the form
\begin{eqnarray}
\rho({\bf r},\omega) = -\int A ({\bf r}, {\bf r'},\omega)
{e_{q}} {V^{0}} ({\bf r',\omega})
d^{3}r'- \hspace{1.0in}\\
\int A ({\bf r}, {\bf r_{1}},\omega) F ({\bf r_{1},r_{2}}) {\rho
({\bf r_{2}},\omega)} d^{3}r_{1}d^{3}r_{2}. \nonumber
 \label{eq:35}
\end{eqnarray}
   The ph propagator $A$, given by
\begin{equation}
A({\bf r, r'},\omega) = \sum_{12} \frac{n_{2} -
n_{1}}{\epsilon_{1} - \epsilon_{2} - \omega}\varphi_{1}^{*}({\bf
r})\varphi_{2}({\bf r'})
 \varphi_{2}^{*}({\bf r})\varphi_{1}({\bf r'}),
\end{equation}
can be rewritten as
\begin{equation}
A({\bf r, r'},\omega) = - \sum_{1}n_{1}\varphi_{1}^{*}({\bf r})
\varphi_{1}({\bf r'})[G({\bf r',r};\epsilon_{1}+\omega)+                        
                                                G({\bf r',r};\epsilon_{1}-\omega)]
\label{eq:37}
\end{equation}
using the formula for the one-particle GF
\begin{equation}
G({\bf r,r'};\epsilon) = \sum_{2}\frac{\varphi_{2}({\bf                        
r})\varphi_{2}^{*}({\bf r'})} {\epsilon - \epsilon_{2}}, \label{eq:38}         
\end{equation}
where $\varphi_{2}({\bf r})$ are the single-particle wave
functions calculated in a mean-field potential.               

The summation in Eq.~(\ref{eq:37}) is over states ${\it below}$ the  
Fermi surface, i.e. the single-particle continuum is already
contained in Eq.~(1.38).
On the other hand, the coordinate part of this GF                    
$G_{lj}=g_{lj}/rr'$  can be  expressed in closed form in terms of
the regular $y^{(1)}_{lj}$ and irregular $y^{(2)}_{lj}$ solutions
of the one-dimensional Schr$\ddot o$dinger equation as
\begin{equation}
g_{lj}(r,r';\epsilon) = \frac{2m}{\hbar^{2}} y^{(1)}_{lj}(r_{<};
\epsilon)y^{(2)}_{lj}(r_{>};\epsilon) /W_{lj}(\epsilon),
\end{equation}
where $r_{<}$ and $r_{>}$ denote the lesser and the greater of $r$   
and $r'$, respectively and $W$ is the Wronskian of the two
solutions. The irregular solution  $y^{(2)}_{lj}$ is determined by
the boundary conditions at $\infty$; e.g., for neutrons              
\begin{equation}
y^{(2)}_{lj}(r \to \infty) \sim exp(-kr) \nonumber
\end{equation}
for negative energies $\epsilon < 0$ and
\begin{equation}
 y^{(2)}_{lj}(r \to \infty) \sim exp[\imath (kr - \frac{\pi l}{2} + \delta_{lj})]
\end{equation}
for positive energies $\epsilon > 0$, where $k =
\sqrt{2m|\epsilon|}/{\hbar}$ and $\delta_{lj}$ is the   scattering
phase for the mean  nuclear potential considered.

Thus, the functions $y_{lj}^{(1,2)}$ are calculated numerically if
the mean potential is known. For $\epsilon < 0$ the functions
$g_{lj}$ have no imaginary part; that is, the 1p1h states have      
automatically no width if the smearing parameter $\Delta = 0$.

Inclusion of the single-particle continuum makes it possible to     
obtain a physical envelope of the resonance without using a
smearing parameter, that is, to obtain directly the escape width
$\Gamma^{\uparrow}$.

In the complex configuration problem under consideration, using     
the representation of the single-particle wave functions
($\lambda$-representation) gives matrices of a very large rank
especially, in the case of treatment of the ground state
correlations caused by complex configurations. Using the
coordinate representation affords a big numerical advantage in      
this problem because the rank of matrices is determined not by the
number of configurations but by the number of mesh points used in
solving the corresponding integral equation.

\subsubsection{Main results of the GMR description within the TFFS}

In order to distinguish GMR from other collective excitations,     
one can define them as follows:
\begin{enumerate}
\item The form and the width $\Gamma$ of the resonance depend
rather weakly on $A$; as a rule, the dependence $\Gamma \sim
A^{-2/3}$ is used.
\item The resonance mean energy $E$ also depends weakly
on $A$; usually one uses $E \sim A^{-1/3}$.
\item The resonance width is small compared with its excitation energy.
\item The resonance exhausts a large fraction of its energy weighted sum
rule (EWSR)---usually more than 50 percent.
\end{enumerate}
The last of these is the most quantitative characteristic of the
resonance and justifies its name ``giant''.

In 1971-72 in inelastic electron \cite{pit71,fukuda72} and proton  
\cite{lewis72} scattering, giant multipole resonances (GMR) were
detected that were different from the well-known isovector
electric dipole resonance. That was the starting point of  a
period, sometimes called a renaissance of giant resonance physics,
of very rapid and intensive development. A large amount of
experimental data on the GMR in stable 
nuclei---principally their energies, total strengths, widths and
resonance envelopes---has been accumulated  and discussed since
the mid-seventies.  Currently there exists experimental
information on more than 20 different types of GMR that were
detected in a large number of nuclei in a broad range of
excitation energies. A detailed review, with experimental results
up to the end of the eighties, is presented in ref.~\cite{speth}.
Further information can be found in the proceedings of the last
four international conferences devoted to giant resonances
\cite{gr1,gr2,gr3,gr4}, and in a recent monograph \cite{HW} that
gives an excellent review of the present experimental situation.
Conventional theoretical methods such as the RPA and QRPA, and
their comparison with the data, have been the subject of many
review articles and books (e.g.,
refs.~\cite{HW,vg92,speth,bohrmot} and
\cite{kaev97,ktts97,ost91,drozdz90,bbb83,SWW77,Kaev79}). For
nuclei with $A\,>\,40$, the experimental situation is essentially
settled. There is no longer any major controversy over the
centroid energies and the total strengths of GMR, and the
theoretical interpretation of these data within the framework of
the conventional theoretical methods is also clear. These methods,
however, do not allow description of the widths and the fine
structure of the resonances, nor do they offer any possibility to
analyze complex spectra with overlapping resonances. This is the
main subject of our review and will be presented in the following
sections.

In the following we illustrate and briefly discuss typical results
 that have been                                                           
obtained within the continuum TFFS (CTFFS), which is, as mentioned
before, formally identical to the continuum RPA (CRPA).  We show
in Figs.~1.2-1.4 and Table~1.1 the hadron  strength functions for
the E2 isoscalar (IS) and isovector (IV) resonances in $^{208}$Pb
and the E2 IS resonance in $^{40}$Ca and some of their integral
characteristics. The calculations have been performed in the
coordinate representation within the CTFFS, i.e., using
Eqs.~(1.35),
 (1.37) and (1.39). In order to simulate the
finite experimental resolution, we introduced a smearing parameter
$\Delta$ with a value of 250 keV. As the TFFF is not
self--consistent, as discuss in the beginning, we have to
determine the parameters of the effective interaction, Eqs.~(1.10)
and (1.11), from experiment by fitting some specific theoretical
results to experimental data. In our calculations we always used,
with the exception of $f_{ex}$,  the following Landau-Migdal
interaction parameters, which were
adjusted previously to various experimental quantities ~\cite{fayns88,borzov84}:          
\begin{equation}
\begin{array}{rcl}
f_{in} & = & - 0.002 \nonumber \\
f_{ex}^{\prime} & = & 2.30  \nonumber \\
f_{in}^{\prime} & = & 0.76  \nonumber \\
g & = &  0.05 \nonumber \\
g^{\prime} & = & 0.96  \nonumber \\
C_{0} & = & 300\;{\rm MeVfm^{3}}. \label{eq:41}                                 
\end{array}
\end{equation}

For the nuclear density $\rho_{0}(r)$ in the interpolation formula
(1.11) we chose the theoretical  ground state density distribution
of the corresponding nucleus,
\begin{equation}
\rho_{0}(r) = \sum_{\epsilon_{i} \leq \epsilon_{F}} \frac{1}{4\pi}
(2j_{i} + 1)
           R^{2}_{i} (r),
\label{eq:42}
\end{equation}
where $R_{i}(r)$ are the single-particle radial wave functions         
of the particular Woods-Saxon potential used. For other details of
the calculations, including the definitions used in Table~1.1, see
section~3.1.3

For the parameter $f_{ex}$ we have used the values $f_{ex} = -1.9$  
and -2.2 for $^{208}$Pb and $^{40}$Ca, respectively. These
parameters were adjusted to reproduce the energies of the first
excited $2_{1}^{+}$ level in $^{208}$Pb and $3_{1}^{-}$ level in
$^{40}$Ca. We will see in sections 3 and 4  that the same
parameters (\ref{eq:41}) can also be used if more complex
configurations are considered explicitly. Even the parameter
$f_{ex}$ changes only slightly in the latter case.
We have used these parameters in all our calculations for stable
and unstable closed shell nuclei from $^{16}$O  to  $^{208}$Pb,
where we investigated many different types of GMR. In our opinion,
the reasonable agreement with experiment we obtained confirms the
assumed universality \cite{AB} of the parameters of the
Landau-Migdal interaction, Eq.~(1.10).

In Figs.~1.2-1.4 the CTFFS results are given together with the     
results of the calculations without taking the effective
interaction into account (``free response''). As the parameters
$f_{in}, f_{ex}$ ($f'_{in}, f'_{ex}$) for the isoscalar
(isovector) resonance are negative (positive), the IS (IV)
resonances are shifted to lower (higher) energies, when the
interaction is included, compared with the free response. For the
latter we have the  shell model estimate for $^{208}$Pb $E \simeq
2\times41A^{-1/3}$ = 13.8~MeV while, according to Table~1.1,
$\overline{E}_{is}$ = 8.1~MeV, $\overline{E}_{iv}$ = 18.2~MeV  or,
if one uses another definition for $\overline{E} = E_{2,0}$ (see
section~3.1.3), $\overline{E}_{is}$ = 10.2~MeV,
$\overline{E}_{iv}$ = 19.3~MeV. In both cases---with and without
interaction---reasonable depletion values (90--100 percent for
large energy intervals) of the corresponding EWSR have been
obtained (see Table~1.1). The depletion is in satisfactory
agreement with the corresponding experimental values of the EWSR.
(See sections 3 and 4 and, in particular, Tables~3.3 and 3.7.) In
Table~1.1 we give also the experimental values of the mean
energies. It should be noted, however, that the experimental data
were obtained, as a rule,  for  intervals which are  much smaller
than those given in Table~1.1, not to mention that the
experimental mean energies may be determined in different ways
that may be important for unstable nuclei for which taking           
the single-particle continuum into account is necessary.               
Therefore the comparison with experiment should be made more
carefully, and this will be done in section~3, where we will also   
discuss the width question.

Thus, we have obtained a relatively good description of the mean  
energies and total strengths of the GMR under consideration. These
are typical results of the CRPA.

In Figs.~1.2-1.4 we have chosen the smearing parameter noticeably
larger than the experimental resolution in order to simulate at
least some of the decay width not included in the present
approach.
Nevertheless, one cannot see in Figs.~1.2-1.4 any                   
resemblance to an observed resonance
 because the smallest width   %
among the three resonances under discussion is the one for the E2
IS resonance in $^{208}$Pb, which has an experimental width of
3.1~$\pm$~0.3~MeV \cite{Bra}. In other words, the widths of
resonances, which are among the most important characteristics,
are not reproduced within the CRPA. Even in medium--mass nuclei,
where the role of the continuum (escape width)is much larger, the
theoretical widths are still in disagreement with the experimental
ones. The reason is well known: the one particle--one hole
configurations describe only the escape widths, which---in
general---are only a small fraction of the total widths. For a
realistic description one has to include the spreading widths,
which are due to the coupling of the one particle--one hole
configurations to more complicated configurations. One possible
solution to this problem will be discussed next.

\subsection{Physical arguments for extending the standard approach}
We have seen that the Landau-Migdal theory is based an a
microscopic many--body theory with, however, important elements
taken from experiment. For that reason it is quite natural that an
extension of that successful theoretical frame work is also based
on experimental facts. It is well known from nuclear spectroscopy
that in odd--mass nuclei that differ from a magic nucleus by one
nucleon (or hole), the coupling of the low--lying phonons of the
even nucleus to the odd particle (or hole) play an important role.
It gives rise to a strong fragmentation of the corresponding
single-particle (hole) strength over a range of the phonon energy.
This observation gives us the possibility to include in the
standard TFFS the coupling of the 1p1h states to more complex
configurations. We will show that with this coupling to the
low--lying phonons one includes those 2p2h configurations that
give rise to the strongest fragmentation of the giant resonances,
which---together with the coupling to the single-particle
continuum---makes it possible to calculate quantitatively the
strength distribution of the GMR.

\subsubsection{Giant resonances in  nuclei: the width problem} 
It is clear now that GMR are a universal property of nuclei. The
investigations of GMR are not only important for a detailed
understanding of the structure of nuclei, but they are also an
important tool for a better understanding of nuclear reaction
mechanisms involved in the excitation of the different types of
these resonances. Moreover, we obtain from the investigation of
GMR additional information on the Landau parameters. The most
important of these is $f_{in}$, which is connected with the
compressibility of nuclear matter and is therefore of crucial
importance in astro-physics. The isoscalar electric monopole
resonance (breathing mode), on the other hand, is closely related
to this parameter and therefore one would need to know this
resonance in nuclei far from stability in order to obtain the
dependence of the compression modulus on the number of protons and
neutrons. The magnetic resonances are related to the parameters
$g$ and $g'$ of the spin- and isospin-dependent parts of the
forces.  The latter is related to pions in nuclei and is of
special interest in connection with the possibility of pion
condensation.  Phenomenologically, GMR  inform us about                         
the nuclear shape (splitting of the E1, E2and E0 resonances in
deformed nuclei), and about volume, surface and other kinds of
vibrations.

The understanding of the widths of GMR is obviously connected with
the damping of small amplitude vibrations in finite systems, as we
shall soon discuss. Thus the general problem of how energy from
highly ordered excitations is dissipated in nuclei, including the
question of transition from order to chaos, can be clarified through GMR studies      
\cite{DrozdSpeth94}.

Many ideas from GMR physics have been used in other applications,
such as the recent investigations of metallic clusters
\cite{metcl} and the fullerene molecules \cite{fullerene}.  See
also ref.~\cite{HW}, chapt. 11.                                     

  \begin{center}
          {\it Necessity of inclusion of complex configurations\\
                  and single-particle continuum}
  \end{center}

As discussed in section~1.2.5, the standard continuum TFFS
or the continuum RPA in closed shell 
nuclei  are  able to describe only two
integral characteristics of GMR: their mean energies and  
total strengths. The quantity that is of equal importance, the
strength distribution of the GMR (i.e., their widths) cannot be
reproduced within this approximation. The reason for this failure
has been already been indicated above: the coupling of the 1p1h
configurations to more complex configurations, which gives rise to
the spreading width, has so far been neglected. The escape width,
which is included in the present continuum approaches, represents
in the giant resonance region only the lesser part of the total
width. There are several reasons why one would like do describe
theoretically the widths of the GMR quantitatively:
\begin{enumerate}
\item the intellectual challenge to develop a microscopic theory that
gives a quantitative explanation for the collective motion in
strongly interacting finite Fermi systems;
\item the new insight into the fine structure of the GMR due to the
rapid improvement of the experimental resolution to $\Delta E<$ 10
keV, which needs to be understood (see, for example,
ref.~\cite{richter93});                                                  
\item the need for microscopically derived strength distributions
that quantitatively reproduce the date on resonances in the medium
mass region, where the various multipole resonances overlap in
energy. The conventional analyses with phenomenological transition
densities are no longer applicable because they introduce strong
uncertainties, as we shall discuss in the following sections.
\end{enumerate}

  The inclusion of the single--particle continuum gives the physical
envelope for processes at excitation energies higher than the
nucleon separation energy. For giant resonances this gives the
escape width $\Gamma^{\uparrow}$, the magnitude of which depends
significantly on the mass number, the excitation   
energy, the multipolarity, etc., so that it is not justified,
especially in the calculations of such delicate properties as fine
structure and
decay characteristics, to simulate the role of the continuum by  
a constant smearing parameter. The exact microscopic treatment of
the continuum is therefore crucial to a realistic theory of giant
resonances.

 In addition to this, a realistic microscopic theory of
collective motion in nuclei has also to consider more complex
configurations than those included in the RPA.
There are new data, e.g. \cite{kneissl96,hartmann2002,aumann2001},    
  and the largely unsatisfactory explanation
of the older results concerning the low-lying structures in cross
sections in a wide excitation energy range around the nucleon
binding energy \cite{laszaxel79,dol85,igashira86,belyaev92,
624r7,hofstee95}, see also section 3.2.5.
 An extended theory
will also have implications for the interpretation of experimental
data obtained with modern germanium detectors and gamma
spectrometers such as EUROBALL cluster, EUROBALL and others
\cite{euroball,euroball2001}. Unprecedentedly high resolution and
high efficiency of detecting gamma rays with energies up to 20~MeV
have already given new and very precise information, not only on
deformed nuclei, but also on low-lying levels in odd and even-even
spherical nuclei. In fact, these detectors give direct information
about the low-lying complex configurations containing phonons
\cite{euroball,euroball2001,covello98,reif97,ponomarev2000},
which may be seen again in the fine structure of the giant
resonances. At last, in order to explain the large amount of         
available data on the decay properties of GMR's gained
in experiments with coincidences of secondary particles
\cite{knopflewagner91,HW} it is also necessary to take complex
configurations into account (see, for example, refs. \cite{colo92,colo94}).

It is clear that in the immediate future the number of such data
will increase rapidly and that these results require
improved microscopic approaches for their interpretation.                                 

\begin{center}
{\it Ground state correlations caused by complex configurations}
\end{center}

The ground state correlations (GSC) problem has a long history.    
(See, for example, references in \cite{ringschuck,karvorcat93} and
also the article \cite{kaevtk82}.) It started with the
Hartree-Fock approximation (HF), where the effects of the Pauli
principle was included in the calculation of the ground state of
fermion systems. Some specific ground state correlations are taken
into account if one calculates excited states within the RPA. The
most important consequence of these ground state correlations is
that the energy-weighted sum rule for the transition probabilities
is conserved. This is not so in the Tamm-Dankoff approximation,
which starts from the uncorrelated HF ground state. During the
past ten years, in connection with the development of the extended
TFFS, where configurations beyond the 1p1h states are included in
the excited states, the question arose as to how far one has also
to
consider the same configurations in the ground state.              
\cite{karvorcat93,ks96,kst97} There is a fundamental difference
between the effects of GSC in the RPA and
their effects in models where more complex configurations are included   
explicitly. The RPA GSC do not lead to the appearance of new
transitions compared with the TDA, but only shift the energies and
redistribute the transition strengths.

The GSC induced by the more complex configurations, on the other
hand,
 lead not only to a redistribution of strength but also to
 new transitions, which give rise to a change of the EWSR
\cite{adachilip88,kst97}. Thus these GSC are at least as important
and physically interesting as the RPA GSC.  Actually, their
consequences are much richer and far-reaching. The present
approach is an extension of the previously developed Extended
Second RPA (see, for example refs.\cite{adachilip88,drozdz90}),           
 in which uncorrelated 2p2h GSC have been considered.

We will see in the following that these effects play a noticeable---sometimes        
decisive---role in the theoretical description of the experimental
data. The most striking example obtained within the GF approach is
the  explanation \cite{kt84,kt89} of the observed M1 excitations
in $^{40}$Ca and $^{16}$O with energies of about 10~MeV and
16~MeV, respectively, solely as a result of ground state
correlations.

\subsubsection{Nuclei far from the stability line}

There is increasing interest in the structure of nuclei far from        
the stability line. The study of these exotic nuclei, is of
importance not only in itself
\cite{tanihata2001,inpc2001,richter93}, but also for its relevance
to astrophysics
\cite{tanihata2001,inpc2001,Thielemann91,Langanke2000,richter93}.
The ETFFS we are discussing here may play an important
role in the analysis of the experiments done at the proposed
radioactive beam facilities. As mentioned earlier, one of the
crucial quantities one wants to know is the breathing mode in
nuclei with very different numbers of protons and neutrons. This
will give us the compression modulus as a function of the proton
and neutron number, which is needed for the extrapolation to
nuclear matter. We may suppose that in nuclei far from stability,
even with closed shells, the high-lying spectra may be as
complicated as in the medium mass nuclei, where the various
multipole resonances overlap and a microscopic theory is necessary
for the analysis in order to obtain reliable nuclear structure
information.

  If we extrapolate our present knowledge of unstable
nuclei to nuclei far away from the stability line we may expect
two characteristic features: (i) there will be
 very
low-lying collective states and (ii) the nucleon separation energy
may also be relatively low. For these reasons a realistic theory
has to treat the continuum in an exact way, and the phonon
coupling is not only important for the analysis of GMR but also
for a quantitative understanding of the low-lying spectrum. In
connection with the application of the present approach, one has
to investigate the extent to which the Landau-Migdal parameters
are dependent on the numbers of protons and neutrons. Our
extensive experience indicates that this dependence may be quite
weak, so we can at least use the present parameters as a good
starting point.  The second important input into the theory
concerns the single--particle spectrum and the single--particle
wave functions.  Here one may use self-consistent approaches,       
 for example, those with the density functional, in order to        
obtain a reliable quasiparticle basis. Investigations in these
directions are in progress.


It should be emphasized that only a reliable inclusion of the           
single-particle continuum can make it possible to do calculations for
nuclei with the separation energy near zero.This is important for
understanding drip-line nuclei and for astrophysical sdudies.
  For neutron-reach nuclei with the separation energy near zero,
this is of prime interest because of the absence of the Coulomb barrier.
 The CRPA calculations in $^{28}$O  have shown that the
strength distributions of the E2 \cite{hamamotosagawa98} and
isovector E1 \cite{vitturi98} resonances are very different from
those for $^{16}$O; the resonances are more spread out, shifted
down and have a noticeable low-lying  strengths.  The effect of
complex configurations is also noticeable, at least for the
isoscalar E2 resonance in $^{28}$O \cite{ghielmetti96}.  It should
be pointed out, however, that except for ref.~\cite{ghielmetti96}
and the calculations we shall present in section~3.3, there exists
almost no theoretical information about  the role of complex
configurations in unstable nuclei \cite{HW}.

\subsubsection[Implications of experiments for and status
of the theory]{Implications of experiments for and current status
of the microscopic theory}                                           

We can summarize our discussion so far by asserting that a
microscopic theory that is able to describe quantitatively the
structure of collective excitations in nuclei
and which is based on the mean-field approximation has to include
four major effects:
\begin{enumerate}
\item the 1p1h RPA, which creates collectivity out
of the uncorrelated ph states, as a starting point;
\item complex configurations beyond the 1p1h states, which give
rise to a fragmentation of the collective states derived from the
RPA;
\item the single-particle continuum;
\item ground state correlations induced by the complex configurations
under consideration.
\end{enumerate}
In addition, one should not use separable forces, because one then
must use different forces for each mulipolarity, which strongly
reduces the predictive power of the theory. Indeed one needs an
interaction that is universal for the whole periodic table, or at
least that changes only very little with the mass number, and
which should be adjusted to  quantities other than those that one
is going to calculate. As we shall see,
the GF approach that we are going to discuss in the what                
follows allows the inclusion of all these effects simultaneously.

It is obvious that, compared with the simple  1p1h     
configuration problem, the present task is much more
difficult---both theoretically and numerically. In addition we
shall develop and apply various stages of sophistication of our
theory to the nuclear structure problem in order to clarify the
different effects. At present there exist several other approaches
that have considered some of the effects mentioned above. These
are reviewed in refs.~\cite{sw91,drozdz90} (``pure'' 2p2h
configurations) and \cite{vg92,bbb83,ktts97} (configurations with
phonons).

In the past, microscopic theories of GMR have been developed using    
two different approaches: RPA + continuum on the one hand and RPA
+ complex configurations on the other. As we have seen, however,
both extensions of the RPA are need to explain the data. The first
of these can now be considered solved, and there exist several
numerical methods for it. One, which was mentioned in
section~1.2.4, uses the GF method. There one considers the
one-particle continuum exactly (for a contact interaction) by
transforming the RPA equation into the coordinate representation.
Other methods for solving this problem have also been developed
that even admit the use of nonlocal forces
\cite{giai90,buballa90}.  As for the problem of including complex
configurations, the most advanced approach is the
quasiparticle-phonon model for magic and non-magic nuclei by
Soloviev and his co-workers \cite{vg71,vg92}. These authors,
however, used separable forces in order to reduce the numerical
difficulties of the problem, and they leave out the
single-particle continuum. In addition, the ground state
correlations are included only partially, that is,                     
mainly  on the RPA or QRPA level.

The microscopic theory for GMR that satisfies the requirements      
mentioned above turned out to be quite difficult to formulate and
especially to realize numerically if one uses non-separable
forces, as shall will do.

There have been some successful developments in this direction in
the past ten years. The first attempts of this kind, which
simultaneously consider RPA configurations, the single-particle
continuum (escape width $\Gamma \uparrow$) and more complex (2p2h
\cite{adachi87,migli91} or 1p1h$\otimes$phonon \cite{kt91,colo92})
configurations (spreading width $\Gamma \downarrow$)  using
non-separable forces were made in
refs.~\cite{adachi87,migli91,kt91,colo92}
for some closed shell nuclei. These authors investigated various
types of GMR using, of course, different approximations and
methods.

The model developed in ref.~\cite{colo92} used only            
1p1h$\otimes$phonon configurations and it considered only a
particle-phonon interaction. It is based on the Bohr-Mottelson
model for the strength function of the phonons. This model was
also successfully used \cite{colo94} to calculate partial
branching coefficients of the proton decay of the isobar--analog
and Gamow--Teller resonances in $^{208}$Bi.
The papers                                                                
\cite{colo92,colo94} and \cite{adachi87}
were the first articles in which all
three  (that is, the above-mentioned items 1, 2 and 3) microscopic
mechanisms of GMR formation were used to explain such delicate
properties as the decay characteristics of the GMR. It was also
shown in refs.~\cite{colo92,colo94} that the complex
1p1h$\otimes$phonon configurations noticeably improve
the description of the decay characteristics.
 The advantage
of this method \cite{colo92,colo94} is a self-consistency (on the       
RPA level), that is, the phonons that are used in the extended
theory have been obtained in RPA using the calculated interaction. In   
this development, however, ground state correlations due to
complex configurations have been ignored.

The most extensive investigations of GMR,
which include the above-mentioned effects,                           
 were performed within  our ETFFS                                     
approach, where calculations for stable and unstable closed shell
nuclei have been made. The theory is based on the consistent use
of the GF method
\cite{kstt93,ktu92,kstw93,kstvar,kts94,kstprl,ks96,kst97,k...richter97,kstepj2000}.
The ETFFS simultaneously takes into account 1p1h, complex
1p1h$\otimes$phonon configurations, the single-particle continuum
and ground state correlations both of the RPA type and of those
caused by the complex configurations under consideration. In
addition, in its final equations it includes explicitly both the      
effective particle-hole interaction and the quasiparticle-phonon
interaction.

\subsection{Qualitative discussion of the extension}

A theoretical approach that takes into account the 2p2h
configuration space including the full 2p2h interaction is
numerically hardly solvable if one also uses a realistically large
configuration space. For that reason the main approximation in
ETFFS concerns the selection of the 2p2h configurations. In our
approach, guided by experimental observations, we include the most
important correlations in the 2p2h space by coupling phonons
(correlated 1p1h states) to a one-particle and one-hole state.
With this procedure we obtain effectively
1p$\otimes$phonon and 1h$\otimes$phonon
configurations. If we then couple an additional hole and particle,
respectively, to the previous configurations we obtain
1p1h$\otimes$phonon configurations
 where part of the 2p2h interaction is included. As
one can see from the applications, these configurations are indeed
the most important ones for the understanding of the spreading
width of GMR. The  configurations with phonons also nicely explain a
part of the low-lying spectrum in the neighboring odd mass nuclei.
configurations with phonons
are used in many theoretical approaches~\cite{bohrmot,vg92,bbb83}).

There is, however, an additional fact that greatly simplifies the  
problem, and that is the existence of a small parameter for closed
shell nuclei \cite{bohrmot}:
\begin{equation}
\alpha = \frac{\langle 1\Vert {\rm g} \Vert 2\rangle^{2}}{(2j_{1}
+ 1)\omega ^{2}_{s}} < 1,
\end{equation}
where $\langle 1\Vert {\rm g} \Vert 2\rangle$ is a reduced matrix
element of the amplitude for low-lying phonon creation with the
energy $\omega_{s}$ and 1 represents the set of single-particle
quantum numbers $n_{1}$, $l_{1}$ and $j_{1}$ for spherical nuclei.
Henceforth, when we refer to the g$^{2}$ approximation, it will be
understood that the dimensionless $\alpha$ is small. Using this
small parameter affords following advantages:
\begin{enumerate}
\item We obtain a general principle for selecting terms:
as $\alpha$ is small, we may restrict ourselves to
1p1h$\otimes$phonon configurations, which correspond to second
order in g (two-phonon configurations correspond to terms of order
g$^{4}$). Because we use the g$^{2}$ approximation in the
propagators of our integral equations, our approach is not the
usual perturbative theory in g$^{2}$.
\item  For the widths of GMR
the most important contribution comes from low-lying phonons,
which give rise to a strong energy dependence in the energy range
of the high lying collective (1p1h) RPA solutions. Therefore we
may confine ourselves to the most collective low-lying phonons,
which are restricted in number. The effects of the other phonons
are effectively already included through the phenomenological
parameters of our approach.
\item The restriction to a small number of collective phonons
noticeably reduces the numerical difficulties. This is especially
important for the present approach, in which non-separable forces
and the GSC induced by complex configurations are considered.
\item As some of the 1p1h$\otimes$phonon configurations are treated  
 explicitly
in the ETFFS, one should expect that the Landau-Migdal parameters
that are determined within the 1p1h approximation may change. As
we restrict ourselves to the g$^{2}$ approximation and the collective   
low-lying phonons, this effect in
the actual calculations is found to be small.
\end{enumerate}

The ETFFS approach is, like the original TFFS, a semi-microscopic
theory.  As our approach is based on the TFFS, we actually do not
need additional experimental input beyond that used already in the
TFFS.  We must, however, ``correct'' some of the parameters in
order to avoid double counting. Such corrections can be performed
fully consistently within our approach. The most important
corrections refer to the single--particle energies which are
taken---as far as possible---from experiment, or else from a shell
model potential that is carefully adjusted to the corresponding
closed shell nucleus. The single-particle wave functions are also
taken from that model. These quantities contain contributions from
the same phonons that enter the complex configurations under
consideration. In order to avoid double counting due to these
phonons, the single--particle model  has to be ``refined'' from
this mixing. The procedure for this will be described in
section~3.1.2. The complex configurations that we treat explicitly
in our extended approach are also included implicitly in the force
parameters of the original TFFS approach. Therefore, in principle,
one has also to correct the Landau-Migdal parameters.

There exists so far no self-consistent theoretical approach that
includes all the effects discussed above. In such an approach one
would start with an effective two-body interaction that would
allow  to determine the single-particle energies and wave                 
functions and the ph interaction.  As within such a procedure the
phonon effects are not included, our extended theory would be the
natural formalism in which to do it. In our extended version of
the TFFS we did not include the so-called tadpole graphs with the
low-lying phonons under consideration that have been used in the
self-consistent version of the TFFS~\cite{khs82}.  Their
contribution is contained effectively in our ``refined'' mean
field.

\setcounter{equation}{0}
\section{Framework of the extended theory}
\subsection[General description of collective excitations]{General description of collective excitations,\\
 including the particle-hole and quasiparticle-\\
 phonon interaction}                                    
As mentioned before, the original TFFS allows to calculate only the   
centroid energies and total transition strength of giant resonances
because the approach is restricted to 1p1h configurations.
In order to describe more detailed nuclear structure properties
one has to include higher configurations.
Here we describe the derivation of the main ETFFS equations. These
equations contain both the quasiparticle-phonon interaction and
the effective ph interaction in a general form.

\subsubsection{General relations}

Equation~(\ref{eq:1}) can be considered a definition of the         
response function $R$ only if the quantities $\Sigma$ and $U$ are
known. In order to obtain realistic numerical results, however, we
have at our disposal only model approximations of these quantities
because it is not possible to solve the whole system of               
nonlinear equations for $R$, $\Sigma$ and $U$ for a realistic case
\cite{AB,agd}. For that reason Landau introduced phenomenological
elements into his microscopic theory.  In Landau`s approach
$\Sigma$ is irreducible in the one-particle (one-hole) channel and
$U$ is irreducible in the ph channel. As the strongest energy
dependence in Eqs.~(1.19) and (1.21) is considered explicitly
through  the one-particle (one-hole) and the ph propagators,
respectively, the irreducible parts are weakly energy dependent
and are parameterized in an energy--independent way. In our
extended theory the complex configurations give rise to a strong
energy dependence in the previously weakly energy--dependent
irreducible quantities $\Sigma$ and $U$, which we now have to
consider explicitly. In order to do so we represent $\Sigma$ and
$U$ as a sum of two terms, in which the first terms are again
assumed to depend only weakly on the energy. As we shall see in
the next section, these terms are irreducible with respect to the
complex configurations that are considered explicitly and can, for
the same reasons as in Landau´s original theory, be parameterized.
\begin{equation}
\Sigma_{12}(\epsilon) = \tilde{\Sigma}_{12} +
\Sigma_{12}^{e}(\epsilon),
\end{equation}
\begin{equation}
U_{12,34}(\omega,\epsilon,\epsilon') = \tilde{U}_{12,34} +
U^{e}_{12,34}(\omega,\epsilon,\epsilon'),
\end{equation}

Using Eq.~(2.2) one can transform our main equation,        
Eq.~(\ref{eq:1}), into the symbolic form
\begin{equation}
R = R^{e} - R^{e}\tilde{U}R,
\end{equation}
where the quantity $R^{e}$ satisfies the equation
\begin{equation}
R^{e} = -GG + GGU^{e}R^{e}. \label{eq:2.4}
\end{equation}
Equation~(2.3) is more convenient in the sense that it contains
only the energy-independent amplitude $\tilde{U}$, which allows us
to apply the known renormalization procedure~\cite{AB} to it.

Furthermore, we rewrite the Dyson Eq.~(1.21) as
\begin{equation}
G_{12}(\epsilon) = \tilde{G}_{12}(\epsilon) + \sum_{34}
\tilde{G}_{13}(\epsilon)\Sigma^{e}_{34}
(\epsilon)G_{42}(\epsilon),
\end{equation}
where $\tilde{G}$ is the solution of the Dyson equation with the
mass operator $\tilde{\Sigma}$,
\begin{equation}
 \tilde{G}_{12}(\epsilon) = G^{0}_{12}(\epsilon) +
 \sum_{34} G_{13}^{0}(\epsilon)\tilde{\Sigma}_{34}\tilde{G}_{42}(\epsilon).
\end{equation}

The quantities denoted with tildes, $\tilde{\Sigma}$ and
$\tilde{U}$, as mentioned before, are the analogs of the
quantities $\Sigma$ and $U$ in the standard TFFS.  The
energy-dependent quantities $\Sigma^{e}$ and $U^{e}$ introduce the
effects of the complex configurations into the formalism. The
corresponding single-particle basis $\{\tilde{\varphi}_{\lambda}$,
$\tilde{\epsilon}_{\lambda}\}$,
 which is defined by $\tilde{\Sigma}$, is a new or ``refined''
 basis and should be obtained from the  basis
 $\{\varphi_{\lambda},\epsilon_{\lambda}\}$ used in the TFFS.

\subsubsection[Renormalization of the general equations]{Renormalization of the general equation for the response
function and the equation for the density matrix.}

Equation~(2.3), as well as Eq.~(1.19), contains summation and     
integration over all states, including those that are far away
from the Fermi surface and which, in actual calculations can be
taken into account only effectively. For that reason it is
necessary to perform the Landau-Migdal renormalization procedure
in order to obtain an equation with the summation restricted to
the vicinity of the Fermi energy. This is performed in analogy
with the renormalization of the equation for the response function
within the standard TFFS described in section~1.2.3. \cite{KTs86}.

In analogy with Eqs.~(1.25) and (1.26),  we introduce the
generalized propagator $A$, which will  explicitly be given by the    
complex configurations under consideration                            
 and the quantity $B$, which contains all the rest:
\begin{equation}
  \int^{\infty}_{-\infty} \frac{d\epsilon}{2\pi\imath}R_{12,34}^{e}(\omega,\epsilon) =
A_{12,34}(\omega) + B_{12,34}.
\end{equation}
It is assumed that the quantity $B$ depends only weakly on
$\omega$ compared to $A$.
One obtains from Eq.~(2.3) the renormalized response function
$\tilde{R}(\omega)$, which is connected with our initial response
function by the relation
\begin{equation}
R_{12,34} = \sum_{5678} (\tilde{e}_{q}^{+})_{12,56}
\tilde{R}_{56,78}(\omega)(\tilde{e}_{q})_{78,34} +
               \sum_{56} B_{12,56}(\tilde{e}_{q})_{56,34},
\end{equation}
where $\tilde{R}(\omega)$ satisfies the equation
\begin{equation}
\tilde{R}_{1234}(\omega) = A_{12,34}(\omega) - \sum_{56,34}
A_{12,56}(\omega)\tilde{F}_{5678}\tilde{R}_{7834}(\omega).
\label{eq:50}
\end{equation}
Here the new effective charge operator $\tilde{e}_{q}$ and the
effective ph interaction amplitude $\tilde{F}$ are given by
\begin{equation}
(\tilde{e}_{q})_{12,34} = \delta_{13}\delta_{24} -
\sum_{56}\tilde{F}_{12,56}B_{56,34}
\end{equation}
\begin{equation}
\tilde{F}_{12,34} = \sum_{56}(\tilde{e}_{q})_{12,56}
\tilde{U}_{56,34}
\end{equation}

 Equation~(\ref{eq:2.4}) still contains the full GF and, in fact,
according to Eq.~(2.7), determines the desired propagator
$A(\omega)$ in Eq.~(2.9).  It is useful to transform it to a more
convenient form.  Acting on both sides of Eq.~(\ref{eq:2.4}) with
the operator $(G^{-1}\tilde{G})(\tilde{G}G^{-1})$ and using
Eq.~(2.5), we obtain, in the time-representation,
\begin{equation}
R^{e}(12,34) = \tilde{R}^{0}(12,34) + \sum_{5678;t}
\tilde{R}^{0}(12,56)
                                      W^{e}(56,78)R^{e}(78,34),
\label{eq:2.12}
\end{equation}
where
\begin{equation}
\tilde{R}^{0}(12,34) = -\tilde{G}(3,1)\tilde{G}(2,4)
\end{equation}
\begin{equation}
W^{e}(12,34) = W^{e}_{0}(12,34) - \imath\Sigma^{e}(31)\Sigma^{e}(24)           
\end{equation}
\begin{equation}
W^{e}_{0}(12,34) = U^{e}(12,34)+ \imath\Sigma^{e}(3,1)\tilde{G}^{-1}(2,4)    
                    +  \imath\tilde{G}^{-1}(3,1)\Sigma^{e}(2,4).
\end{equation}

Because the quantity $B$ only weakly depends on the energy as
compared with $A$, it follows from Eq.~(2.8) and Eq.~(1.18) that
$S(E) =\lim_{\Delta\rightarrow+0}S(E,\Delta)$, with                     
\begin{equation}
S(E,\Delta) =
 \frac{1}{\pi}  \; {\rm Im}\;\sum_{1234}(\tilde{e}_{q}V^{0})^{ \ast}_{21}\tilde{R}_{12,34}
       (E+\imath\Delta)(\tilde{e}_{q}V^{0})_{43}
\end{equation}
\begin{equation}
(\tilde{e}_{q}V^{0})_{12}  =
\sum_{34}(\tilde{e}_{q})_{21,43}V^{0}_{34}.
\end{equation}
We see that the strength function is completely determined by the
renormalized response function $\tilde{R}(\omega)$.

The advantage of these transformations is that, as in the original
TFFS, the experimental quantity $S(E)$ is now connected with the
renormalized response function, which can be calculated from
Eq.~(2.9). In contrast to Eq.~(2.4) for the full response function
$R$, which cannot be used for numerical calculations, the equation
for the renormalized response function, Eqs.~(2.9), (2.16) contain only   
 quantities which will be defined in the next section:                   
 the generalized propagator $A$, which can be
explicitly calculated from the complex configurations, and the
effective ph interaction $\tilde{F}$, and the effective charge
operator $\tilde{e}_{q}$ that are treated in the same way as in
the original TFFS approach.

As in the case of the TFFS (section~1.2.3), we also introduce the
change of the density matrix $\rho_{12}$ due to an external field     
 $V^{0}$ for
the generalized case under consideration:
\begin{equation}
 \rho_{12} (\omega) = -\sum_{34}
 \tilde{R}_{1234}(\omega)(\tilde{e}_{q}V^{0})_{43}.
\end{equation}
The equation for $\rho_{12}$ follows directly from Eq.~(2.9) and
has the form
\begin{equation}
\rho_{12}(\omega) = -\sum_{34} A_{12,34}(\omega) (\tilde{e}_{q}
V^{0})_{43} -
                       \sum_{3456} A_{12,34}(\omega) \tilde{F}_{3456}
                       \rho_{56}(\omega),
\label{eq:60}
\end{equation}
and the expression for the strength function can be written as:
\begin{equation}
S(\omega,\Delta) = - \frac{1}{\pi} \; {\rm Im} \;
\sum_{12}(\tilde{e}_{q}V^{0})_{21}^{*}\rho_{12}(\omega+\imath\Delta).
\label{eq:49}
\end{equation}

Equations(\ref{eq:49}) and (\ref{eq:60}) are our main general    
results. We stress that in Eqs.~(2.1), (2.2), (2.4) and (2.14),
(2.15)
we did not yet specify the form of our energy-dependent terms
$\Sigma^{e}$ and $U^{e}$ and the transformation from $R$ to
$\tilde{R}$ is formally exact. So far we have only assumed that
the quantities  $B$ and  $\tilde{U}$, as in the standard TFFS,
depend only weakly on the energy. The same holds for
$\tilde{e}_{q}$ and $\tilde{F}$.
After we   specify the mass operator and the irreducible                  
amplitude in section~2.2, we shall obtain the generalized propagator $A$   
in section~2.3.  With this information we are able to apply the
ETFFS to GMR, the results of which we shall discuss in sections~3
and 4.

Equations  (2.8) and (2.9) for our                                       
renormalized response function $\tilde{R}(\omega)$ and the final         
equations, Eqs.~(2.20) and (\ref{eq:60}), have a structure similar
to the relations in the TFFS given in section~1.2.3. Let us point
out the main differences as compared with the TFFS.

First of all, the generalized propagator $A$, will differ the most
from the TFFS form, Eqs.~(1.23) and (1.26),
because we must now solve an equation, either (2.4) or (2.12)        
to obtain the propagator.                                            
 In the standard theory
one considers the propagation of a particle--hole pair, whereas in
the extended theory the propagator has to include also the
configurations that go beyond the RPA. In our specific case these
will be 1p1h$\otimes$phonon configurations. In the next sections
we shall specify these contribution and derive explicit
expression for $A$.
 Here we already point out,
 that  the final Eqs. (2.9) or (2.19)                                 
will explicitly include both the
ph interaction $\tilde{F}$ and  (in the propagator)                    
the quasiparticle--phonon
interaction. In this respect our formulation differs from the
approaches containing only the quasiparticle--phonon
interaction~\cite{vg92,bbb83},
and we do not use a phonon representation.                          

Secondly, the formulas of the extended theory contain also the quantities $\tilde{F}$ and    
$\tilde{e}_{q}$ that describe the effective interaction and local
charge of our ``refined'' quasiparticles. They play the same role
as $F$ and $e_{q}$ in the TFFS and will be parameterized in a
similar way. In general they should differ numerically from the
quantities $F$ and $e_{q}$ because the ``dangerous'' pole terms
corresponding to the complex configurations, which will be
considered explicitly in the propagator $A$, have been removed
from $\tilde{F}$ and $\tilde{e}_{q}$.  As discussed in
section~1.4, however, these differences turn out to be small, as
we shall see in sections 3 and 4.

Finally, we have written Eqs.~(2.9), (2.20) and (\ref{eq:60}) in a
new shell model representation $\tilde{\lambda}$, which differs
from the representation used in the TFFS.  These new
single-particle wave functions
 $\tilde{\varphi}_{\lambda}$ and energies $\tilde{\epsilon}_{\lambda}$
that represent the new basis $\{\tilde{\varphi}_{\lambda},
\tilde{\epsilon}_{\lambda}\}$ instead of the phenomenological one
 $\{\varphi_{\lambda},\epsilon_{\lambda}\}$,
do not contain the phonon mixing, which is included in the mass
operator $\Sigma^{e}$. In a sense, the basis corresponds to a          
Hartree-Fock basis. The mass operator $\tilde{\Sigma}$ defines the
basis functions of the configuration space. The reason for this is
that in the GF formalism the basis $\{\tilde{\varphi}_{\lambda}$,
$\tilde{\epsilon}_{\lambda}\}$ is chosen in such a way that it
diagonalizes the Greens function $\tilde{G}$, see Eq. (2.6):           
\begin{equation}
\tilde{G}_{12}(\epsilon) =
\delta_{12}\tilde{G}_{1}(\epsilon),\hspace{0.5cm}
\tilde{G}_{1}(\epsilon) = \frac{1}{\epsilon - \tilde{\epsilon}_{1}
+ \imath\sigma_{1}\delta},
                       \;\;   \delta\to+0.
\end{equation}
In this connection the mass operator is nothing else than a
generalized
single--particle potential.
In Eq.~(2.21) we introduced the quantity $\sigma_{1}$, which is equal  
to +1 for particles and --1 for holes, and is related to the level
occupation number by $\sigma_{1} = 1 - 2n_{1}$.  The ``refined''
basis $\{\tilde{\varphi}_{\lambda}$,
$\tilde{\epsilon}_{\lambda}\}$ can be calculated if the old
phenomenological basis
 $\{\varphi_{\lambda},\epsilon_{\lambda}\}$
and the quantity $\Sigma^{e}$ are known. In section~3.1.2 we shall
describe our method of calculating $\{\tilde{\varphi}_{\lambda}$,
$\tilde{\epsilon}_{\lambda}\}$.

\subsection[Approx. for the mass operator and the irreducible amplitude]{Approximation for the mass operator \\ and the irreducible amplitude
in the case of \\ 1p1h$\otimes$phonon configurations}

Let us return to Eq.~(\ref{eq:1}), which is the starting point of   
our approach. We have shown in section~1.2.3. that by an
appropriate choice of the ph propagator $A$ and the effective
interaction $F$, that is, in fact, using the approximations
 $\Sigma = \tilde{\Sigma}$ and $U = \tilde{U}$,                      
 one obtains from this equation Eq.~(1.31) for the
renormalized response functions $\tilde{R}$, which is the basis of
Migdal's theory. Therefore our problem of including more complex
configurations in the generalized renormalized response function
defined by Eq.~(2.9) requires the appropriate choice of the
corresponding quantities $\Sigma^{e}$ and $U^{e}$, which describe the
coupling of ph configurations with the more complex ones
and represent, therefore, the basic input to our extended theory.
Depending on the approximations made for these quantities, we
obtain the various versions of extended RPA theories mentioned in
the introduction to this chapter:

\begin{enumerate}
\item {\it The Extended Second RPA.}  If we include in $\Sigma^{e}$
and $U^{e}$ complex configurations in the lowest order of
perturbation theory in the effective nucleon-nucleon interaction,
we obtain the Extended Second RPA of Ref.~[88]. Here one may or
may not consider GSC. Such a  variant of inclusion of ``pure''
2p2h configurations within the GF  method is briefly discussed in
ref.~\cite{ktts97}. As it is known that such a perturbation series
in the nucleon-nucleon interaction converges very slowly, one has
to consider partial summations that include the interaction to all
orders.

\item {\it The Phonon Coupling Model.}  Such a possibility is given
by the phonon coupling model. Here the interaction between the ph
pairs that build up the phonons is included to all orders.
Moreover, in the case of closed shell nuclei, one can show that
the coupling between the particle (or hole) with the phonons, the
dimensionless                                                       
parameter $g$ (Eq.~(1.44)), is a small quantity.  For that reason
one can restrict oneself to the coupling of one phonon to that
particle (or hole), which is the g$^{2}$ approximation discussed
earlier. In the g$^{2}$ approximation the expressions for
$\Sigma^{e}$ and $U^{e}$ are represented graphically in Fig.~2.1
and have the following analytical form:
\begin{equation}
\Sigma^{e}_{12}(\epsilon) =
\sum_{3,m}\frac{g^{m(\sigma_{3})^{*}}_{13}g^{m(\sigma_{3})}_{23}}
{\epsilon - \tilde{\epsilon}_{3} - \sigma_{3}(\omega_{m} -
\imath\delta)} \label{eq:51}
\end{equation}
\begin{equation}
U^{e}_{12,34}(\omega,\epsilon,\epsilon') = \sum_{\sigma
,m}\frac{\sigma g^{m(\sigma)^{*}}_{31}g^{m(\sigma)}_{42}}
{\epsilon - \epsilon' +\sigma(\omega_{m} - \imath\delta)},
\label{eq:52}
\end{equation}
Here we have introduced the abbreviation:
\begin{eqnarray}
g^{m(\sigma)}_{12} = \delta_{\sigma,+1}g^{m}_{12} +
\delta_{\sigma,-1}g^{m*}_{21}. \nonumber
\end{eqnarray}
where $g^{m}_{12}$ is a matrix element of the creation phonon      
amplitude and $\omega_{m}$  is the phonon energy.                  
Upon substituting Eqs.~(\ref{eq:51}) and (\ref{eq:52}) into
Eqs.~(2.1), (2.2) and (2.4), (2.5), (2.7)
one obtains the propagator $A$ and                                 
Eq.~(2.9) for the
renormalized response function in the ETFFS.

 The simplest propagator $A$, obtained from Eq. (2.12)       
in the g$^{2}$ approximation, is
 given in graphical representation in Fig.~2.2.
The corresponding model, which have used this propagator,            
 was suggested in refs.~\cite{kam79,kam83}                           
and realized in refs.~\cite{kt84,kt86,kt89} for the M1 resonances  
in magic nuclei $^{16}$O, $^{40,48}$Ca and $^{208}$Pb in the      
$\lambda$ representation. A reasonably good description of the         
experimental data, including solution of the long-standing problem
of the isovector M1 resonance in $^{208}$Pb was obtained. In
addition, as mentioned in section~1.3.1, the model also reproduced
the experimentally known 1$^{+}$ levels in  $^{16}$O and $^{40}$Ca
as a result of including the complex 1p1h$\otimes$phonon
configurations in the ground state. Such an approach, however,
leads to the following problem: the propagator $A$ contains in
this approximation poles of second order in the variable $\omega$
at the points of the simple poles of the RPA--like propagator. The
problem is that the second order poles do not have the same
physical meaning as the simple poles of the exact response
function. This can result in distortion of the energy dependence
of the strength function near these poles. It was shown, however,
that for the M1 resonances this distortion was negligible
\cite{kt86,kt89}.

The way out of this difficulty is to  perform the summation of the
g$^{2}$ terms in the propagator. In ref.~\cite{KTs86} the method
of summation was developed both for
the linearized mass operator (Eq.~(2.22)) and for a nonlinear     
case, which uses the full GF  $G$ in the mass operator instead of
$\tilde{G}$. Even so, the numerical difficulties are still rather
serious in both cases. There exist, however, a simpler method,
which treats the summation approximately. The main idea in this
method is to perform a summation of an infinite series of {\it
some} of the g$^{2}$ terms in the propagator.
\end{enumerate}
\subsection[Constructing the generalized propagator]{Constructing the generalized propagator: the\\
 method of chronological decoupling of diagrams}                           

In this section~we consider the method of constructing our
generalized ph propagator $A(\omega)$ \cite{Tselyaev}, where the
problem of second-order poles of the model \cite{kam83,kt89} has
been solved by a partial summation of the diagrammatic series in
the propagator.

\subsubsection{The basic idea.}                                 
Here we present the technical aspects of the method we use to
construct the generalized ph propagator $A(\omega)$. As in
section~1.2.2, we denote the set of one-particle quantum numbers
and the time variable by 1 = $\{_{1},t_{1}\}$ and so forth. In
particular, in the time representation the Green function
$\tilde{G}$ given by Eq.~(2.21) becomes
\begin{equation}
\tilde{G}(1,2) =
-\imath\sigma_{1}\delta_{12}\theta(\sigma_{1}t_{12})
 e^{-\imath\tilde{\epsilon}_{1}t_{12}},
\end{equation}
where $t_{12} = t_{1} - t_{2}$ and $\theta$ is the Heaviside
(step) function.

Let us consider an  operator in the time representation whose
matrix elements in the basis $\{\tilde{\varphi}_{\lambda}$,
$\tilde{\epsilon}_{\lambda}\}$ are given by
\begin{equation}
\tilde{D}(12,34) = \delta_{\sigma_{1},-\sigma_{2}}
                   \theta(\sigma_{1}t_{41})
                  \theta(\sigma_{1}t_{32})
                    \tilde{G}(3,1)\tilde{G}(2,4).
\end{equation}
Multiplying the right-hand side by the sum
$\theta(\sigma_{1}t_{12}) + \theta(\sigma_{1}t_{21})=1$ and using
Eq.~(2.24) we find
\begin{equation}
\begin{array}{c}
   \tilde{D}(12,34) = \delta_{\sigma_{1},-\sigma_{2}}\delta_{13}\delta_{24}
   e^{-\imath(\tilde{\epsilon}_{1}t_{31} + \tilde{\epsilon}_{2}t_{24})} \hspace{0.5in} \\[4mm]
   \times  \left[\theta(\sigma_{1}t_{12})\theta(\sigma_{1}t_{41})\theta(\sigma_{1}t_{31}) +
   \theta(\sigma_{1}t_{21})\theta(\sigma_{1}t_{32})\theta(\sigma_{1}t_{42})]\right.
\end{array}
\end{equation}
Here we have used the easily verified identity ($\sigma = \pm1$),
\begin{equation}
 \theta(\sigma t_{13})\theta(\sigma t_{12})\theta(\sigma t_{23}) = \theta(\sigma t_{12})
                  \theta(\sigma t_{23}).
\end{equation}

We further introduce, in accordance with Eq.~(\ref{eq:1}) and
ref.~\cite{KTs86}, the time difference variables $\tau_{1} =
t_{3}-t_{1}, \tau_{2} = t_{2}-t_{1}, \tau_{3}=t_{3}-t_{4}$, so
that
\begin{equation}
\tilde{D}(12,34) = \tilde{D}_{12,34}(\tau_{1},\tau_{2},\tau_{3}),
\end{equation}
and transform to the energy representation:
\begin{equation}
\tilde{D}_{12,34}(\omega,\epsilon,\epsilon') =
\int_{-\infty}^{+\infty}d\tau_{1}d\tau_{2}d\tau_{3}
                                          e^{\imath(\omega\tau_{1}+\epsilon\tau_{2}+\epsilon'\tau_{3})}
\tilde{D}_{12,34}(\tau_{1},\tau_{2},\tau_{3}).
\end{equation}
Substituting Eq.~(2.26) into Eq.~(2.29) and using the well-known
formulas
\begin{equation}
\int_{-\infty}^{+\infty}d\tau
e^{\imath\omega\tau}\theta(\sigma\tau) = \frac{\imath\sigma}
                                                                       {\omega+\imath\sigma\delta},
                                        \;\;   \sigma = \pm1, \;\;\delta\to+0,
\end{equation}
we find
\begin{equation}\begin{array}{c}
\tilde{D}_{12,34}(\omega,\epsilon,\epsilon') =
\delta_{\sigma_{1},-\sigma_{2}}\delta_{13}\delta_{24}
               \fr{\imath\sigma_{1}}{(\epsilon'+\omega-\tilde{\epsilon}_{1}+\imath\sigma_{1}\delta)
                                       (\epsilon'-\tilde{\epsilon}_{2}+\imath\sigma_{2}\delta)} \times \\[4mm]
                \left(\fr{1}{\epsilon +\omega-\tilde{\epsilon}_{1}+\imath\sigma_{1}\delta}-
                       \fr{1}{\epsilon-\tilde{\epsilon}_{2}+\imath\sigma_{2}\delta}\right).
\end{array}\end{equation}
>From this, using the definition Eq.~(2.21), we obtain the final
result
\begin{equation}
\begin{array}{c}
\tilde{D}_{12,34}(\omega,\epsilon,\epsilon')
 = -\imath \delta_{\sigma_{1},-\sigma_{2}}\delta_{13}\delta_{24}
  \sigma_{1}(\omega-\tilde{\epsilon}_{12}+\imath\sigma_{1}\delta)
  \\ \nonumber
 \hspace{1.0in} \times \,  \tilde{G}_{1}(\epsilon+\omega)\tilde{G}_{2}(\epsilon)\tilde{G}_{3}(\epsilon'+\omega)\tilde{G}_{4}(\epsilon').
\end{array}
\end{equation}
We see from this relation that the dependence of the function
$\tilde{D}_{12,34}(\omega,\epsilon,\epsilon')$ on the variables
$\epsilon$ and $\epsilon'$ is separable. This is essential for the
subsequent model transformations. We note that the simple product
of two Green functions $\tilde{G}(3,1)\tilde{G}(2,4)$ in
Eq.~(2.13) does not possess this property in the energy
representation:
\begin{equation}
\tilde{R}^{0}_{12,34}(\omega,\epsilon,\epsilon') =
-\delta_{13}\delta_{24}2\pi
                        \delta(\epsilon-\epsilon')\tilde{G}_{1}(\epsilon+\omega)\tilde{G}_{2}(\epsilon).
\end{equation}

In order to illustrate the effect of this separability, let us
consider a function $F(12,34)$ which, in the time representation,
is given by
\begin{equation}
F(12,34) = \sum_{56,78;t}V^{L}(12,56)\tilde{D}(56,78)V^{R}(78,34),
\end{equation}
where $V^{L}$ and $V^{R}$ are different functions whose dependence
on the time arguments reduces to a dependence on three time
difference variables, e.g.,
\begin{equation}
V^{L}(12,34) = V^{L}_{12,34}(t_{31},t_{21},t_{34})
\end{equation}
and  $\sum_{12...;t}$  denotes summation over the one-particle
indices of the basis wave functions $\tilde{\varphi}_{1}$ and
integration over the time variables. Then, in the energy
representation, Eq.~(2.34) is given by
\begin{equation}
F_{12,34}(\omega,\epsilon,\epsilon') =
\sum_{5678}\int_{-\infty}^{+\infty}\frac{d\epsilon_{1}d\epsilon_{2}}
                       {(2\pi)^{2}}
                       V^{L}_{12,56}(\omega,\epsilon,\epsilon_{1})
                        \tilde{D}_{56,78}(\omega,\epsilon_{1},\epsilon_{2})
                          V^{R}_{78,34}(\omega,\epsilon_{2},\epsilon').
\label{eq:36}
\end{equation}
Substituting Eq.~(2.32) into this, we see that, owing  to the
separability of the energy dependence of
$\tilde{D}_{56,78}(\omega,\epsilon_{1},\epsilon_{2})$, the
integrations over the variables $\epsilon_{1}$ and $\epsilon_{2}$
in Eq.~(\ref{eq:36}) decouple. From the technical point of view
this is due to the presence of the additional (compared to
$\tilde{R}^{0}$) factor of two $\theta$ functions on the
right-hand side  of Eq.~(2.25) written in the time representation.
The replacement of $\tilde{R}^{0}$ by -$\tilde{D}$ in the
expressions determining the generalized ph propagator is the main
element of the method of constructing the propagator in our
approach. The physical meaning of it will become clear with the
specification of the quantities entering into equations such as
Eqs.~(2.34) and (2.36) and in terms of Feynmann diagrams. This
also provides the name of the method of constructing of the
generalized propagator, which we refer to as the method of
chronological decoupling of diagrams (MCDD) \cite{Tselyaev}.

\subsubsection{Formulation of the method}
Using symbolic notation one can write the solution of Eq.~(2.12)
in the form
\begin{equation}
R^{e} =
 \tilde{R}^{0} + \imath \tilde{R}^{0}\Gamma^{e} \tilde{R}^{0},
\end{equation}
where the amplitude $\Gamma^{e}$ satisfies the equation
\begin{equation}
\Gamma^{e} = W^{e} + \imath W^{e}\tilde{R}^{0}\Gamma^{e}.
\label{eq:38}
\end{equation}
We introduce the new amplitude $\Gamma^{e}$ defined by the
equation
\begin{equation}
\tilde{\Gamma}^{e} = \tilde{W}^{e} + \imath
\tilde{W}^{e}(-\tilde{D})\tilde{\Gamma}^{e},
\end{equation}
which is obtained from Eq.~(\ref{eq:38}) by the replacement
\begin{equation}
\tilde{R}^{0}\to -\tilde{D},  W^{e}\to \tilde{W}^{e}.
\end{equation}
Here the function $\tilde{D}$ is given by Eq.~(2.25) and the
amplitude $\tilde{W}^{e}$ is given by
\begin{equation}
\tilde{W}^{e}(12,34) = W^{e}_{0}(12,34) + W^{comp}(12,34),
\end{equation}
which differs from Eq.~(2.14) by the replacement of
$-\imath\Sigma^{e}\Sigma^{e}$ by $W^{comp}$. The amplitude
$W^{comp}$ in Eq.~(2.41)
plays the same role as the term $-\imath\Sigma^{e}\Sigma^{e}$ in
Eq.~(2.14) and will be defined below. We now  replace in
Eq.~(2.37) the amplitude $\Gamma^{e}$ by $\tilde{\Gamma}^{e}$:
\begin{equation}
\tilde{R}^{e} = \tilde{R}^{0} +
\imath\tilde{R}^{0}\tilde\Gamma^{e}\tilde{R}^{0}.
\end{equation}
In the energy representation this function defines the desired
propagator,
\begin{equation}
A_{12,34}(\omega) =
\int^{+\infty}_{-\infty}\frac{d\epsilon}{2\pi\imath}\tilde{R}^{e}_{12,34}
                                                                       (\omega,\epsilon).
\end{equation}

 Let us discuss the physical meaning of these model transformations.
It is clear that $W^{comp}$ contains g$^{4}$ terms (see also
Eq.~(2.44), below) and that the solution of the integral equation
(2.39) (after the substitution of Eqs.~(2.22) and (2.23)) gives
the function $\tilde{R}^{e}$, and therefore the propagator $A$,
which contains an infinite sum of the terms of higher order in the
g$^{2}$ terms. An additional---and physical---condition, whose
fulfilment must be verified, is that this sum must correspond to a
subset of the series of Feynmann graphs that correspond to the
initial quantity $R^{e}$. In this sense the replacement of
$\tilde{R}^{0}$ (Eq.~(2.13)) by $-\tilde{D}$ (Eq.~(2.25)) is
justified  by the fact that the quantity $-\tilde{D}$
contains $\theta$ functions, which are projection operators in the
space of time variables; that is, the graphs corresponding to
different combinations of arguments of the function
$-\tilde{D}(12,34)$ belong to the set of graphs corresponding to
the function $\tilde{R}^{0}(12,34)$. Physically, however, it is
very important that the replacement of $\tilde{R}^{0}$ by
$-\tilde{D}$ {\it eliminates} from the diagrammatic expansion of
$R^{e}$ the terms corresponding to processes in which
configurations more complex than 1p1h$\otimes$phonon are excited
while leaving most of the 1p1h$\otimes$phonon configurations. In
our  model with 1p1h$\otimes$phonon configurations the discarded
graphs include those in which a time cut through two or more
phonon lines is possible in the time representation. As an
example, in Figs.~2.3a, 2.3b and 2.3c we show three graphs of
order g$^{4}$ that are excluded from the expansion, and in
Figs.~2.3d, 2.3e and 2.3f, we show three similar graphs
that remain in the diagrammatic expansion of $\tilde{R}^{e}$.

The sum of the contributions of higher order in g$^{2}$, which is   
contained in the function $\tilde{R}^{e}$, corresponds to the sum
of chains of graphs similar to those shown in Figs.~2.3d, 2.3e and
2.3f. The contributions of the graphs shown in Figs.~2.3a, 2.3b
and 2.3c are excluded due simply to the $\theta$ functions in the
definition of $\tilde{D}$, Eq.~(2.25).

The statements that in the method under consideration all the        
contributions of  configurations more complex than
1p1h$\otimes$phonon are excluded, while all 1p1h$\otimes$phonon
contributions are included, are completely  valid only when GSC
are ignored. The role played by their effect will be discussed
below. These effects correspond to the so-called backward--going
diagrams which, on the one hand, take us beyond the
1p1h$\otimes$phonon (or 2p2h) approximation (see Fig.~2.4) but, on
the other hand, not all the 1p1h$\otimes$phonon contributions
related to them (those of order g$^{4}$ and above) are included in
the MCDD.  Since g$^{2}$ is a small parameter, however, these
neglected contributions of higher order are small.
Our task in constructing the generalized propagator is to avoid
the second order poles, that is, to perform approximately an
infinite summation of the graphs of the type shown in Figs.~2.3d,
2.3e and 2.3f and, simultaneously, to take into account the
remaining graphs---at least within the accuracy of g$^{2}$.

The last formula, which is necessary to obtain our generalized
propagator, is an expression for $W^{comp}$ in Eq.~(2.41). The
role of this quantity
is to remove the multiple counting of graphs with self-energy
insertions $\Sigma^{e}$.
This problem therefore arises only for the terms of order g$^{4}$   
and above and for the inclusion of the GSC induced by
1p1h$\otimes$phonon configurations because,
 if there are no backward-going graphs like those shown in Fig.~2.4,
the multiple counting is eliminated by the $\theta$ functions in
the definition of the function $\tilde{D}$ (Eq.~(2.25)), and in
that case W$^{comp}$ = 0. The expression for $W^{comp}$ was found
from the condition of cancellation of the double-counting of
self-energy contributions obtained after the first iteration of
Eq.~(2.39) \cite{Tselyaev}:
\begin{equation}\begin{array}{c}
W^{comp}(12,34) = -\imath
              \delta_{\sigma_{1},-\sigma_{2}}\delta_{\sigma_{3},-\sigma_{4}}
              \delta_{\sigma_{1},\sigma_{3}} \times\hspace{3.0in}\\[4mm] \nonumber
           \sum_{1'2'3'4';t} \sum_{5'6'7'8';t}
           \tilde{G}^{-1}(1',1)\tilde{G}(5',1')\tilde{G}^{-1}(2,2')\tilde{G}(2',6')\,
           \times \hspace{1.0in}
           \\[4mm] \nonumber
          \Sigma^{e}(7',5')\Sigma^{e}(6',8') \tilde{G}(3',7')\tilde{G}^{-1}(3,3')\tilde{G}(8',4')\tilde{G}^{-1}(4',4)\Theta
           (1'2',3'4'),\hspace{1.0in}
\end{array}\end{equation}
where $\Theta(1'2',3'4') =
\theta(\sigma_{1'}t_{4'1'})\theta(\sigma_{1'}t_{5'8'})\theta(\sigma_{1'}t_{3'2'})\theta(\sigma_{1'}t_{6'7'})$.
This expression was used in our calculations (the explicit form of
it is given in the Appendix). An improved form of this term was
discussed in ref.~\cite{ktts97}.   All these effects, however, are
of order of g$^{4}$, so that the difference is small in our case.

\subsubsection[The generalized propagator in the energy representation]{The generalized propagator in the energy representation and its properties}
Equations~(2.25), (2.39), (2.41), (2.42), (2.43) and (2.44)
completely determine the generalized ph propagator $A(\omega)$ in
the MCCD. Since most of them are written in the time
representation it is necessary to transform to the energy
representation. After some algebraic effort we obtain          
\begin{equation}\begin{array}{c}
A_{12,34}(\omega) =  \nonumber\\
 \sum_{56,78}[\delta_{15}\delta_{26} + Q_{12,56}^{(+-)}(\omega)]A_{56,78}^{(--)}(\omega)
 [\delta_{73}\delta_{84} + Q_{78,34}^{(-+)}(\omega)] +
 P_{12,34}^{(++)}(\omega),
\label{eq:A}
\end{array}\end{equation}
where
\begin{equation}
Q^{(+-)}_{12,34}(\omega) = -\delta_{\sigma_{1},\sigma_{2}}         
\delta_{\sigma_{3},-\sigma_{4}} A^{[1]}_{12,34}(\omega)\sigma_{3}
(\omega-\tilde{\epsilon}_{34})
\end{equation}

\begin{equation}
Q^{(-+)}_{12,34}(\omega) = -\delta_{\sigma_{1},-\sigma_{2}}       
\delta_{\sigma_{3},\sigma_{4}} A^{[1]}_{12,34}(\omega)\sigma_{1}
(\omega-\tilde{\epsilon}_{12})
\end{equation}

\begin{equation}
P^{(++)}_{12,34}(\omega) = \delta_{\sigma_{1},\sigma_{2}}
\delta_{\sigma_{3},\sigma_{4}} A^{[1]}_{12,34}(\omega)
\end{equation}

\begin{equation}\begin{array}{c}
A^{[1]}_{12,34}(\omega) = - \int^{+\infty}_{-\infty}\fr{d\epsilon     
d\epsilon'}{(2\pi\imath)^{2}}
                  \tilde{G}_{1}(\epsilon +\omega)\tilde{G}_{2}(\epsilon)U^{e}_{12,34}(\omega,\epsilon,\epsilon')
                   \tilde{G}_{3}(\epsilon' +\omega) \tilde{G}_{4}(\epsilon' )-\\[4mm]
                    \int^{+\infty}_{-\infty}\fr{d\epsilon }{(2\pi\imath)}[\delta_{31}
                     \tilde{G}_{1}(\epsilon +\omega) \tilde{G}_{2}(\epsilon)
                     \Sigma^{e}_{24}(\epsilon) \tilde{G}_{4}(\epsilon)+ \\[4mm]
                    \delta_{24}\tilde{G}_{3}(\epsilon +\omega)\Sigma^{e}_{31}(\epsilon +\omega)
                  \tilde{G}_{1}(\epsilon +\omega)\tilde{G}_{2}(\epsilon)],
\end{array}\end{equation}
with $\tilde{\epsilon}_{12} = \tilde{\epsilon}_{1} -
\tilde{\epsilon}_{2}$
 and the GF $\tilde{G}_{1}(\epsilon)$ given by Eq.~(2.21).

The  only equation that must be solved to find the propagator  
$A(\omega)$ is the equation for the ph-ph part $A^{(--)}_{56,78}$
of the propagator:
\begin{equation}
A^{(--)}_{12,34}(\omega) = \tilde{A}_{12,34}(\omega) -
\sum_{56,78}\tilde{A}_{12,56}(\omega)\Phi_{56,78}(\omega)A^{(--)}_{78,34}(\omega),
\end{equation}
where the propagator
\begin{eqnarray}
\tilde{A}_{12,34}(\omega) = -\delta_{13}\delta_{24}\sigma_{1}
                           \delta_{\sigma_{1},-\sigma_{2}} /(\omega-\tilde{\epsilon}_{12})
\end{eqnarray}
is the ``refined'' RPA propagator and
\begin{equation}
\Phi_{12,34}(\omega) = \tilde{\Phi}_{12,34}(\omega) +
\Phi_{12,34}^{comp}(\omega)
\end{equation}
\begin{equation}
\tilde{\Phi}_{12,34}(\omega) = -\delta_{\sigma_{1},-\sigma_{2}}        
\delta_{\sigma_{3},-\sigma_{4}}\sigma_{1}(\omega-\tilde{\epsilon}_{12})
A^{[1]}_{12,34}(\omega)\sigma_{3} (\omega-\tilde{\epsilon}_{34}).
\end{equation}
The expression for $\Phi^{comp}(\omega)$ is given in the Appendix.

Properties of the generalized propagator $A(\omega)$ have already     
been discussed in detail in ref.~\cite{ktts97}. Here we enumerate
only the most important ones. Let us first introduce for
convenience a new terminology. Let us represent our propagator in
the form
\begin{equation}\begin{array}{c}
A(\omega) = A^{(--)}(\omega) + A^{(-+)}(\omega) + A^{(+-)}(\omega)+ A^{(++)}(\omega) \\[4mm] 
A^{(-+)}(\omega) = A^{(--)}(\omega)Q^{(-+)}(\omega),\;\; A^{(+-)}(\omega) = Q^{(+-)}(\omega)A^{(--)}(\omega),\\[4mm]
A^{(++)}(\omega) = Q^{(+-)}(\omega)A^{(--)}Q^{(-+)}(\omega) +
P^{(++)}(\omega)
\end{array}\end{equation}
and  refer to the term $A^{(--)}$ as the {\it unlike} component
(e.g., the ph-ph component), and to the remaining three terms as
the {\it associated} components.

First we consider the question of the second order poles, which is   
the main technical problem that the MCDD is designed to solve. It
is obvious, in fact, that $A(\omega)$ does not contain
second-order poles at the points $\omega = \pm
(\tilde{\epsilon}_{p} - \tilde{\epsilon}_{h})$, which are simple
poles of the RPA propagator $\tilde{A}(\omega)$. Moreover it
follows from Eq.~(2.50) that the function $A^{(--)}$, and
consequently $A(\omega)$ too, is regular at these points, as would
be expected if the matrix $[\tilde{A}^{-1}(\omega) + \Phi
(\omega)]$ is invertible at these values of $\omega$, as is true
except for accidental cases. One can also prove \cite{ktts97}
that: (i) in solving the problem of constructing the MCCD
propagator in the complete one-particle basis, all its components
are regular at singular points of the function $A^{[1]}(\omega)$
(Eq.~(2.49)), and (ii) when the full one-particle basis
 $\{\tilde{\phi}_{1},\tilde{\epsilon}_{1}\}$
is used, all the poles of the propagator $A(\omega)$ coincide with
the poles of its unlike component $A^{(--)}(\omega)$.

As for the poles of the function $A^{(--)}(\omega)$ itself, the  
number of poles of $A^{(--)}(\omega)$ can be greater than the
total number of poles of the functions $\tilde{A}(\omega)$ and
$\Phi (\omega)$ due to splitting of the poles of $\Phi (\omega)$
in the complete or partial lifting of their degeneracy in the
solution of Eq.~(2.50). These effects are of order~g$^{2}$.


Let us also point out a role of the term
$Q^{(+-)}A^{(--)}Q^{(-+)}$ in Eq.(2.54) for the MCDD propagator,
although formally this term is of order of g$^{4}$. The analysis
of its poles shows that they coincide with the poles of the unlike
component $A^{(--)}$ and can ensure the removal of the
singularities of the function $A^{(++)}$, which coincide with the    
poles of $P^{(++)}$ of Eq.~(2.54). This absence of ``extra'' poles
provides an indirect confirmation of our  model approximation.

Using Eqs.~(2.45)--(2.49) or  formulas from the Appendix, one
obtains for the MCDD propagator
\begin{equation}
\sum_{1}A_{11,34}(\omega) = \sum_{3}A_{12,33}(\omega) = 0.
\end{equation}
It follows from Eq.~(2.9) that analogous identities are valid for
the renormalized response function $\tilde{R}(\omega)$. We then
find that the relation
\begin{equation}
\sum_{1}\rho_{11} = 0
\end{equation}
holds in our approach. This means that an external field does not   
change the total number of particles; that is, we have the
particle number conservation law. Were the problem to be solved
exactly, this would be obvious from very general conditions. As in
many other models, however, the validity of the law is not obvious
{\it a priori}. This implies that the graphs with self-energy
insertions $\Sigma^{e}$ and phonons in the cross ph channel
$U^{e}$ must be taken into account simultaneously.

\subsubsection[Ground state correl. induced by 1p1h$\otimes$phonon configs.]{Ground state correlations induced by 1p1h$\otimes$phonon configurations}
As discussed earlier, in the GF graphical language any GSC            
correspond to the so-called backward-going diagrams. In  Fig.~2.5
we show some examples of the graphs of order g$^{2}$ that
correspond to the GSC caused by 1p1h$\otimes$phonon configurations
(GSC$_{phon}$). In  the approach under discussion there are two
types of GSC$_{phon}$ effects: those whose inclusion, as in the
RPA GSC case, affect only the location and the magnitude of the
residues of our ph propagator A($\omega$) and those of new type
that can give additional poles of the function $A(\omega)$. In the
latter case our approach is qualitatively different from the RPA.
Some of the corresponding graphs are shown in Figs.~2.5e--h. These
new poles are caused by the associated components of the MCDD
propagator, which are directly connected with GSC$_{phon}$
effects. The same components can also give GSC$_{phon}$ effects
that are not connected with the new poles and, in this sense, are
analogous to the RPA GSC.

Another important feature of the associated components of the MCDD
propagator is that the solution of Eqs.~($\ref{eq:50}$) or (2.19)    
using this propagator also contains associated components.
Therefore---and contrary to the RPA case, according to the
spectral expansion of the response function, non-zero components
of the transition densities $\rho_{pp'}^{n0}$ and
$\rho_{hh'}^{n0}$ appear, where $\rho_{12}^{n0} =
<n|a_{1}^{+}a_{2}|0>$. These effects are caused by the
GSC$_{phon}$ corresponding to the above-mentioned new poles.
Physically, the corresponding graphs describe processes of
creation from the vacuum or annihilation of two ph pairs
simultaneously, i.e, 2p2h configurations. In other words, the
external field can directly excite 2p2h configuration from the
correlated ground state in such a way that a particle of the 2p2h
configuration makes a transition from one state to another ($pp'$
transition) or a hole of this configuration makes a transition to
another hole state ($hh'$ transition). This means that new
components, $\rho_{pp'}^{n0}$ or $\rho_{hh'}^{n0}$, appear in the
spectral expansion of the response function in addition to the
usual $\rho_{ph}^{n0}$ and $\rho_{hp}^{n0}$ components.  The
inclusion of these effects leads to a change of the sum rules for
the moments of the strength function. Within the ESRPA, that is,
for the case of ``pure'' 2p2h configurations, this question has
been treated analytically in ref.~\cite{adachilip88} and in
ref.~\cite{drozdz90}, where calculations for the M1 and
Gamov-Teller resonances in $^{48}$Ca have been reported. Our
calculations (see section~3), where perturbation theory on the ph
interaction was not used, show the quantitative importance of the
inclusion of the associated components of the propagator.

We can already see from this brief discussion of the GSC$_{phon}$ 
effects that they manifest themselves in a considerably more
complicated way than the RPA GSC effects. This question therefore
merits further study.

\subsection{Inclusion of the single-particle continuum}
In the coordinate representation the equation  for the density
matrix is obtained from Eq.~(2.19):
\begin{equation}
\begin{array}{c}
\rho({\bf r},\omega) = -\int A ({\bf r}, {\bf r'},\omega)
{\tilde{e}_{q}} {V^{0}} ({\bf r'},\omega)
d^{3}r' - \hspace{1.75in}\\
\int A ({\bf r}, {\bf r_{1}},\omega) \tilde{F} ({\bf r_{1},r_{2}})
{\rho ({\bf r_{2}},\omega)} d^{3}r_{1}d^{3}r_{2}. \nonumber
\end{array}
\end{equation}

 The present equation for the change of the density matrix is formally identical with the corresponding equation (1.35) for the case of the standard
TFFS. However, the quantities which enter
$\{\tilde{e}_{q},\tilde{F}\}$ and the propagator $A$ of the
extended theory are defined differently compared with the
quantities $\{e_{q},F\}$ and $A$ of the standard theory. The
propagator of the ETFFS differs the most from the standard RPA
propagator. It consists of two parts: the  RPA-like part and the
part that contains much more complicated physics caused by
1p1h$\otimes$phonon configurations, including GSC$_{phon}$.

As discussed for the case of  the standard TFFS (section~1.2.4),    
use of the coordinate representation allows accounting for the
single-particle continuum exactly. The effects associated with
resonance decay into the continuum (the width $\Gamma\uparrow$),
that is, with nucleon emission from the nucleus, are thereby
automatically included in the width.

The systematic use of this technique for the case of complex
configurations leads to great computational difficulties. (See,
for example, the general expression for $A({\bf r,r'},\omega)$ in
the g$^{2}$ approximation obtained in ref.~\cite{kt86}.) At this
stage we therefore included the continuum  only in the RPA part of
the propagator, using the idea of the so-called combined (${\bf
r},\lambda$) representation that was developed within the TFFS for
nuclei with pairing \cite{platonov87}. In our ETFFS the expression
for the propagator $A$ is given by
\begin{equation}
\begin{array}{l}
A({\bf r},{\bf r'},\omega)  = {\tilde A}_{cont}^{RPA} ({\bf r},
{\bf r'},\omega)\; + \\ 
 \hspace{0.875in}\sum_{1234}
(A_{1234}(\omega) - {\tilde A}_{1234}^{RPA}(\omega))                   
 {\tilde {\varphi}^*_{1} ({\bf r}}) {\tilde
{\varphi}_{2} ({\bf r}}) {\tilde {\varphi}_{3} ({\bf r'}}) {\tilde
{\varphi}^*_{4} ({\bf r'}})\,, \label{eq:26}
\end{array}
\end{equation}
where $ {\tilde A}_{cont}^{RPA}$ is the refined RPA propagator in
which the single-particle continuum is taken into account exactly.
In the following applications the summation in Eq.~(\ref{eq:26})
is performed over two shells above and all shells below the Fermi
level. If one would sum over all configurations, the terms with
$\tilde{A}^{RPA}$ in Eq.~(\ref{eq:26}) would cancel each other and
we would have the exact expression for $A({\bf r},{\bf
r'},\omega)$.  It is necessary to subtract the
$\tilde{A}^{RPA}_{1234}$ term in order to avoid double counting as
the full propagator $A_{1234}$ contains the RPA part already.
Thus, in this method the single-particle continuum is taken into
account completely only in the RPA part of the generalized
propagator. Our calculations show that this approximation is
satisfactory. The expression for the propagator $A_{1234}$ in  the
``refined'' single-particle basis
$\{\tilde{\varphi}_{\lambda},\tilde\epsilon_{\lambda}\}$, is given
explicitly in the Appendix.

The distribution of the transition strength is then given by
\begin{equation}
S(\omega,\Delta) = - \frac{1}{\pi} \; {\rm Im} \; \int d^{3}r
[\tilde{e}_{q} V^{0}({\bf r},\omega)]^{*} \rho({\bf r},
\omega+\imath\Delta)\,,
\end{equation}
from which one can easily obtain the transition probabilities and
the EWSR summed in an energy interval.

Equations~(2.57), (2.58) and (2.59) are the basic relations used   
in the calculations reported below.

\setcounter{equation}{0}
\section{Application to giant resonances}
\subsection{Scheme of the calculations}
In the previous section the basic equation (2.57) of the ETFFS for  
the case of closed shell nuclei has been formulated in coordinate
space for the change of the density matrix in an external field
$V^{0}({\bf r},\omega)$ with the energy $\omega$.
 The strength function (Eq.~(2.59))  gives
the energy distribution of the excitation strength under
consideration:
\begin{equation}
S(\omega, \Delta) = \frac{dB(EL)}{d\omega} = - \frac{1}{\pi} \;
{\rm Im} \; \Pi(\omega + \imath\Delta),
\end{equation}
where
\begin{equation}
\Pi(\omega + \imath\Delta) = \int d^{3}r [\tilde{e}_{q} V^{0}({\bf
r})]^{*} \rho({\bf r}, \omega+\imath\Delta)
\end{equation}
is the polarizability propagator and  $\Delta$ is a smearing
parameter. By using this we take into account phenomenologically
those complex configurations that are not explicitly treated and
we simulate the finite experimental resolution. The use of a
sufficiently large $\Delta$ also greatly reduces numerical
difficulties in the calculations. From the solutions of this
equation one obtains
the transition probabilities
and the EWSR, summed over the energy interval [$E_{1}$, $E_{2}$],
---e.g., the linear EWSR:                                              
\begin{equation}
S_{L} = \sum_{[E_{1},E_{2}]} E_{i} B_{i}(EL)\uparrow \; = \;
\frac{1}{2\pi\imath} \oint d\omega  \;\omega  \Pi(\omega).
\end{equation}
Here the integration contour in the complex plain intersects the
real energy axis at the points
$E_{1}$ and $E_{2}$.

\subsubsection{Electric sum rules}

In the following we discuss electromagnetically induced E0, E1 and
E2 excitations. In order to avoid  possible confusion we will
define all relevant quantities explicitly. The electric operators
are given as:
\begin{equation}
Q_{LM} = e \sum_{i = 1}^{Z}  r^{L}_{i} Y_{LM}(\Omega_{i}),
\hspace{2cm} L \geq 2
\end{equation}

\begin{equation}
Q_{00} = e \sum_{i = 1}^{Z}  r^{2}_{i}
\end{equation}

\begin{equation}
Q_{1M} = \frac{eN}{A} \sum^{Z}_{i = 1}  r_{i} Y_{1M}(\Omega_{i}) -
         \frac{eZ}{A} \sum^{N}_{i = 1}  r_{i} Y_{1M}(\Omega_{i}),
\end{equation}
where the electric isovector dipole operator contains the
well-known kinematic charges $e_{p} = Ne/A$, $e_{n} = -Ze/A$. The
linear (full) energy--weighted sum rules for these operators        
$$
S_{L} \equiv  EWSR = \sum_{k} (E_{k} - E_{0}) B_{k}(EL)\uparrow
$$
are then given by \cite{624r20}                                   
\begin{equation}
S_{L} = \frac{\hbar^{2}e^{2}}{8\pi m_{p}} L(2L + 1)^{2} Z <
r^{2L-2} >_{p}\,,  \hspace{2cm}
 L \geq 2,
\end{equation}

\begin{equation}
S_{0} = \frac{2\hbar^{2}e^{2}}{m_{p}} Z < r^{2} >_{p},
\end{equation}

\begin{equation}
S_{1} = \frac{9\hbar^{2}e^{2}}{8m_{p}} \frac{NZ}{A},
\end{equation}
where the radial average is
usually                                                        
taken over the proton distribution in
the ground state. The electric dipole EWSR,  Eq.~(3.9),
corresponds to the well-known Thomas-Reiche-Kuhn sum rules for
photoabsorption cross section,
$$
\sigma_{0}^{cl} =  \int \sigma(E) dE =
\frac{2\pi^{2}e^{2}\hbar}{mc} \frac{NZ}{A}.
$$

The sums  $S_{2}$ and $S_{0}$ include both the isoscalar ($\Delta
T$ = 0) and isovector ($\Delta T$ = 1) contributions, where $T$ is
the isospin of the nucleus.
These contributions may be separated for the E0 and
 EL (L$ \geq 2$) transitions                                      
by means of the additional physical approximation $<\tau_{3}>$ =
($N-Z$)/$A$~\cite{624r20} so that
$$
Q_{LM} = Q_{LM}^{0} + Q_{LM}^{1},
$$
where for $L \geq 2$
\begin{equation}
Q^{0}_{LM} = \frac{eZ}{A}  \sum_{i = 1}^{A}  r^{L}_{i}
Y_{LM}(\Omega_{i})
\end{equation}
and
\begin{equation}
Q_{LM}^{1} = \frac{eN}{A}  \sum_{i = 1}^{Z} r^{L}_{i}
Y_{LM}(\Omega_{i}) -
             \frac{eZ}{A}  \sum_{i = 1}^{N} r^{L}_{i} Y_{LM}(\Omega_{i})
\end{equation}
 and for $L = 0$
\begin{equation}
Q_{00}^{0} = \frac{eZ}{A}  \sum^{A}_{i = 1}  r^{2}_{i}
\end{equation}
\begin{equation}
Q_{00}^{1} = \frac{eN}{A}  \sum_{i = 1}^{Z}  r^{2}_{i} -
             \frac{eZ}{A}  \sum_{i = 1}^{N}  r^{2}_{i}.
\end{equation}
 The effective charges in Eqs.~(3.11) and (3.13)
coincide with the kinematic charges for isovector E1 excitations,
but they have another physical origin.

These operators give the following isoscalar and isovector
electric EWSR
\begin{equation}
S_{L}^{0} = \frac{\hbar^{2}e^{2}}{8\pi m_{p}} L(2L + 1)^{2}
\frac{Z^{2}}{A}
             < r^{(2L - 2)} >_{p}
\end{equation}
\begin{equation}
S_{L}^{1} = \frac{\hbar^{2}e^{2}}{8\pi m_{p}} L(2L + 1)^{2}
\frac{ZN}{A}
             < r^{(2L - 2)} >_{p}
\end{equation}
for $L \geq 2$
 and
\begin{equation}
S_{0}^{0} = \frac{2\hbar^{2}e^{2}}{m_{p}} \frac{Z^{2}}{A} < r^{2}
>_{p}
\end{equation}
\begin{equation}
S_{0}^{1} = \frac{2\hbar^{2}e^{2}}{m_{p}} \frac{NZ}{A} < r^{2}
>_{p}
\end{equation}
for $L = 0$.

We have calculated our strength distributions $S(E)$ with the      
electric operators (3.4), (3.5) and (3.6) for the full $S(E)$,
(3.10) and (3.12) for the isoscalar $S(E)$ and (3.11) and (3.13)
for the isovector $S(E)$.

 \subsubsection{Input quantities for the ETFFS}
As discussed earlier, as in the standard TFFS, the initial
numerical input to the ETFFS  are two sets of phenomenological
parameters that describe (i) the Woods-Saxon single-particle
potential and (ii) the effective interaction of the Landau-Migdal
type.

In all of the  approaches that include complex configurations in   
this way, two problems concerning  double counting of the complex
configurations (in our case, 1p1h$\otimes$phonon) arise.  They
concern
 the necessity to extract the contribution of the complex
configurations that are treated explicitly from the
``old'' phenomenological quantities, that is,(i) from the mean field   
 and (ii) from                                                         
the effective  interaction and  local charges of
quasiparticles.

As far as the procedure for the mean field is concerned,  this       
``refining'' from our phonons
 goes as follows:  To obtain the new (``refined'') single-particle basis
$\{\tilde\epsilon_{\lambda}, \tilde\varphi_{\lambda}\}$
 we must subtract the contribution of the phonon mixing from the "old"
 phenomenological Woods-Saxon energies $\epsilon_{\lambda}$
 \cite{kam79,kam83},
 \begin{eqnarray}
 \tilde\epsilon_{\lambda} = \epsilon_{\lambda} - \Sigma^{e}_{\lambda \lambda}
(\epsilon_{\lambda}),
\end{eqnarray}
where
 $\Sigma^{e}$ is  the  energy--dependent part of the  mass operator
 in Eq.~(2.1). In our g$^{2}$ approximation it is defined in Eq.~(2.22).
More detailed expressions are given in the Appendix.

As for the effective interaction and                                    
 local charges of quasiparticles, in all our equations                   
the new effective interaction $\tilde F$ and local charges $\tilde
e_{q}$ enter and  play the same role as the corresponding
quantities $F$ and $e_{q}$ of the TFFS. By definition they should
not contain the complex configurations that are considered
explicitly and, therefore, the parameters describing in this
approximation $\tilde F$ and $\tilde e_{q}$ have to be determined,
in principle, from experiment. However, as discussed in
section~1.4, we use only a small number of low-lying phonons,
which give the main contribution to the giant resonances
characteristics under consideration. The corrections to the
quantities $F$ and $e_{q}$ due to these phonons must be irregular
and of a long-range character. Therefore one can hope that these
non-local corrections of the  local quantities $F$ and $e_{q}$ are
not very important and that we can use the old parameters of $F$
and $e_{q}$ instead of the new ones. A detailed investigation of      
the Landau-Migdal parameters shows indeed that such changes,
except in the case of $f_{ex}$, are negligibly small.

Thus, although  additional configurations,  are treated in the
ETFFS explicitly, it is possible to use the the parameters of the
TFFS approach. This means that  all but one, of the input
quantities of the ETFFS are, at least to a good approximation,
already known.

\subsubsection{Numerical details}

For the reasons mentioned above, we use in the following
application of the ETFFS the same Landau-Migdal interaction as we
had used previously in the standart TFFS, where we also
investigated low- and high-lying collective states. The parameters
are given in Eq.~(1.42). The only exception is the parameter
$f_{ex}$. Because of the non-self-consistency of the Landau-Migdal
approach, $f_{ex}$ should be chosen in such a way that the
spurious $1^{-}$ state lays at zero energy. In our calculations
we used for the density interpolation in Eq.~(1.11) the
theoretical ground state density  $\rho_{0}(r)$ given by
Eq.~(1.43). In our opinion, such a choice of $\rho_{0}(r)$ instead
of the usual Woods-Saxon form---which we have used starting with
ref.~\cite{kstprl}---makes our calculations more consistent. (See
also the discussion of previous values of $f_{ex}$ in
ref.~\cite{ktts97}.)

For the medium mass nuclei $^{40}$Ca, $^{48}$Ca and $^{56}$Ni,
using Eq.~(1.43) gives noticeably different results
(mainly for the IS E0 resonances)                                      
 from those
obtained using the Woods-Saxon form of $\rho_{0}(r)$ as compared
with the difference for $^{208}$Pb.  In particular,                   
 this change in the numerical scheme                                  
has shown  that the  effects of GSC$_{phon}$ there were numerically               
overestimated by us \cite{ekstw94}                                    
although their qualitative features                       
were left unchanged~\cite{kst97}.                                              

The phonon characteristics, that is, their energies and transition
probabilities, which have been calculated within the standard TFFS
\cite{AB} with the same parameters (Eq.~(1.42)),
 are  given in Table~3.1 and in ref.~\cite{ekstw94}.
The criteria for the selection of these phonons were that they
should have the strongest coupling parameters---and therefore
the largest transition       
probabilities, and that the energies of the corresponding
configurations should be in                                        
 the energy region of the giant resonance.

The influence of the different number of phonons under
consideration was  investigated for $^{208}$Pb. In
ref.~\cite{ktt91} it was shown that the role played by the
``unnatural parity'' states (e.g., $2^{-}_{1}, 3^{+}_{1},
4^{-}_{1}$ and 5$^{+}_{1}$) is very small. A similar  result for a
much larger number of unnatural parity phonons has been obtained
within a simpler model \cite{wml82}. In  refs.~\cite{ktt91,kstt93}
the contribution of the four 2$^{+}$ phonons corresponding to the
isoscalar E2 resonance at an energy of about 10~MeV in $^{208}$Pb
has also been checked. It turned out that their contribution was
only noticeable in the high-energy tail of the E1 resonance.

The single-particle levels and wave functions were calculated with
the standard Woods-Saxon potential \cite{chep67}. In order to get
good agreement with the experimental single-particle energies, the
well depth of the central part of the potential was adjusted by     
changing the depth parameter by less than 5 percent.  The
single-particle energies thus obtained are given in refs.
\cite{kstw93,kst97}.

Numerical values of all the EWSR used in the calculations are
presented in Table~3.2. The quantities $<r^{2}>_{p}$ were obtained
using the proton density distribution in the ground state,
Eq.~(1.43), calculated from the single-particle scheme described    
above.

The mean energies  and  dispersion $D$ are                      
determined by
\begin{equation}
E_{1,0} = \frac{m_{1}}{m_{0}}, \hspace{3mm} E_{3,1}                
 = (\frac{m_{3}}{m_{1}})^{1/2}, \hspace{3mm}
E_{2,0} = (\frac{m_{2}}{m_{0}})^{1/2}, \hspace{3mm} D =                   
\sqrt{\frac{m_{2}}{m_{0}} - (\frac{m_{1}}{m_{0}})^{2}},
\end{equation}
where the energy moments $m_{k}$ corresponding to the energy       
interval $\Delta E = E_{max} - E_{min}$  have been calculated
using
\begin{equation}
m_{k} = \int_{E_{min}}^{E_{max}}dE \; E^{k} S(E).
\end{equation}

\subsection{Electric and magnetic resonances in magic nuclei}


\subsubsection{Photoabsorption cross sections; E1 excitations.}   
The E1 absorption cross section has been calculated using the
formula
\begin{equation}
\sigma_{E1}(\omega) = 4.022\,\omega S_{E1}(\omega),
\end{equation}
where $\omega$ is taken in MeV, $S$ in fm$^{2}$-MeV$^{-1}$ and
$\sigma$ in mb.

The results of these calculations in $^{40}$Ca, $^{48}$Ca and     
$^{208}$Pb and the comparisons with experiment are presented in
Figs.~3.1--3.3 and Table~3.3 \cite{kstt93}. The label
1p1h+continuum in Table~3.3 denotes the calculations done within
the CRPA (or CTFFS) and the label 1p1h+2p2h denotes those without
the continuum but with the smearing parameter $\Delta \ne 0$. In
order to obtain the integral characteristics of the resonances we
used a Lorentz function to approximate the resonance curves, as is
usually done in analyzing experiments. The parameters of this
function---the mean energy $\overline{E}$, the maximum value of
the cross section $\sigma_{max}$ and the resonance width
$\Gamma$---were found from the condition that the three lowest
energy moments in Eq.~(3.19) ($k = 0,1,2$) coincide for the exact
and approximate resonance curves. The characteristics of the
fitted lorentzians---the mean energy $\overline{E}$,
 the width $\Gamma$ and  the integral cross sections $\sigma_{0}$
---were obtained for the same experimental energy interval
 and  are compared with the corresponding experimental
 values in Table~3.3.  For details see ref.~\cite{kstt93}.

In the case of CRPA  we see that the mean energy values        
$\overline{E}$ of the E1 resonances and the widths do not agree
with experiment. The energies are too low by 1 to 2 MeV, where as
the widths are off by more than a factor of two. The inclusion of
the 1p1h$\otimes$phonon configurations noticeably improves the
agreement with experiment as compared with the CRPA. As can be
seen from Figs.~3.1-3.3, the complex correlations lead to a large
reduction of the maxima of the photoabsorption cross sections and
to a better description of the E1 resonance tails.

The final results (denoted in Table~3.3 as 1p1h+2p2h+continuum)
show that inclusion of the continuum changes the values of
$\overline{E}$ by 0.6--0.8~MeV in the required direction as
compared with the 1p1h+2p2h approximation, except for the case of
$^{40}$Ca. The theoretical mean energies for $^{48}$Ca and $^{208}$Pb    
 are now in good agreement with the data.

The most important result of these calculations, however, is the now       
satisfactory explanation of E1 widths.
  To be precise, they are                                               
explained to an accuracy of the value of the smearing parameter used,        
which is smaller than the widths.                                
 The two new (as compared
with the standard RPA) elements, namely the 1p1h$\otimes$phonon
configurations and the single-particle continuum, increase the
values of $\Gamma$ by about a factor of two. As should be
expected, the contribution of the continuum to $\Gamma$ is small
for the heavy nucleus $^{208}$Pb and somewhat larger for the
lighter ones. Specifically, its contribution to the calculated
total width is 14 percent for $^{40}$Ca and 28 percent for
$^{48}$Ca, but only 7 percent for $^{208}$Pb. Now we underestimate
$\Gamma_{exp}$ by only 14 percent for $^{40}$Ca, 11 percent for
$^{48}$Ca and 4 percent for $^{208}$Pb. These small differences
may be connected with the approximate treatment of the
single-particle continuum.

In Figs.~3.1--3.3  the theoretical curves were approximated by   
lorentzians in the energy interval in which there are experimental
points and, correspondingly, the integral characteristics of
Table~3.3 were calculated for the same interval. In order to study
the EWSR questions and the role of the GSC$_{phon}$ it is
necessary, at least at first, to increase the energy interval
under consideration. Therefore, in our later calculations
\cite{kst97} for $^{40}$Ca, $^{48}$Ca and $^{56}$Ni, considerably
larger energy intervals were studied and the corresponding
theoretical curves are presented there.  Comparing these results
with experimental curves for $^{40}$Ca \cite{ahrens75} and
 $^{48}$Ca \cite{o'keefe87}, one can conclude that we have also
obtained agreement with experiment for mean energies, total widths
and maximum values of photoabsorption cross sections, although the
agreement is better for the previous analysis, as should be
expected.  This agreement is  mainly obtained by  inclusion of
1p1h$\otimes$phonon configurations, but the role of our
GSC$_{phon}$ is noticeable, for example, in the integral quantity E1  
EWSR (see Table~3.4. and section~3.2.4.).
We have also calculated the depletion of EWSR $\sigma_{0}$ in the
experimentally studied intervals 10.0--32.0~MeV
($\sigma_{0}^{exp}$ = 637.7~mb-MeV) for $^{40}$Ca and
11.0--27.5~MeV ($\sigma_{0}^{exp}$ = 836.6~mb-MeV) for $^{48}$Ca
and obtained reasonable agreement with experiment: 599.2~mb-MeV
(99.9 percent of $\sigma_{0}^{cl}$) and 635.3~mb-MeV (90.8 percent
of $\sigma_{0}^{cl}$), respectively.

The integral characteristics calculated for the large intervals    
are given in Table~3.4. The depletion of the corresponding EWSR
(in percentages of $S_{1}$, Eq.~(3.9)) in Table~3.4 were obtained
using the values of the reference  EWSR from Table~3.2. We have
obtained  97--101 percent of the E1  EWSR for the RPA case as well
as for the RPA+1p1h$\otimes$phonon configurations                    
 (without GSC$_{phon}$).
The case with GSC$_{phon}$ will be discussed in section~3.2.4.

\subsubsection{E0 and E2 resonances}

The  isoscalar electric monopole (IS E0) giant resonance  in
nuclei  is  a unique source of information on  the compressibility
and equation of state of nuclei. The extrapolation to  nuclear
matter and neutron  stars requires that the energy of the
resonance be known over a wide range of the mass number $A$ and
for very different numbers of protons and neutrons. Here we
discuss ETFFS results simultaneously for the  E0 and  E2             
resonances because these resonances have rather similar excitation
energies and overlap to a large extent. The results of isoscalar
(IS) and also isovector (IV) E0 and E2 resonances are given in
Figs.~3.4, 3.5, Tables 3.5, 3.6 for $^{40}$Ca, $^{48}$Ca and
$^{56}$Ni and in Figs.~3.6, 3.7 and Table~3.7 for $^{208}$Pb.

\begin{center}
{\it E0 and E2 resonances in $^{40}${\rm Ca} and $^{48}${\rm Ca}}
\end{center}

The main part of the IS E0-resonance in $^{40}$Ca                   
---65 percent of S$_{0}^{0}$
as taken from Table 3.2, is                                          
in the 11.0--23.0~MeV interval (see Fig.~3.5).
In the 6.0--30.0~MeV interval we have 85.5 percent..
 The comparison with experiment for
isoscalar E0 and E2 excitations\footnote{ In our comparisons with
experiments there is an uncertainty because in the experimental
works, as a rule, there is no exact
information about numerical values of the reference EWSR used.     
Therefore there may be a discrepancy due to  the different values
of $<r^{2}>_{p}$ used here and in the experimental articles.}
in $^{40}$Ca has been made in ref.~\cite{kstprl}. It was shown   
there that in order to explain the electron scattering experiments
\cite{624r7} where both of the resonances (E0 and E2) are excited,
it was necessary to take into account GSC$_{phon}$ for both of
them. This can be seen in Fig.~3.4 for the full electromagnetic E0
and E2 strengths in $^{40}$Ca and also in Fig.~3.5. See also the
discussion in section~4.

For the mean energies of the isoscalar E0 resonance in $^{40}$Ca
we obtained $E_{1,0}^{th}$ = 18.4~MeV (Table~3.5), compared with the      
experimental value of $E_{1,0}^{exp}$ =
18.89~$\pm$~0.11~MeV~\cite{er4}. Our theoretical results agree
with experiment reasonably well not only for the integral
characteristics, but we also for     
gross structure of the isoscalar E0 strength function in Fig.~3.5
and for of the cross section in refs.~\cite{er4,lui2001}, as we
shall see in section~4.

It was first observed in ref.~\cite{624r5} that the full isoscalar E2    
strength in $^{40}$Ca is located in the 0--22~Mev region and is
divided into approximately equal parts around 13.5~Mev and 18~Mev.
This splitting was confirmed later \cite{624r6,624r7}. Our
calculations give 73.8 percent of S$^{0}_{2}$  for the 0--22~MeV
region, 18.3 percent  for the 12--15~MeV region and 27.9 percent
in the broader 10.5-16.5~MeV interval. For the 16.5--19.5~MeV and
15--21~MeV regions,  we obtain 25 percent and 44.6 percent,
respectively. It is therefore difficult to speak about the {\it
equal} distribution between the regions around 13.5~Mev and
18~MeV, but the trend is the same and corresponds roughly to
experiment. The splitting can be seen in Fig.~3.5. In agreement
with the experiment in ref.~\cite{624r7}, we obtained also the
third maximum at 12~MeV.

As can be seen from Fig.~3.5, the isovector E0 as well as
isovector E2 resonances are spread out over larger regions than
the corresponding isoscalar E0 and E2 resonances, for one obvious
reason: the attractive isoscalar interaction        
shifts the isoscalar strength down and that reduces the (escape)
width; the repulsive isovector interaction has the opposite
effect. There are also
noticeable low-lying tails of isovector strength in the regions of  
the isoscalar E0 and E2 resonances. (Table~6  of ref.~\cite{kst97}
contains  numerical results for all the nuclei under
consideration.) The total isovector E0+E2 contributions to the
main region of the isoscalar E0 resonance are about one-half for
$^{40}$Ca and $^{56}$Ni and about one-fourth for $^{48}$Ca. These
results may be important for electron scattering on these nuclei;
at least a similar effect has been obtained in
$^{28}$Si($e,e'\alpha$) coincidence scattering \cite{624Knopfle}.

Recently, the comparison of our calculations for the low-energy     
isoscalar E2 strength in $^{40}$Ca (in the 10--18~MeV interval)
and $^{48}$Ca (in the 11--14~MeV interval) with the strength that
was extracted from ($p,p'x$) reactions ($x=\alpha_{0}$, $p_{0}$
and $n_{0}$) has been performed in ref.~\cite{schweda2001}. In
order to do this the authors multiplied the calculated E2 strength
by the ground state branching ratios deduced from statistical
model calculations. They obtained good overall agreement with our
calculations, which probably also supports the need to take the
GSC$_{phon}$ into account. On the other hand, a very large percentage of   
the E2 EWSR observed in both $^{40}$Ca and $^{48}$Ca
\cite{ottini99} between 6 and 12~MeV (about 40 percent and 23 percent,     
 respectively), which means in fact a large
shift of these  E2 resonances---especially in $^{40}$Ca---to the
low-energy region, is in strong conflict with the results of our
calculations shown in Fig.~3.5. In these experiments a heavy ion
$^{86}$Kr beam of 60~MeV/nucleon was used to excite the Ca
isotopes.

\begin{center}
{\it E0 and E2 resonances in $^{208}${\rm Pb}}
\end{center}

The E0 and E2 hadronic response functions in $^{208}$Pb have been
calculated here for both the IS and IV resonances  within the
ETFFS and for large energy intervals (see Figs.~3.6 and 3.7 and
Table~3.7). The parameter $f_{ex}$ = --2.2  was obtained by
fitting to the energy of the  2$^{+}_{1}$ level. The difference
from the value of --1.9 obtained in the CTFFS calculations
(section~1.2.5) is caused by the complex configurations that have
been included here. This small difference of the parameters shows
the numerical effect of inclusion of our  complex configurations
for this low-lying level in $^{208}$Pb.

It is natural (and in accord with the experiment) that, as one can  
see from Table~3.7,  the mean energy values and depletion of the
EWSR depend rather sensitively on the energy interval used in the
calculation.
In most cases we obtained reasonable agreement with the
experimental data available for these integral characteristics if
the energy intervals under consideration are comparable. (A recent
summary of the experimental data on the E0 and E2 resonances in
$^{208}$Pb is presented in ref.~\cite{HW}.) As for the cases of
$^{40}$Ca and $^{48}$Ca, there is a noticeable amount of the IS
strength in the main region of the IV strength and vice-versa for
both E0 and E2 resonances.

Comparison  of the E2 results shown in Fig.~3.7 and Table~3.7 with    
those of the CTFFS in Figs.~1.3, 1.4  and Table~1.1, calculated
with the same smearing parameter, $\Delta$ = 250~keV, shows the
role of 1p1h$\otimes$phonon configurations: (i) its inclusion
gives the resonance widths that are numerically  similar to the
observed values of 3.1~$\pm$~0.3~MeV \cite{Bra} and
5~$\pm$~0.5~MeV \cite{DLA} for the IS E2 and IV E2, respectively
and (ii) as for the isovector E1, the mean energy of the IV E2
resonance is changed in the desired direction (see Table~3.7).       

In the calculations with the smearing parameter $\Delta$ = 250~keV
shown in Fig.~3.7, a gross structure of the IS E2 hadronic          
response was obtained that is absent, of course, in the
calculations with $\Delta$ = 800~keV, shown also in Fig.~3.7. Thus
one can see a natural and noticeable difference between these two
calculations that could be checked in hadron experiments. The
calculated fine structure of the {\it electromagnetic} response
function is given in Fig.~3.9 and discussed below.

For the E0 IS and IV resonances in $^{208}$Pb calculated with
$\Delta$ =~250~keV, shown in Fig.~3.6 and Table~3.6, we obtained
approximately the same results as for the E2 case: agreement of
the integral characteristics with experiment, as well as a gross
structure that probably has not been observed.

\begin{center}
{\it On the fine structure of E1 and E2 resonances}
\end{center}
Generally speaking, the description of the giant resonance fine
structure \footnote{ For definiteness we will refer to an observed   
structure as {\it fine} if the experimental resolution is less
than 100~keV and {\it gross} if the resolution is more than
100~keV.} is a natural step once the single-particle continuum and
some complex configurations have been taken into account. A
principal motivation for such a step is the rapidly improving
experimental resolution. The phenomenological smearing parameter
used in our and many other calculations  in practice unifies (or
simulates, to be more precise) two quite different effects:
realistic experimental resolution and complex configurations that
are not treated explicitly in the approach under consideration.

To exclude the influence of smearing and to imitate, in a sense,
the results of future E1 experiments with very good resolution, we
have repeated the calculations for the E1 resonance in $^{208}$Pb
with $\Delta$~=~10~keV.  The results are shown in Fig.~3.8. Of
course, it is unrealistic to expect to observe the fine structure
obtained on the high-lying slope of the E1 resonance,
but the one on the low-lying slope could probably be observed if
the ``strongest'' complex configurations were chosen correctly.  A
more detailed discussion of the low-energy strength is presented        
in section~3.2.5.

The results of the calculations of the E2 resonance fine structure   
are presented in Fig.~3.9 for the electromagnetic IS E2 strength
function in $^{208}$Pb. They improve our earlier calculations
\cite{k...richter97} in the following way: (i) as in the previous
calculations under discussion,                                       
 our GSC$_{phon}$ have been taken into account; (ii)
as in Fig.~3.7, the value of the $f_{ex}$ = --2.2 has been
adjusted to obtain the energy of the 2$^{+}_{1}$ level and (iii)
as mentioned above, Eq.~(1.43) for the nuclear density in the
ground state has been used. The smearing parameter was taken as
$\Delta$~=~40~keV so as to be comparable with the experimental
resolution in the $(e,e')$ and $(p,p')$ experiments, results of
which are presented in Fig.~3.9.

As can be seen from Fig.~3.9 and corresponding Fig.~1 in
ref.~\cite{k...richter97}, we obtained here better agreement with
experiment than in ref.~\cite{k...richter97}. In that work it was
shown that a large part of the fine structure under consideration
was due to complex configurations. The same result can be  clearly
seen  in Fig.~3.9.  All of this confirms the decisive  role  of
our complex 1p1h$\otimes$phonon configurations in the formation of   
the fine structure---at least for this resonance. On the whole,
however, the situation is very involved, due in particular to the
fact that the escape width depends strongly on the excitation
energy and that there are very many non-collective
configurations. If we do not yet include additional complex           
configurations (which is, of course, quite possible to do within
the GF formalism using the general prescription described in
section~2.1), then one of the first steps towards clarifying the
situation is probably to take into account the energy dependence
of the smearing parameter and/or to use a ``refined'' optical
potential that effectively takes into account the complex
configurations that we do not consider explicitly
\cite{kaev97,ks96}.

 In any case, the explanation of the giant
resonance fine structure is a challenge for any microscopic
theory. It is clear that after improving the experimental resolution  
it will be possible to observe more complex configurations
in the fine structure. In this case it will be necessary
to improve the ETFFS by the addition of new configurations,
which should be done following the methods of construction of
the ETFFS described in section 2.                                      
 One can also  expect successes in this direction for
the low-lying structures in the nucleon separation energy region
(see section~3.2.5).

\subsubsection{M1 resonances}
There is a rich history of study of M1 excitations in nuclei
\cite{raman91}. The calculations within the standard TFFS with the
Landau-Migdal interaction were able to reproduce the excitation
energies of the  isoscalar and isovector M1 excitations with the
universal values of spin interaction parameters $g$ and $g'$
defined in Eq.~(1.42)
\cite{ringspeth73,tkachev76,borzovtkachev77,borzov84}.

 The observed total transition
strengths were also reproduced in these calculations because the
standard TFFS contains the universal spin local charges determined
from~\cite{AB}:
\begin{eqnarray}
e^{p}_{q}V^{0p} = (1 - \xi_{l}){\bf j}^{p} + [(1 -
\xi_{s})\gamma^{p} + \xi_{s}\gamma^{n}
               + \frac{1}{2}\xi_{l} - \frac{1}{2}]{\bf {\bfsigma}}^{p} \\ \nonumber
 e^{n}_{q}V^{0n} =  \xi_{l}{\bf j}^{n} + [(1 - \xi_{s})\gamma^{n} + \xi_{s}\gamma^{p}
             + \frac{1}{2}\xi_{l} ]{\bf {\bfsigma}}^{n},
\end{eqnarray}
where $\gamma^{p} = 2.79 \mu_{0}$, $\gamma^{n} = - 1.91 \mu_{0}$,
$\mu_{0} = \frac{e \hbar }{2m_{p}}$ and
\begin{equation}
\xi^{p}_{s} = \xi^{n}_{s} = 0.1,\hspace{0.5cm} \xi^{p}_{l} =
\xi^{n}_{l} = - 0.03,
\end{equation}
as obtained earlier \cite{borzov84,SWW77}. These values yield for   
the spin local charges $e^{p}_{q} = 0.64\gamma^{p}$ and $e^{n}_{q}
= 0.74\gamma^{n}$, which explains the observed quenching of M1
strength. This quenching was one of the reasons for the
long-standing problem of the `` missing'' M1 strength in heavy
nuclei like $^{208}$Pb, which was greatly clarified by the
polarized photon scattering experiments of Laszewski et al.
\cite{lasz88}. (For details see the reviews \cite{sw91,raman91}
and also ref.~\cite{kt89}.)

Such an approach, however, fails to explain the resonance widths    
that, especially in heavy nuclei, are caused by a coupling to more
complex configurations than those accounted for in the RPA. In
other words, the RPA or TFFS calulations did not answer the
important question of why there are so many 1$^+$ levels with
B(M1)$\uparrow \le(2-3)\mu^{2}_{0}$ observed in $^{208}$Pb
\cite{raman91,lasz88}. That was another part of the problem of the
``missing'' M1 strength.

The existence of deviations from the RPA predictions was confirmed  
by the discovery of the low-lying M1 resonance in the magic nuclei
$^{16}$O (at about 15~MeV) \cite{kuechler83} and $^{40}$Ca (at
about 10~MeV) \cite{pringle82} because, according to the RPA, no
M1 resonance with a similar energy can exist if the $p$ shell in
$^{16}$O or the $sd$ shell in $^{40}$Ca is fully occupied. In
order to illustrate this,  we show in Fig.~3.10 the results of
calculations of the M1 resonance in $^{40}$Ca \cite{kt84,kt89}
performed within a simpler model of taking 1p1h$\otimes$phonon
configurations into account and without inclusion of the
single-particle continuum, as was briefly discussed in
section~2.2.  As was shown in refs.~\cite{kt84,kt89}, only the
terms that correspond to the ``backward-going'' graphs---that is,     
the GSC$_{phon}$ graphs---are responsible for the effect. We found
that inclusion of the GSC$_{phon}$ eliminated the prohibition of
the existence of the low-lying M1 excitations in $^{16}$O and
$^{40}$Ca  and, as one can see in Fig.~3.10, reasonably explained
at least the strong M1 excitations observed in
ref.~\cite{pringle82}. One should note, however, that the total M1
strength calculated is substantially larger than that observed
\cite{pringle82} and that the calculations were performed using
the Woods-Saxon form of the nuclear density in the ground state
and with the phonons treated within the Bohr-Mottelson model.

In Fig.~3.11 and Table~3.8 the results of the ETFFS calculations    
using Eqs.~(3.22) and (3.23) for both isoscalar and isovector M1
excitations in $^{208}$Pb are presented. In spite of the
above-mentioned quenching due to the spin local charges, it was
necessary to include the complex 1p1h$\otimes$phonon
configurations as well as our GSC$_{phon}$ in order to obtain
agreement with experiment. The width, which can be deduced from
the curve, agrees with the experimental value of 1~MeV
\cite{lasz88} (smearing parameter $\Delta$ = 100~keV). The
$\Sigma_{i}B_{i}(M1)\uparrow $ value of 11.57 $\mu^{2}_{0}$ in the
6.3--8.7~Mev interval and the mean energy $\overline{E}$ = 7.7~MeV
agree reasonably well with $\Sigma_{i}B_{i}(M1)\uparrow_{exp}
\approx $ 15.6 $\mu^{2}_{0}$ and $\overline{E}_{exp}$ = 7.3~MeV
\cite{lasz88} .                                                      
 A reasonable agreement for the isoscalar 1$^{+}$ level at
$E_{exp}$~=~5.85~MeV was obtained. A more detailed discussion of
the M1 calculations, including those for $^{48}$Ca and $^{54}$Fe,
can be found in refs.~\cite{kstw93,kt89,kt91}.

\subsubsection{Effects of GSC$_{phon}$}

We define the case of absence of GSC$_{phon}$ as that when the    
associated components of the full propagator $A(\omega)$,
Eq.~(2.54) or---in other words---when the quantities $Q^{(-+)},
Q^{(+-)}, P^{(++)}$,
Eq. (2.45) are equal to zero.                                     
In principle also the unlike components A$^{(--)}$ give rise to     
GSC$_{phon}$ but it turns out that this effect is very small,       
as indicated in Tables 3.4, 3.5 and 3.6, where the third row        
includes the effects of the A$^{(--)}$ only. These correlations     
 are similar to the RPA GSC in the sense that they affect only
the locations and values of the residues of the poles
but do not change the EWSR.
 We will discuss in the following only                              
 GSC$_{phon}$, which  are created by
the associated components of the full propagator and which gives
some new effects, as were discussed in section~2.3.4.
and demonstrated for the case of a ``pure'' GSC$^{phon}$ effect       
in section 3.2.3.                                                     


In all the calculations for $^{40,48}$Ca and $^{56}$Ni in the       
large energy intervals, which are given in Tables~3.4, 3.5 and
3.6, we obtained  97--102 percent  of the corresponding EWSR  for   
the RPA case as well as for that of the RPA +
1p1h$\otimes$configurations (without the above-mentioned
GSC$_{phon}$). Taking into account our GSC$_{phon}$, however,
increases the EWSR by 4--7 percent as a rule.
  This result is in
accord with the result obtained analytically in
ref.~\cite{adachilip88}, although our model and the ESRPA used in
that work differ greatly. The main differences are that we use
complex configurations with collective phonons while in the ESRPA
``pure'' 2p2h configurations are used and, in contrast to the
ESRPA, we do not use perturbation theory in the effective
interaction. The complex configurations in the ground state give      
rise to
an increase of   $<r^{2}>_{p}$ and, therefore, of the EWSR, and
perhaps to changes in other ground state characteristics.                  

The role of GSC$_{phon}$ is especially noticeable in the low-lying   
energy region of the E0 and E2 resonances in the nuclei under
consideration. We have from Table 3.2 for the 5--12~MeV interval
in $^{40}$Ca 7 percent of our S$_{0}^{0}$              
and 7.4 percent of S$_{2}^{0}$.
 The contribution of GSC$_{phon}$ to these figures is
3.6 percent and 4.8 percent, respectively---i.e, more than half.
(For comparison, the contributions to this interval obtained
within the RPA are 1.0 percent and 0.4 percent). There is also
additional low-lying strength due to GSC$_{phon}$ in $^{56}$Ni
(Fig.~3.4 ). There it is 4.8 percent of S$_{0}^{0}$ in the            
5.0--14.0~MeV interval as compared with 2.5 percent without
GSC$_{phon}$---i.e., about half. For higher energies the role of
GSC$_{phon}$ is diminished; for example, for the 10--20.5~MeV
interval in $^{40}$Ca their contribution to the full E2 EWSR
decreases it by about one-fourth.

For M1 resonances, as can be seen from Table~3.8 and Fig.~3.12 for
$^{56}$Ni and $^{78}$Ni and Fig.~3.11 for $^{208}$Pb, the role of    
GSC$_{phon}$ is also noticeable, although probably not to the
extent that it is for the electric resonances under discussion.
This is in agreement  with the results obtained for M1 resonances
within the ESRPA \cite{drozdz90}. Their role in the fine structure
of the E1 strength in $^{208}$Pb has been calculated \cite{ks96}
and will be discussed briefly in the next subsection.

\subsubsection{The pygmy resonance and low-lying structures}  

One of the impressive examples of the gross or fine structure in a   
broad region near the nucleon separation energy is the so-called
pygmy resonance. At present this subject is the subject of active
discussion in connection with the general  interest in nuclei with
a large neutron excess because, phenomenologically, the pygmy
resonance is described as the vibration of the neutron excess
against the core with $N\,=\,Z$. Thus, properties of the pygmy
resonance should strongly depend on the  $N/Z$ ratio
\cite{adams96}.

Resonances of the same name were observed in ($n,\gamma$)
spectra at neutron energies from 10~keV up to several MeV in   
many nuclei \cite{igashira86,kopecky90}. They were described by
means of an additional lorentzian in the radiative strength
functions with the fitted parameters $\overline{E} \approx$
2--6~MeV, $\Gamma \approx$ 1.0--1.7~MeV and an integral strength
of 0.1--1.0 percent of the classical sum rule \cite{igashira86}.
In the ($\gamma,n$) cross sections in $^{208}$Pb there are
well-known structures in the 7.6-~12.0~MeV interval
\cite{belyaev92} and between 9.9 and 11.2~MeV \cite{bell82}, which
were partly manifest earlier in the photoabsorbtion cross
section~\cite{veyssiere70} too.

In different experiments for many nuclei with $A$ = 58--208,   
resonance-like structures were observed  as  ``the low-lying
52A$^{-1/3}$ MeV dipole resonance'' \cite{dol85}. Its
characteristics are  $\Gamma$ = 1.2--2.0~MeV and an integral
strength of 1.65--2.5 percent of the experimental integral
strength in the region of the isovector E1 resonance. There are
also other low-lying structures in the broad energy region near
the nucleon separation energy; see, for example,
ref.~\cite{kaev97}.

In the broad energy region under consideration, one should observe  
at least E2, M1 and---mainly---E1 transitions.  As yet, even for
$^{208}$Pb, there are no consistent microscopic calculations with
non-separable forces performed within the {\it same} calculational
scheme for all three multipoles. It is clear that the usual RPA or
QRPA calculations are not able to explain these structures. The
calculations  should take into account more complex configurations
and the single-particle continuum for the energies above the
threshold.   Moreover, as calculations  in $^{208}$Pb  have shown
\cite{ks96}, the role of GSC$_{phon}$ is also essential. The E1
photoabsorption cross section in the 4.5--7.5~MeV interval has
been calculated there within the ETFFS, both with and without
GSC$_{phon}$ (with smearing parameter $\Delta$ = 80~keV), and a
very noticeable difference between these two cases was found. In
particular, the known structure at 5.5~MeV was explained just by
the GSC$_{phon}$ in these calculations. The experimental value of
$\Sigma B(E1) = 1.338\,e^{2}$-fm$^{2}$ \cite{laszaxel79} for six
1$^{-}$ levels is in good agreement with the corresponding
theoretical value of 1.304~$e^{2}$-fm$^{2}$. See also a discussion
of this question in ref.~\cite{kstt93}.

The results of nuclear resonance fluorescence experiments should   
significantly clarify the situation because they are able in
principle to solve the long-standing problem of the identification
of separate $1^{-}, 2^{+}$ and $1^{+}$ states \cite{kneissl96}. In
connection with the numerous and successful results of these
experiments, and with the interest in neutron-rich nuclei, the
term ``pygmy dipole resonance'' (PDR) is used now for the E1 part
of the ``old'' pygmy resonance.

The new experiments in the neutron-rich oxygen isotopes $^{18}$O,   
$^{20}$O and $^{22}$O, which used electromagnetic excitation in
heavy ion collisions at beam energies of about 600 MeV/nucleon,
gave up to 10 percent of the classical E1 sum rule in the region
up to 15~MeV \cite{aumann2001}, which is in sharp contrast to the
dipole response of stable nuclei where the experiments usually
give about 1 percent. It is clear that the role of the
single-particle continuum is essential here but, as was discussed
earlier, the role of complex configurations and, for non-magic
nuclei, of a new (phonon) mechanism of pairing \cite{avekaev99}
should be important for such delicate  properties of nuclei,
especially of unstable ones. Because there are already many
experimental data \cite{kneissl96,hartmann2002} for stable nuclei,
the comparison with the theory, which uses a non-separable
nucleon-nucleon interaction and takes into account simultaneously
the single-particle continuum and complex configurations, will
enable the choice of a reasonable variant of the microscopic
theory to apply to calculations in unstable nuclei.

\subsection{Giant resonances in unstable magic nuclei}        
Because the ETFFS takes into account  the single-particle
continuum and uses the universal parameters of the effective
interaction and local charges, which are the minimum necessary for
calculating properties of unstable nuclei (see sections 1.3.2 and
1.3.3), this approach is applicable to such calculations. On the
other hand, the corresponding measurements could be a convincing
verification of the universality of the ETFFS parameters.

The results of the calculations of E0 and E2 resonances in the
unstable nucleus $^{56}$Ni are presented in Figs.~3.4 and 3.5
and in Tables 3.5 and 3.6. The E1 calculations for this nucleus are     
given in Table~3.4  and in ref.~\cite{kst97}. These results are
largely similar to those  for $^{40}$Ca and $^{48}$Ca discussed
above.

As  discussed in section~1, the general interest in unstable   
nuclei is connected first of all with the neutron-rich nuclei.  We
have therefore calculated the isovector E1 \cite{kstvar} and the
isoscalar and isovector M1 resonance \cite{kstvar,kstw93}              
 (see Fig.~ 3.12) in the very
exotic nucleus $^{78}$Ni.

The calculations of the E1 resonance in refs.~\cite{kstvar,kstw93} were  
performed using our  Woods-Saxon potential to calculate                  
the density in the nuclear  ground state in the interpolation
formula (1.11), and thus with another value of the parameter
$f_{ex}$, but that is not critical---not, at least, for integral
characteristics. It was found that the isovector E1 resonance in
$^{78}$Ni had a width of 5.2~MeV. Its mean energy $\overline{E}$ =
15.2~MeV is much less than that obtained from the empirical
formula $\overline{E} = 78A^{-1/3}$ (= 18.3~MeV). In this case the
influence of the 1p1h$\otimes$phonon configurations is less than
in other, ``less exotic'' nuclei: within the CTFFS we obtained
$\overline{E}$ = 14.7~MeV and $\Gamma$ = 4.8~MeV.

As can be seen                                                         
from Fig.~3.12, the M1 resonance in $^{78}$Ni has an asymmetric
shape with a width of about 1~MeV, whereas this resonance in
$^{56}$Ni and $^{132}$Sn has no fragmentation width.

The M1 resonance in the neutron-deficient $^{100}$Sn nucleus was   
split into two major peaks, with $\overline{E}_{1}$ = 9.8~MeV and
$\overline{E}_{2}$ = 10.5~MeV (note the smearing parameter
$\Delta$ = 100~keV). This result was obtained without
GSC$_{phon}$, however in ref.~\cite{ave96} they were taken into
account in a simpler model  and  the result did not change. We see
from Fig.~3.12 that these features for $^{78}$Ni and $^{100}$Sn
were caused by the inclusion of the complex configurations and for
$^{100}$Sn also by the small value of the proton separation energy
$B_{p}$ = 2.91~MeV (for $^{78}$Ni $B_{n}$ = 5.98~MeV was used).


\setcounter{equation}{0}
\section[Microscopic transition densities and cross sections]{Microscopic transition densities and the calculations of cross sections}

\subsection[Comparison of microscopic and phenomen. transition densities]{Comparison of microscopic and phenomenological transition densities}


The transition densities ${\rho_{tr}^{L}}$,  which are necessary
to describe the nuclear structure in calculations of cross
sections, are  simply connected with our density matrix
$\rho_{L}(r,E + \imath\eta)$ determined in Eq.~(2.57):
\begin{equation}
\rho_{tr}^{L}(r, \Delta E) = \frac{1}{\pi\sqrt{\Sigma B(EL)}} \;\;
{\rm Im} \; \int_{E_{min}}^{E_{max}} dE \rho_{L}(r, E +
\imath\eta),
\end{equation}
where $\Sigma B(EL)$ is the $B(EL)$ value summed over the interval
$\Delta E$.


 The isovector E1 transition densities calculated
in our approach for  the large energy interval were obtained in
ref.~\cite{kst97}. There is no significant difference between the
continuum RPA and our full calculation. In all cases there are
maxima of proton and neutron transition densities on the nuclear
surface with approximately equal amplitudes and opposite signs.
This corresponds  to the isovector nature of these giant
vibrations. The existence of the maxima on the nuclear surface
corresponds to the usual phenomenological models used, especially
the Goldhaber-Teller model for $^{40}$Ca and $^{48}$Ca. We pointed
out also that the transition densities defined in Eq.~(4.1) are also
large inside the nucleus, especially for the protons in $^{48}$Ca
and for the neutrons in $^{56}$Ni, which may be important in the
analysis of ($e,e'$) experiments done at large momentum transfer.

In the experiments, however, a smaller energy interval is usually   
investigated. In order to understand better the dependence of
$\rho(r)$ on the energy region, we have  calculated it for the
small energy intervals of 5~MeV \cite{kst97} and 2~MeV (see
Fig.~4.1). There is a considerable difference in the radial form
of the transition densities
as compared with the large interval, as well as with the different  
intervals under consideration.  In this respect the microscopic
${\rho_{tr}}$ are very different from the phenomenological ones
used in  conventional analyses, where they  are taken to be
independent of energy  over the whole energy interval \cite{satchler83}. 
 Such a
strong dependence on the energy region needs to be checked
experimentally.  It is clear that this dependence may be important
for the description of the gross structure of cross sections.

In Fig.~4.1 we compare the phenomenological transition densities
for the IS E0 resonance in $^{40}$Ca with the theoretical
transition densities obtained for two 2~MeV intervals in the main
part of the resonance.
We see noticeable differences for the two intervals, both between
the theoretical and phenomenological densities and between             
the theoretical densities themselves.  The behavior of  the
microscopic and phenomenological densities near the nuclear
surface, to which the ($\alpha,\alpha'$) cross sections are most
sensitive, is very different in the two intervals considered.
Inside the nucleus the behavior of the transition densities also
differs strongly for the two intervals; for example, for the
17--19~MeV interval  the neutron  microscopic density has three
nodes, whereas the IS E0 phenomenological transition density has
always just one node.


Differential  ($\alpha,\alpha'$) cross sections calculated with
the transition densities of Fig.~4.1 are shown in Fig.~4.2
(details of the calculations are given in section~4.2.2, below).
The values of the  ratio ${M_{n}/M_{p}}$ of the nuclear transition
momenta,
 which are taken to be 1
for $^{40}$Ca in the phenomenological approach, are  also given in   
Figs.~4.1 and 4.2. In the microscopic approach there is a
noticeable  difference in the ${M_{n}/M_{p}}$ ratios, both with
respect to themselves for the two intervals and between the
phenomenological and microscopic densities. The cross sections
calculated with  phenomenological densities taken at microscopic
values of ${M_{n}/M_{p}}$ are also shown. The various cross
sections differ most around zero degrees, where our theoretical
cross sections are smaller by 5--8 percent compared to the
phenomenological analysis.

In order to demonstrate  the role of the size of the energy
interval, we compare two theoretical $^{40}$Ca($\alpha,\alpha'$)
differential cross sections for the IS E0 giant resonance in
Fig.~4.3.  The dashed line is the result of using a transition
density averaged over the whole energy range, 12--30~MeV; the full
line corresponds to the differential  cross sections that were
obtained by the procedure discussed, performed for the 2~MeV bin   
and summed over  the same energy range.  As one can see, around
zero degrees the two cross sections calculated with microscopic
densities differ by nearly 25 percent. The difference between
the results with the phenomenological \cite{satchler83} and our 2 MeV microscopic   
transition densities is even larger.

In addition to the differences in energy dependence, there are     
further important differences between the microscopic and
phenomenological transition densities: (i) In the microscopic
approach, the simple relations for the proton and neutron
components of the  nuclear transition momenta (as discussed above)
and transition densities, such as the ones with the ratio $N/Z$,
do not exist.   This results in a specific energy dependence of
the Coulomb-nuclear  interference in total transition potentials,
which may even change a destructive interference into a
constructive one. (ii) The multipole decompositions of the cross
sections  are also  very different in the phenomenological
approach as compared to the microscopic one. In the first case one
tries to extract the corresponding multipole composition by
fitting to various total and differential experimental cross
sections, whereas in the microscopic approach these multipole
decompositions are determined within the  theoretical model with
known parameters. Indeed, our microscopic  results deviate
appreciably from the ones derived from phenomenological
approaches.

\subsection[Microscopic analyses of inelastic electron and alpha scattering]{Microscopic analyses of inelastic electron and alpha scattering
experiments}    

\subsubsection{Electron scattering}  
In order to compare with electron scattering experiments, in which  
neither of the isoscalar E0 and E2 resonances
 are disentangled, it is                                           
necessary to sum the E2 and E0 strengths in the same proportion
that they enter into the longitudinal electron form factor
\cite{Bol}. Thus the quantity $[dB(E2)/d\omega +
(25/16\pi)dB(E0)/d\omega]$ should  be calculated and compared with
experiment.

In our calculations of electron scattering in $^{40}$Ca
\cite{kstprl} we obtained a good description of the experiments
\cite{624r7}.   As can be seen in Fig.~4.4, the description of the
splitting of the isoscalar (E2+E0) strength into three peaks
observed in ref.~\cite{624r7} at about 12 MeV, 14 MeV and 17 MeV
was obtained.
The value of the E2+E0 theoretical isoscalar EWSR in the observed
interval 10.0--20.5~MeV, which is equal to [6581+(25/16$\pi$)3729]
= 8436 $e^{2}$-fm$^{4}$MeV, agrees with the experimental value
(7899~$\pm$~1580)~$e^{2}$-fm$^{4}$~MeV.                               

Our calculations of the electromagnetic E2 strength in $^{40}$Ca \cite{kst97} 
gave a nearly uniform distribution in the 12.0--22.0~MeV interval.
If we take the  isoscalar part of it, however, the E2 strength
shows more structure, which is washed out by the isovector
contribution. In earlier experiments \cite{624Brandenburg} the
contribution of (23.5 $\pm$ 4.7)\% of the isoscalar E0 EWSR in the
10.5--15.7~MeV was determined by  two independent methods:
($\alpha,\alpha\prime$) and ($\alpha,\alpha\prime\alpha_{0}$)
reactions. A contribution of (30 $\pm$ 6)\% in the 10.5--20.0~MeV
interval just from ($\alpha,\alpha\prime$) reactions was also
found. We have obtained good agreement with experiment for the  
first interval (20.7 percent of our S$^{0}_{0}$ in the
10.5--15.5~MeV region) and somewhat more (50.2 percent of
S$^{0}_{0}$) in the second interval. See also a discussion of the
IS E2 and E0 resonances in section~3.2.2.

Thus the ETFFS describes reasonably well not only the summed        
strengths but also the gross structure of the E2 and E0 strength
extracted from electron and alpha scattering experiments with
$^{40}$Ca. A noticeable influence of our GSC$_{phon}$ was also
found \cite{kstprl}.  We shall discuss below some of the ETFFS
 results for cross sections.

\subsubsection{$\alpha$ particle scattering}

The  calculations of the cross sections were carried out with the
modified code DWUCK4. The modification was done in such a way that
the multipole transition potentials were constructed by
single-folding the complex density--dependent Gaussian effective
$\alpha$-nucleon interaction \cite{er3} with  our microscopic
${\rho_{tr}^{L}}$ (Eq.~(4.1)) following the prescription given in
ref.~\cite{eSL}. The four parameters of the                           
 strength and range of the real and imaginary
parts of the effective interaction were adjusted to the
experimental cross sections of the low-lying 2$^{+}$ and 3$^{-}$
collective states \cite{Youngblood96}. In  Fig.~4.5 we show the
results of these calculations for the $3^{-}_{1}$ and $2^{+}_{1}$
levels in $^{58}$Ni.  The agreement between the theory and
experiment allows us to conclude that the choice of the
above-mentioned four parameters is good, so that they have also
been used in the cross section calculations in the 10--35~MeV
interval. The parameters of the optical model potential were taken
from ref.~\cite{er14}.  As in the experimental analyses of
refs.~\cite{Youngblood96,lui2000}, contributions of the IS and IV
E1 and IS E0, E2, E3 and E4 resonances, which were calculated
within our approach in the region under consideration, have been
taken into account.

In our analysis of the ($\alpha,\alpha\prime$) cross sections the
following procedure was used: The theoretical transition densities
for each of the six EL resonances were analyzed over 5~MeV
energy intervals for $^{58}$Ni and these
densities were used as input into the DWUCK4 code. For every
energy interval the DWBA cross sections
$d\sigma_{L}(\overline{E}_{L},\theta)/d\Omega$ were calculated.
The inelastic $\alpha$ spectrum was then obtained from the
expression
\begin{equation}
\frac{d^{2}\sigma}{d\Omega dE} (E, \theta) = \sum_{L}\frac{2}{\pi
\Gamma_{L}}\frac{(\Gamma_{L}/2)^{2}} {(E - \overline{E}_{L})^{2} +
(\Gamma_{L}/2)^{2}} \frac{d\sigma_{L}}{d\Omega} (\overline{E}_{L},
\theta).
\end{equation}
Here the summation runs over the six multipolarities  considered
and the parameters $\overline{E}_{L}$ and $\Gamma_{L}$ were
obtained from a lorentzian fit to the calculated strength function  
$S_{L}(E)$ defined in Eq.~(3.1).

For the cross section calculations in $^{40}$Ca the same method     
was used, but it was necessary to use 2~MeV bins because the IS E0
strength and the total spectra have much more structure. As in the
case of $^{58}$Ni, the contributions of the same six resonances
were taken into account.

In the calculations of the giant resonances we used a standard
Woods-Saxon single-particle basis. The residual Landau-Migdal
interaction given in Eqs.~(1.10), (1.11) and (1.43), with the
parameters of Eq.~(1.42), was used.  Additional information about
numerical details can be found in section~3.1. In all these
calculations the smearing parameter $\Delta$ = 500~keV was used.

\subsection[Calculation of the $^{58}$Ni  and $^{40}$Ca $(\alpha,\alpha\prime)$ cross
sections]{Calculation of the $^{58}$Ni  and $^{40}$Ca $(\alpha,\alpha\prime)$ cross
sections; comparison with experiment.}  


A general problem in nuclei with $A\,<\,90$ is that the isoscalar      
monopole resonance is very broad and no longer concentrated in one
single peak. In moving to lighter nuclei the role of the surface
becomes more important than in heavy  nuclei and the
($\alpha,\alpha\prime$) reaction is very sensitive to the nuclear
surface.  Moreover, for light and medium mass  nuclei the
single-particle continuum becomes very important.   All of this
has to be considered if  theoretical models for giant resonances
in lighter nuclei are to be developed. Therefore the conventional RPA
approach to collective states has to be extended by the inclusion
of surface modes and the single-particle continuum.

Until recently there were several open questions concerning the
experimental information about the IS E0 resonance in nuclei with   
$A\,<\,90$ \cite{HW,624r2}.  In some cases the magnitude of the
detected strength was much smaller than the EWSR limit.  An
important example in this connection is $^{58}$Ni, where
originally only 32 percent of the EWSR was observed with
inelastically scattered $\alpha$-particles \cite{Youngblood96}.
These authors used standard data analysis techniques with the same
phenomenological transition densities ${\rho_{tr}^{L}}$ for
different excitation energies.
For comparison, the same type of experiment in $^{40}$Ca and a
similar analysis by the same authors showed (92 $\pm$ 15)\% of the
EWSR \cite{er4}. Such uncertainties  might  have serious
consequences for nuclear matter compressibility and its
applications to astrophysics.

Our analysis within the ETFFS \cite{ekst,kstacta} of the same
experimental data, where microscopic transition densities were
used instead, gave 71.4 percent of the IS E0 EWSR in the observed
12.0--25.0~MeV interval, compared with the 32 percent of
ref.~\cite{Youngblood96}. In these calculations the energy
interval considered was divided into 5~MeV bins, for which the
calculations were performed separately. The results of the
calculations allowed the assumption \cite{ekst,kstacta} that a
part of the IS E0 strength in $^{58}$Ni might be hidden in the
experimental background.

In the meantime the 12--25~MeV energy range of the original
experiment in $^{58}$Ni has been extended \cite{lui2000}. Here the
authors present  results from the 12.0--31.1~MeV excitation
region.
Compared with the original analysis  \cite{Youngblood96}, two new   
ingredients were included by the same authors in their new work
\cite{lui2000}: (i) a nuclear reaction description following the
methods  of ref. \cite{er3} was used and (ii) the giant resonance
peak obtained after subtraction of the continuum
was divided into 15 energy intervals from 1 to 3~MeV, each of
which was analyzed separately.

In what follows we discuss the ETFFS analysis of the               
($\alpha,\alpha'$) experiments in $^{58}$Ni and $^{40}$Ca,
including the newest ones \cite{lui2000,lui2001} for both nuclei,
where the isoscalar giant resonance region was investigated. This
analysis includes both of the necessary effects mentioned at the
beginning of this subsection.  Within this  model we calculated
the distribution of isoscalar strength of the isoscalar E0--E4
resonances and the corresponding transition densities. From these
transition densities  we obtained, in the standard way,
($\alpha,\alpha'$) cross sections that we compare with the
experimental data in $^{40}$Ca in the observed 8.0--29.0~MeV
interval~\cite{er4}
 and for the isotope $^{58}$Ni in several energy
regions: 12--25~MeV \cite{Youngblood96}, 12.0--31.1~MeV
\cite{lui2000}  and  12--35~MeV \cite{eVDW}.

\subsubsection{$^{58}$Ni results}

In Fig.~4.6 we compare our microscopic calculations with the
newest experimental results, given in ref.~\cite{lui2000}, for the  
IS E0 and E2 strength distributions.  For  the observed
12.0--31.1~MeV interval we obtained a value of the mean energy
(defined as $m_{1}/m_{0}$) of the IS E0 resonance of 19.9~MeV and
find 81.5 percent of the  EWSR. The experimental data are
(20.30$^{+1.69}_{-0.14})$~MeV and (74$^{+20}_{-6}$)\%,
respectively. For the ``old'' 12.0--25.0~MeV interval
the new phenomenological analysis finds (58$\pm$6)\% of the EWSR
\cite{lui2000}, which is now much closer to the microscopic value
of 71.4 percent \cite{ekst} obtained for the same interval. Thus
both analyses now give very similar results.

As one can see from the lower part of Fig.~4.6, however, for the    
IS E2 strength the two approaches still arrive at quite different
results.
Our E2 resonance mean energy value and the depletion of the IS E2
EWSR defined in the 10.0--20.0~MeV energy interval are 19.1 MeV
and 47 percent, while the  phenomenological analysis gives
16.1~MeV and (115$\pm$15)\%. There is also a disagreement over the
IS E1 and E3 strengths. Our results  are presented in Fig.~4.7.
(See also the discussion  at the end of this subsection.)

In Figs.~4.8  and 4.9 we compare the same experimental data  with  
our theoretical results in a slightly different way. There we
obtain in both cases good agreement between our theory and the
data for the total cross sections.

In Fig.~4.8 we  show the role of the background and the
contribution of the various giant resonances to the total cross
section. We compare our theoretical results (full line with dots)
with the experimental cross section at $\theta = 1.08^{\circ}$
(histogram) of ref.~\cite{lui2000}. We obtained these data by
subtracting an instrumental background (denoted by ``Backgr. (2000)'')   
from the original experimental spectrum. The theoretical cross
section is the sum of six different multipoles, of which we only
show E0 and E2. The straight line in the lower part of the figure
denoted by
``Backgr.~(1996)'' corresponds to the analysis of                       
ref.~\cite{Youngblood96}, where the considered energy interval,
12--25~MeV,  was smaller. In the original analysis with
phenomenological transition densities, only 32 percent of the IS
E0 EWSR limit was found. With the improved---but still
conventional---analysis, the authors of  ref.~\cite{er3} obtained
about 50 percent.

In Fig.~4.8, we also compare the 12--25~Mev interval and the
previous background subtraction with the new, extended interval
and the experimental data, where a different background has been
subtracted. We find here that the total IS E0 resonance cross
section in the 12--25~MeV energy range is  138.3~mb/sr and
corresponds to 71.4 percent of EWSR. The area under the old
background line in this region, which is included in our 71.4
percent of the EWSR, corresponds to 22 percent of the IS E0 EWSR
limit, or 42.6~mb/sr. This strength had been subtracted as a part
of the background in ref.~\cite{Youngblood96} and was therefore
not included in the analysis of ref.~\cite{er3}. If we extend the
analysis to the larger 12--35~MeV interval our theoretical model
predicts 89.6 percent of the IS E0 EWSR limit. For the comparison
with experiment see ref.~\cite{eVDW}.

In Fig.~4.10 we compare our theoretical cross sections with the  
data at $\theta$~=~$4.08^{\circ}$. We reconstructed  the
experimental cross section  from  Fig.~4 and Fig.~1 of
ref.~\cite{Youngblood96}.  Good agreement between the theory and
experiment \cite{Youngblood96} is obtained. We also see that at
this angle the IS E2 resonance and higher multipoles contribute
most  to the cross section, whereas the monopole contribution is
small.

Thus one can conclude that the new experimental data for the IS E0  
resonance in $^{58}$Ni are in good agreement with the microscopic
calculations, which do not contain any fitting parameters for the
nuclear structure part.  The values of integral characteristics
correspond now to the known experimental systematics.  The
disagreement between our microscopic results and  the
phenomenological analysis \cite{lui2000} for the IS E2 resonance
only confirms our earlier conclusion \cite{ekst} regarding the
necessity of using microscopic transition densities in the
experimental analysis.  The same is true for the IS E1 resonance,
where the two approaches also lead to very different conclusions.
The authors of ref.~\cite{lui2000}  obtained only 41 percent of
the IS E1 EWSR and this strength was spread more or less uniformly
from 12~MeV to 35~MeV. Our distribution  of this strength is shown
in Fig.~4.7. One can see that the distribution is not uniform;
rather, it has a resonance structure.  We obtained 89 percent of
the IS E1 EWSR  and  a value  of 25.0~MeV for the mean energy in
the interval under consideration.  These figures are consistent
with the results of ref.~\cite{ecly99} for other nuclei. Our E3 IS
strength is more uniformly distributed.

\subsubsection{$^{40}$Ca results}

Our improved calculations of the IS E0 resonance in $^{40}$Ca,
presented in ref.~\cite{kst97},  show that it has a more compact
form than in our earlier calculations \cite{ekstw94}, but it     
remains strongly structured and spread out over a large energy
interval: 65 percent of the EWSR is in the 11--23~MeV interval and
106.7 percent is in the 5.0--45.0~MeV interval.
Thus it is important that the large energy interval,
8.0--29.0~MeV, was studied in ref.~\cite{er4} and that, in fact,
most of the IS E0 strength was found in their analysis.

Our theoretical distributions of the IS E0 and E2 strengths in
$^{40}$Ca are shown in Fig.~4.11. Compared with the corresponding  
results of the analysis of ref.~\cite{er4}, which were obtained
from the difference between spectra taken at
$\theta$~=~$1.1^{\circ}$ and $\theta$~=~$2.4^{\circ}$ (see Fig.~6
of ref.~\cite{er4}), a reasonable agreement for the central part
(10--23~MeV) of the IS E0 resonance  was obtained. The percentages
of the IS E0 EWSR in the four observed intervals 7.5--12.5~MeV,
12.5--22.5~MeV, 22.5--28.8~MeV and 7.5--28.8~MeV are
6.0(7.6$\pm$0.2)\%, 60.0(50.0$\pm$1.4)\%, 16.0(34.7$\pm$1.7)\% and
81.6(92$\pm$2)\%, respectively. (The quantities in parentheses are
the results of ref.\cite{er4} with statistical errors only.) The
final result of the analysis in ref.~\cite{er4} of the IS E0 EWSR
for $E_{x}$ between 8 and 29~MeV is (92$\pm$15)\%, which agrees
with our value of 81.6 percent.  There is, however, noticeable
disagreement in the low-lying and high-lying regions of the
excitation spectrum.

In ref.~\cite{er4} the authors find (33$\pm$4)\% of the IS E0 EWSR and   
(57$\pm$6)\% of the IS E2 EWSR at a peak energy of
17.5$\pm$0.4~MeV.  Our results for the IS E0 and E2 resonance mean
energies, obtained (as $m_{1}/m_{0}$) from averaging over the
observed 8.0--29.0~MeV interval, are 17.2~MeV and 17.1~MeV,
respectively. It is impossible to compare the experimental  IS E0
EWSR depletion value with our value of 81.6 percent, one of the
reasons being that  the experimental value of (33$\pm$4)\% does
not contain the continuum~\cite{er4}.  For the IS E2 resonance,
however, the difference between our value of 88.0 percent,
obtained for the 8.0--29.0~MeV interval, and the 57 percent in
ref.~\cite{er4} is smaller than in the IS E0 case, which  may
indicate that there is less 
E2 strength  in the experimental  continuum.

A clearer comparison with experiment is shown in Fig.~4.12 for the  
double-differential cross section for $^{40}$Ca. Our full
calculations reproduce the experimental gross structure reasonably
well but, again, we see differences at low and high energies. The
general difference between the values of  the theoretical and
experimental cross sections may be hinting that multipoles other
than the IS E0 contribute to that cross section.

In order to understand better the role of complex configurations,   
we also show in Fig.~4.12 ($\alpha,\alpha'$) cross sections
obtained with the  microscopic transition density calculated
within the continuum RPA only; that is, without inclusion of our
complex 1p1h$\otimes$phonon configurations.  It can easily be seen
that there is a big difference between the two theoretical curves.
We conclude that the gross structure of the IS E0 resonance in
$^{40}$Ca is caused by the complex 1p1h$\otimes$phonon
configurations. A similar conclusion for the fine structure of the
IS E2 resonance in $^{208}$Pb was reached in our earlier
calculations \cite{k...richter97} and in section~3.2.2.

As seen in section~4.3.1 and Figs.~4.8, 4.9 and 4.10, the
calculations for $^{58}$Ni explained reasonably well not only the
IS E0 resonance in this nucleus but also the total spectra.
We therefore calculated the  total spectrum of the
$^{40}$Ca($\alpha,\alpha'$) reaction observed in ref.~\cite{er4}.
It has a detailed  structure but, unfortunately, is given there
only as counts. We obtained the following results, shown in
Fig.~4.13:
\begin{enumerate}
\item The ETFFS describes reasonably  well the gross structure of total   
spectra, except for the region below 9~MeV, which may be connected
with the excitations of the $\alpha$+$^{36}$Ar system.
\item The difference between the two horizontal axes in Fig.~4.13               
gives an instrumental background of the experiment under
discussion. This difference is about 7 mb/sr-MeV, so that the
integrated cross section is about 7$\times$25 = 175~mb/sr.  (These
numbers are, of course, very approximate because they were
obtained by imposing the experimental curve on the theoretical
one). Integration of the theoretical curve gives 422~mb/sr and of
the experimental one 676~mb/sr.  Thus we should compare (676-175)
= 501~mb/sr (experiment) and 422~mb/sr (theory).
\item As can be seen in Fig.~4.13, the experimental background             
shown by the dot-dashed line contains a noticeable contribution
from giant resonances, including that from the IS E0.
\end{enumerate}

Thus, in the calculations presented in Fig.~4.12 and 4.13, a
reasonably good description of the gross structure, both in the          
0$^{\circ}$ cross section for the IS E0 strength and in the total
spectra in $^{40}$Ca, has been obtained.  One can, however, see           
some disagreement with the experiments around 19~MeV. Fortunately,
the results of the new and improved experiments \cite{lui2001}
have been published recently, and these should  be analyzed within
the ETFFS.
\begin{center}
{\it Description of the improved E0 cross section data \\
in $^{40}${\rm Ca($\alpha,\alpha'$)} at E$_{\alpha}$ = 240~MeV.}
\end{center}

According to the authors of ref.~\cite{lui2001}, the new
experimental results were obtained using an analysis technique        
that unambiguously identifies multipole strength, whereas the old
spectrum subtraction technique \cite{er4} is very sensitive,
particularly to experimental background. In ref.~\cite{er4} a
definite assignment could be made only for (33$\pm$4)\% of the E0
strength  in the peak.  What is also important is that the new
method allows considerable extension of the observed energy
interval to obtain thereby some additional E0 strength in the high
energy tail region, so that
a new experimental background must be taken. In other words, the
authors have extracted  an additional E0 strength from the old
background, as was assumed in our work on $^{58}$Ni beginning in
1996~\cite{ekst,kstacta}. In the new $^{40}$Ca
measurements~\cite{lui2001} (97$\pm$11)\% of the EWSR was observed
in the 10--55~MeV energy interval, whereas the original interval
was 4--27~MeV. This result agrees with our calculations: 81.6 percent  
in the (7.5-28.8) Mev interval \cite{kstepj2000} and 106.7 percent in  
the (5-45) MeV interval \cite{kst97}.                                  

A possible reason for the above-mentioned  disagreement
with the experiment around 19 MeV                                   
could be                                                            
that the number of low-lying phonons used---the $3^{-}$ and
$5^{-}$)---is insufficient.   We have therefore added a third one.     
Because this additional $2^{+}_{1}$-phonon at $E$ = 3.90~MeV has
another (probably 2p2h) nature as compared with those used, we had
to make
further approximations for this specific phonon. We have therefore
calculated, first of all, the distribution of the EWSR for the IS
E0 resonance with three phonons and compared it with experiment in
Fig.~4.14. In Fig.~4.15 we compare the new \cite{lui2001} and old
\cite{er4} experimental results for the E0 cross section. One can
see a noticeable difference between the experimental results of
1997 and 2001. In Fig.~4.15 we show  also our previous
calculations within the ETFFS \cite{kstepj2000} with two phonons,
which  were used earlier in the calculations for $^{40}$Ca.

With the inclusion of the third phonon, very good agreement with
the new experimental results was obtained, both for the EWSR
distribution (Fig.~4.14) and for the IS E0 cross section                
(Fig.~4.16).\footnote{Our estimates of the contribution of the
fourth, i.e. $0^{+}_{1}$,  low-lying phonon at 3.35 MeV, have shown     
that its contribution is small.}

We conclude that experiments of this type can shed some light on
low-lying excitation spectra and, for a microscopic theory, can
demonstrate the necessity of accounting for---and identification
of---additional 1p1h$\otimes$phonon configurations, or even of
inclusion of  more complex configurations than the
1p1h$\otimes$phonon.

\section{Conclusion}

In this review we presented a new microscopic many-body theory        
for the structure of closed shell nuclei. The extended theory of
finite Fermi systems  (ETFFS) is based on the Landau-Migdal theory
of finite Fermi systems (TFFS) and includes in a consistent way
configurations beyond the 1p1h level. A large part of this review
is concerned with the application of this new approach to giant
resonances in closed shell nuclei. As in the standard TFFS
\cite{AB}, we formulate the theory within
 the framework of many-body Green functions. As in the original
 approach, one makes an ansatz for the propagators in the equation
for the response function that takes into account these higher
configurations and one renormalizes the resulting equation in the
standard way. After a long mathematical procedure one ends up with
an equation that includes effective charges and an effective
interaction which are parameterized in the same way as in the
standard theory. The corresponding parameters are universal for
all the nuclei, investigated so far. The ETFFS extends the
standard TFFS in the following directions:
\begin{enumerate}
\item In addition to the 1p1h configurations, it considers
in a consistent way
complex  configurations of the 1p1h$\otimes$phonon type.
The corresponding formulation is a natural extension of
the conventional RPA.
\item It includes explicitly the single-particle continuum.
For that reason one is able to obtain
a microscopically determined envelope of the resonance
 without
 using a purely phenomenological
smearing parameter. (For numerical convenience  we still use
 a small smearing parameter of a few hundred keV.)
 This feature of the ETFSS is especially important for the        
nuclei with the nucleon separation energy near zero.
\item The
approach takes into account in a consistent way ground state
correlations  caused  by the conventional RPA and by the more
complex configurations. It is demonstrated that the latter ones
(GSC$_{phon}$) are at least as important as the conventional RPA
ground state correlations.
\end{enumerate}

Another  important  feature of the ETFFS is that the final             
equation explicitly contains both the effective ph interaction and
the quasiparticle-phonon interaction.
 This new interaction, however,
does not introduce additional parameters. It is completely determined
by the original ph interaction  that enters into the conventional
RPA equation. From this equation one obtains the structure of the phonons
and the corresponding quasiparticle-phonon interaction.

We have reviewed the main results obtained within the extended
theory and  discussed in detail the following  physical
results:
\begin{enumerate}
\item Quantitative explanation of the  widths of GMR.
 It is shown that
the 1p1h$\otimes$phonon configurations give the most important
contributions to the widths of the GMR. In addition, a large fraction
of the observed gross and fine structure can be directly traced back to  
specific 1p1h$\otimes$phonon configurations. Detailed results have been
presented for $^{40}$Ca, $^{48}$Ca, $^{56}$Ni, $^{58}$Ni and $^{208}$Pb.
These calculations give not only the solution to a long-standing problem,    
but they are also crucial for the analyses of the experimental data
in medium mass nuclei.
\item Necessity of using {\it microscopic} transition densities:           
In order to extract the parameters of GMR from the experimental
cross sections  one needs the corresponding form factors. So far,
in nearly all the analyses of hadron and electron scattering
experiments, phenomenological transition densities derived from
some macroscopic models have been used. In heavy mass nuclei,
where the GMR
 are relatively narrow and in general well separated from each other,
such a procedure seems to work. For nuclei with A $<$ 90, where
the giant resonances show pronounced structure and are spread out
over a large energy interval and where various multipoles overlap,
such a procedure is no longer appropriate. First of all, the
microscopic transition densities are energy dependent and vary
over the ranges of the GMR in such nuclei. Secondly, as the
resonances overlap, it seems difficult with the  conventional
method to extract in an unique way the width  and the absolute
strength of the individual GMR. Those effects were  discussed in
some detail for $^{58}$Ni and $^{40}$Ca. It was shown in
particular that overlapping multipole resonances give rise to a
very smooth angular distribution that looks like an experimental
background. Such a misinterpretation gives rise to ``missing
strength,'' as occurred some years ago. The analysis of the
corresponding experimental cross sections with microscopic
transition densities derived from the ETFFS has clarified that
point.
\item It has been shown that the ground state correlations induced
by complex 1p1h$\otimes$phonon configurations give rise to a 5-7
percent contribution to the EWSR. This additional strength
 violates the sum rule. One should note that this important question
 still lacks a completely satisfactory answer.
\end{enumerate}

In the  next stage of  the development of this new microscopic
nuclear structure theory one has to include pairing. Such a
generalization will be crucial for the application of the ETFFS to
open shell nuclei. One may expect some new effects for the theory
of nuclear pairing  if one takes the quasiparticle-phonon
interaction
 into account~\cite{avekaev99}.
Such new effects could probably be observed in experiments with
high experimental resolution. The first steps towards extending
the ETFFS in this direction have already been taken \cite{klz2001}
and the corresponding calculations are in progress.

As mentioned earlier, the ETFFS and the original TFFS start with
phenomenological single-particle potentials and parameterize the
residual ph interaction; i.e., the theories are not
self-consistent. For the applications of these theories to nuclei
far from stability the self-consistency, however, is crucial. Here
one starts with, for example, a density functional or some
relativistic Lagrangian, which provides a single-particle scheme
and, simultaneously, the residual interaction. Actually, the ETFFS
is most appropriate for
 a self-consistent approach because it takes into account explicitly
the effects of the phonons on the single-particle energies and
single-particle wave functions as well as in the residual
interaction. In this respect the self-consistent ETFFS will go far
beyond all existing theories for the structure of unstable nuclei.
For closed shell nuclei, initial steps in this direction have
already been taken \cite{tselyaevizv}. We are convinced that for a
quantitative understanding of {\it unstable} nuclei one has to
consider simultaneously non-separable universal forces, complex
configurations, the single-particle continuum and---perhaps---the
GSC$_{phon}$, as discussed in section 1.3.2. We are also convinced
that the new effects, such as the GSC$_{phon}$ (section 3.2.4) and
the second (phonon) mechanism of pairing \cite{avekaev99},
 can manifest themselves in unstable nuclei in a
clearer form than in the stable ones. The advent of modern
radioactive beam accelerators and large gamma detector arrays will
open possibilities to study these exciting new effects.


The general outline for the construction of a theory with complex
configurations used in the formulation of the nuclear ETFFS under
discussion
may be used, of course, for other Fermi systems---with and without
superconductivity, such as metallic clusters and quantum dots,
where phenomena similar to zero sound have already been observed.

Last but not least, all the formulas of the ETFFS can be
straightforwardly
generalized to the finite temperature case because the GF
formalism used  here has a natural generalization  in the
Matsubara temperature technique.

\vspace{1cm}

{\bf Acknowledgements}

We thank our colleagues S.T. Belyaev, P.F. Bortignon, M. Harakeh,
V.A. Khodel, S. Krewald, P. von Neumann-Cosel, Nguyen Van Giai,
 A. Richter, P. Ring,
E.E. Saperstein, A.I. Vdovin, J. Wambach, D.H. Youngblood, V. Zelevinsky
for fruitful discussions.  We also thank V.I. Tselyaev for collaboration
and stimulative discussions. The authors are greatly indebted
 to J. Durso for his  careful reading of the manuscript.
J.S. thanks Tony Thomas for many useful discussions and the
hospitality he enjoyed in Adelaide where part of this review was
written. The Russian authors are very thankful to the Institut
f\"ur Kernphysik and Forschungszentrum J\"ulich for their
hospitality due to which this work could be performed and
finished. S.K. thanks V. Kamerdzhiev and A. Gasparyan for their
help in preparing the manuscript.
 This work
was supported  in part by the DFG Nr.447 Aus 113/14/0.

\newpage
\textwidth 150mm \textheight 231mm \topmargin -2.0cm
\begin{center}
 Table~1.1. Integral characteristics of the E2 IS and IV resonances\\
in $^{40}$Ca,$^{208}$Pb calculated within the Continuum TFFS
\end{center}

\begin{tabular}{|c|c|c|c|c|c|c|}
\hline Nucleus, & Interval, & $\frac{EWSR}{EWSR_{th}}$,\% &
E$_{1,-1}$,
& E$_{1,0}$,& E$_{2,0}$, & ${\bar E}_{exp}$,\\
resonance & MeV & & MeV & MeV & MeV & MeV\\
\hline
$^{40}$Ca, IS & 5-45 & 100 & 20.7 & 21.6 & 22.5 & $\sim$ 18 \cite{BHW}\\
\hline
$^{208}$Pb, IS & 0-45 & 95 & 6.1 & 8.1 & 10.2 & 10.9 $\pm$ 0.3 \cite{Bra}\\
 & & & & & & 10.4 \cite{Bol} \\
\hline
$^{208}$Pb, IV & 0-45 & 87 & 16.6 & 18.2 & 19.3 & 20.2 $\pm$ 0.5 \cite{DLA} \\
 & & & & & & 22.6 $\pm$ 0.4 \cite{BB} \\
\hline
\end{tabular}

\newpage
{\bf Figure captions for section~1}

Fig.~1.1  Graphical representation of the Bethe-Salpeter equation
for the response function in the ph channel.

Fig.~1.2  The hadron strength function of the isoscalar E2
resonance in $^{40}$Ca calculated within the continuum TFFS (solid
line). The dashed line gives the same quantity obtained without
the ph interaction (``free response'').

Fig.~1.3 The same as in Fig.~1.2 but for $^{208}$Pb.

Fig.~1.4 The same as in Fig.~1.2 but for the isovector E2
resonance in $^{208}$Pb.

\newpage
{\bf Figure captions for section~2}

Fig.~2.1 The energy--dependent mass operator $\Sigma^{e}$ and ph
amplitude $U^{e}$ in the g$^{2}$ approximation.

Fig.~2.2 Graphs corresponding to the propagator of the model       
\cite{kam79,kt89} including RPA (a) and 1p1h$\otimes$phonon
configurations (b,c,d).
 The wavy line and the circle denote the phonon Green function and the amplitude $g$,
 and the thin lines denote the Green function $\tilde{G}$.

Fig.~2.3 Typical graphs of fourth order in the
quasiparticle-phonon interaction amplitude $g$ entering into the     
diagrammatic expansion  in the time representation that correspond
to the MCDD propagator used in the ETFFS. The notation is the same
as in Fig.~2.2, but here the direction of the arrow on a fermion
line denotes the particle or hole GF $\tilde{G}$. The dotted line
denotes the time cut at fixed time. Only graphs d, e and f are
explicitly included in the ETFFS.

Fig.~2.4 Graphs of order g$^{4}$ with self-energy insertions in
the time representation that are included in the ETFFS and give a
part of the GSG$_{phon}$.

Fig.~2.5 Typical ``backward-going'' graphs of order g$^{2}$ due to
the GSC$_{phon}$: graphs a, b, c, and d are similar to the RPA GSC
case, whereas graphs e, f, g and h are the GSC$_{phon}$ of new
type that correspond to direct excitation of a complex
configuration by an external field.
\newpage

{\bf Figure captions for section~3}

Fig.~3.1 The E1 photoabsorption cross section for $^{40}$Ca
obtained within CTFFS (dash-dotted), in the
1p1h+1p1h$\otimes$phonon (dashed) and 1p1h+1p1h$\otimes$phonon +
continuum (solid) approximation. The experimental data are taken
from ref.~\cite{ahrens75}. From \cite{kstt93}.

Fig.~3.2 Same as Fig.~3.1 but for $^{48}$Ca. The experimental data
are taken from ref.~\cite{o'keefe87}. From \cite{kstt93}.

Fig.~3.3 Same as Fig.~3.1, but for $^{208}$Pb. The experimental
data are taken from ref.~\cite{veyssiere70}. From \cite{kstt93}.

Fig.~3.4. The full E0 and E2 electromagnetic strength functions
calculated in the 5.0--45.0~MeV energy interval.
The final results are shown by the solid curve. The RPA +
1p1h$\otimes$phonon (without GSC$_{phon}$) and the continuum RPA
results are shown by the dashed and dotted curves, respectively.
>From \cite{kst97}.

Fig.~3.5. The full isoscalar and isovector E0 and E2 strength
functions calculated with taking into account the RPA and
1p1h$\otimes$phonon configurations, the single-particle continuum
and GSC$_{phon}$. From \cite{kst97}.

Fig.~3.6 The E0 isoscalar (full line) and isovector (dashed line)
hadronic strength functions in $^{208}$Pb calculated in the ETFFS
(smearing parameter $\Delta$~=~250~keV).

Fig.~3.7  The E2 hadronic strength  function in $^{208}$Pb
calculated within the ETFFS with smearing parameter
$\Delta$~=~250~keV (upper panel) and 800~keV (lower panel)

Fig.~3.8 The E1 photoabsorption cross section for $^{208}$Pb
calculated in the RPA + 1p1h$\otimes$phonon + continuum
approximation with the smearing parameter $\Delta$~=~250~keV
(heavy line) and $\Delta$~=~10~keV (thin line) compared with
experiment. From \cite{kstt93}.

Fig.~3.9. Upper: high resolution inelastic proton and electron
scattering spectra  in the region of the isoscalar E2 resonance in
$^{208}$Pb~\cite{richter93,k...richter97}. Lower: the isoscalar E2
strength function calculated in the  RPA + 1p1h$\otimes$phonon +
continuum approximation. See text for details.

Fig.~3.10 M1 in $^{40}$Ca calculated in the RPA +
1p1h$\otimes$phonon approximation~\cite{kt84,kt89}. The dashed
lines represent the experiment(s)~\cite{pringle82}. From \cite{kt84}.

Fig.~3.11. Upper: the strength function for the M1 excitation in
$^{208}$Pb in the CRPA (dotted line) and in the RPA +
1p1h$\otimes$phonon + continuum without (dashed line) and with
GSC$_{phon}$~\cite{kstw93} (full line).  Lower: experimental
data~\cite{lasz88}.

Fig.~3.12. The M1 strength function for unstable nuclei calculated
within the CTFFS (dotted curve) and the  ETFFS without (gs-) and
with (gs+) GSP$_{phon}$. From \cite{kstw93}


\newpage
{\bf Figure captions for section~4}.

Fig.~4.1. The microscopic (dotted line) and phenomenological \cite{satchler83} 
(solid line) IS E0 transition densities in $^{40}$Ca,  calculated
for two different energy intervals of 2~MeV each. Results for
proton and neutron microscopic transition densities are also given
(denoted by triangles). The values of the corresponding ratio
$M_{n}/M_{p}$ of the nuclear transition moments are  given in the
legend. From \cite{kstepj2000}.

Fig.~ 4.2 Angular distributions for  the IS E0  resonance in the
same  intervals as in Fig.~1, calculated with phenomenological
 \cite{satchler83}   and                                               
microscopic transition densities. For comparison, the results with
the phenomenological transition density, obtained with the
microscopic value of the ratio $M_{n}/M_{p}$ of the nuclear
transition moments (dashed line) are also shown.  From \cite{kstepj2000}.

Fig.~4.3.  Angular distributions for the IS E0 resonance in
$^{40}$Ca in the 12--30~MeV interval, calculated using two
different methods: the division of the interval into  9 bins of
2~MeV each (solid line) and the interpretation of the 12--30~MeV
interval as one bin (dashed line), respectively. For comparison,
the results obtained with the phenomenological transition density
\cite{satchler83}                                                      
are also given (solid curve with dots).  From \cite{kstepj2000}.

Fig.~4.4  (a) The experimental data \cite{624r7} for the strength
function vs. the theoretical one  for $^{40}$Ca (see text).            
(b) The theoretical
electromagnetic E2 (dashed line) and E0 (dotted line) strength
functions. From \cite{kstprl}.

Fig.~4.5 The experimental \cite{Youngblood96} and theoretical
$\alpha$ particle scattering  differential cross sections
 for the $3^{-}_{1}$ (below) and $2^{+}_{1}$ (above) levels in $^{58}$Ni.

Fig.~4.6. Distribution of the IS E0 and E2 strengths in $^{58}$Ni.
 The experimental data are taken from  \cite{lui2000}. From \cite{kstepj2000}.

Fig.~4.7.  Distribution of the IS E1 and E3 strengths in $^{58}$Ni
as calculated within our theoretical model.  From \cite{kstepj2000}.

Fig.~4.8. Cross sections of $^{58}$Ni(${\alpha,\alpha'}$) at
E$_{\alpha}$~=~240~MeV and $\theta  =  1.08^{\circ}$. The
experimental data (histogram) were taken from Ref.~\cite{lui2000},
where an instrumental background has been subtracted. The solid
curve with dots shows the calculated total (summed) cross
sections, the dashed line (``Rest(EL)'') corresponds to the sum of
the IS and IV E1, and IS E3 and E4 multipoles. The shaded area
shows an additional IS E0 strength which has been subtracted  in
the previous experiments as background~\cite{Youngblood96}. This
area corresponds to 22 percent of the IS E0 EWSR, see text.                 
 From \cite{kstepj2000}.

Fig.~4.9. Cross sections of $^{58}$Ni(${\alpha,\alpha'}$) at
$E_{\alpha}$~=~240~MeV and $\theta  =  1.08^{\circ}$.  The
experimental data (histogram), including  the instrumental
background, are taken from ref.~\cite{lui2000}. The solid curve
with dots gives the calculated total (summed) cross sections. In
the lower part of the picture  the  components of the total cross
sections are shown without the background. In particular, the
dashed line gives the IS E0 contribution.  From \cite{kstepj2000}.

Fig.~4.10. Same as in Fig.~4.9  but for $\theta  =  4.08^{\circ}$.
The experimental data were obtained
by using
the results of ref.~\cite{Youngblood96} (see  text).  From \cite{kstepj2000}.

Fig.~4.11 Distribution of the IS E0 and E2  strengths in $^{40}$Ca
(theory).  From \cite{kstepj2000}.

Fig.~4.12. The $0^{\circ}$ cross section for the IS E0 strength in
$^{40}$Ca(${\alpha,\alpha'}$) at E$_{\alpha}$ = 240 MeV,
calculated with (solid line) and without (dotted line) taking           
into account complex 1p1h$\otimes$phonon configurations. The
experimental data (histogram) are taken from ref.~\cite{er4}. One
can see that the gross structure of the IS E0 resonance is caused
by complex 1p1h$\otimes$phonon configurations.
 From \cite{kstepj2000}.


Fig.~4.13 Cross sections of $^{40}$Ca(${\alpha,\alpha'}$) at
E$_{\alpha}$~=~240~MeV and $\theta = 1.1^{\circ}$.  The
experimental data
 dotted line) and the background (dot-dashed line)
are taken from ref.~ \cite{er4}.  Because in Fig.~2 of
ref.~\cite{er4} there are only counts,  the experimental histogram
was imposed  on the theoretical curve in such a way that the
maxima of the two curves  coincided. In this way it was possible
to estimate
roughly the contribution of the instrumental background as the
difference between two horizontal axes (see text).

Fig.~4.14 Upper: distribution of the IS E0 EWSR calculated with 2
and 3 low-lying phonons. Lower: comparison of the 3-phonon
calculation with experiment~\cite{lui2001}.

Fig.~4.15 E0 cross sections of $^{40}$Ca(${\alpha,\alpha'}$) at
E$_{\alpha}$ = 240 MeV and $\theta  =  1.1^{\circ}$.  The
experimental data 2001, 1997 and the theoretical results with 2
phonons are taken from refs.~\cite{lui2001,er4} and
\cite{kstepj2000}, respectively.

Fig.~4.16 Comparison of the improved experimental data
\cite{lui2001} for cross sections and the theoretical calculations
with 3 phonons.

\newpage

\addcontentsline{toc}{section}{Appendix}
\setcounter{equation}{0}
\appendix
\section*{Appendix.\\ 1.Generalized propagator in the method of chronological decoupling of diagrams}
\setcounter{section}{1}
\renewcommand{\theequation}{\Alph{section}.\arabic{equation}}
The equation for the density matrix $\rho_{12}(\omega)$,
Eq.~(2.19), in the $\tilde{\lambda}$ representation is given by
(the tilde
on $\tilde{\epsilon_{\lambda}}$ and $\tilde n_{\lambda}$ has been
omitted in the equations that follow)
\begin{eqnarray}
 \delta\rho_{12}(\omega) = \delta\rho_{12}^{0}(\omega) - \sum_{56,34}A_{12,34}\tilde{F}_{34,56}
                          \delta\rho_{56}(\omega)  \nonumber\\
\delta\rho_{12}^{0}(\omega) = - \sum_{34}A_{12,34}(\omega)
(\tilde{e}_{q}V^{0})_{43} \label{eq:A1}
\end{eqnarray}
 The generalized propagator A has the form
\begin{equation}
A_{12,34}(\omega) =
 \sum_{56,78}[\delta_{15}\delta_{26} + Q_{12,56}^{+-}(\omega)]A_{56,78}^{--}(\omega)
 [\delta_{73}\delta_{84} + Q_{78,34}^{-+}(\omega)] + P_{12,34}^{++}(\omega).
\end{equation}

For the ph-ph part  $A^{--}_{56,78}$ of the propagator the
following equation must be solved
\begin{equation}
A^{--}_{12,34}(\omega) = \tilde{A}_{12,34}(\omega) -
\sum_{56,78}\tilde{A}_{12,56}(\omega)\Phi_{56,78}(\omega)A^{--}_{78,34}(\omega)
\end{equation}
where the propagator
\begin{eqnarray}
\tilde{A}_{12,34}(\omega) = \delta_{13}\delta_{24}
\frac{n_{1}-n_{2}}{\omega - \epsilon_{12}}
\nonumber\\
 \epsilon_{12} = \epsilon_{1} - \epsilon_{2} , n_{i} = 1 ( \epsilon_{i}<\epsilon_{F}),
 n_{i} =0  ( \epsilon_{i}>\epsilon_{F}),
 \end{eqnarray}
is the RPA propagator and
\begin{equation}
\Phi_{12,34} = \tilde{\Phi}_{12,34} + \Phi_{12,34}^{comp}
\end{equation}
\begin{equation}\begin{array}{c}
\tilde{\Phi}_{12,34} = (n_{2}-n_{1})(n_{4}-n_{3})\sum\limits_{q}
\Bigl\{ \delta_{13}\sum\limits_{6}
\fr{(n_{6}-n_{1})\beta_{24,6}^{q}}{\omega - \epsilon_{16}-(1-2n_{1})\omega_{q}} +\\[4mm]
\delta_{24}\sum\limits_{5}\fr{(n_{2}-n_{5})\beta_{13,5}^{q*}}{\omega - \epsilon_{57}-(1-2n_{1})\omega_{q}} +\\[4mm]
(-1)^{J+J_{q}} \left\{
\begin{array}{ccc}
J&j_1&j_2\\
J_q&j_4&j_3
\end{array}
\right\}
\left[(-1)^{j_{3}-j_{2}}g_{13}^{q}g_{24}^{q*}\left(\fr{(1-n_{1})(1-n_{3})}{\omega
- \epsilon_{32}-\omega_{q}} -
                           \fr{n_{1}n_{3}}{\omega + \epsilon_{41}+\omega_{q}}\right) +\right.  \\[4mm]
\left.
(-1)^{j_{1}-j_{4}}g_{31}^{q*}g_{42}\left(\fr{(1-n_{1})(1-n_{3})}{\omega
- \epsilon_{14}-\omega_{q}} -
                                     \fr{n_{1}n_{3}}{\omega + \epsilon_{23}+\omega_{q}}\right)
                                                          \right]+\\[4mm]
\delta_{13}\sum\limits_{6}\fr{(n_{6}-n_{2})[\omega-\epsilon_{12}-\epsilon_{64}-(1-2n_{1})\omega_{q}]}
                   {[\epsilon_{62}+ (1-2n_{1})\omega_{q}][\epsilon_{64}+ (1-2n_{1})\omega_{q}]}\beta_{24,6}^{q}+\\[4mm]
\delta_{24}\sum\limits_{5}\fr{(n_{1}-n_{5})[\omega-\epsilon_{12}-\epsilon_{35}-(1-2n_{1})\omega_{q}]}
                   {[\epsilon_{15}+ (1-2n_{1})\omega_{q}][\epsilon_{35}+ (1-2n_{1})\omega_{q}]}\beta_{13,5}^{q*}+\\[4mm]
(-1)^{J+J_{q}} \left\{
\begin{array}{ccc}
J&j_1&j_2\\
J_q&j_4&j_3
\end{array}
\right\}
\left[(-1)^{j_{3}-j_{2}}g^{q}_{13}g^{q*}_{24}\left(\fr{(1-n_{1})n_{3}}{\epsilon_{42}+\omega_{q}}+
                                              \fr{(1-n_{3})n_{1}}{\epsilon_{31}+\omega_{q}}\right)+     \right.\\[4mm]
\left.
(-1)^{j_{1}-j_{4}}g^{q*}_{31}g^{q}_{42}\left(\fr{(1-n_{1})n_{3}}{\epsilon_{31}+\omega_{q}}+
                                         \fr{(1-n_{3})n_{1}}{\epsilon_{24}+\omega_{q}}\right) \right]
                                          \Bigr\}  ,
 \end{array}\end{equation}

\begin{equation}\begin{array}{c}
\Phi_{12,34}^{comp} = (n_{2}-n_{1})[(1-n_{1})(1-n_{3})+n_{1}n_{3}][(1-n_{2})(1-n_{4})+n_{2}n_{4}] \times \\[4mm]
                      \sum\limits_{q_{1},q_{2},56}\beta_{31,5}^{q_{1}}\beta_{24,6}^{q_{2}}    \times   \\[4mm]
                          \Bigl\{ \fr{(1-n_{1})(1-n_{5})(1-n_{6})+n_{1}n_{5}n_{6}}
                              {[\epsilon_{62}+(1-2n_{1})\omega_{q_{2}}][\epsilon_{64}+(1-2n_{1})\omega_{q_{2}}]                                                    [\omega-\epsilon_{52}-\epsilon_{64}-(1-2n_{1})(\omega_{q_{1}}+\omega_{q_{2}})]} +\\[4mm]
                           \fr{(1-n_{2})(1-n_{5})(1-n_{6})+n_{2}n_{5}n_{6}}
                              {[\epsilon_{15}+(1-2n_{1})\omega_{q_{1}}][\epsilon_{35}+(1-2n_{1})\omega_{q_{1}}]
                               [\omega-\epsilon_{15}-\epsilon_{36}-(1-2n_{1})(\omega_{q_{1}}+\omega_{q_{2}})]} - \\[4mm]
                            \fr{[(1-n_{2})(1-n_{5})n_{6} + n_{2}n_{5}(1-n_{6})]}
                             {[\epsilon_{15} + (1-2n_{1})\omega_{q_{1}}][\epsilon_{35} + (1-2n_{1})\omega_{q_{1}}]
                          [\epsilon_{62} + (1-2n_{1})\omega_{q_{1}}]}\times \\[4mm]
                  \fr{[\omega - \epsilon_{12} - \epsilon_{34}+\epsilon_{56} - (1-2n_{1})(\omega_{q_{1}} + \omega_{q_{2}})]}
                             {[\epsilon_{64} + (1-2n_{1})\omega_{q_{2}}]}\Bigr\} .

\end{array}\end{equation}
\newpage

   The quantities $Q^{+-}$ and $P^{++}$ are given by
\begin{equation}
 \begin{array}{c}
 Q^{+-}_{12,34} = [(1-n_{1})(1-n_{2}) + n_{1}n_{2}] [(1-n_{3})n_{4} + n_{3}(1-n_{4})]\times \\[4mm]
                 \sum\limits_{q}
                 \left\{
                 \delta_{13}\sum\limits_{6}\left[\left(
                 \fr{1-n_{1}}{\epsilon_{24}(\epsilon_{64}+\omega_{q})}+
                 \fr{n_{1}}{\epsilon_{62}+\omega_{q}}
                 (\fr{1}{\epsilon_{24}}-
                  \fr{1}{\omega+\epsilon_{61}+\omega_{q}}
                 )\right)\right. \right.
                 (1-n_{6})\beta_{24}^{q(+)} + \\[4mm]
                  +\left.\left(
                 \fr{n_{1}}{\epsilon_{42}(\epsilon_{46}+\omega_{q})}+
                \fr{1-n_{1}}{\epsilon_{26}+\omega_{q}}(\fr{1}{\epsilon_{42}}+\fr{1}
                                                  {\omega-\epsilon_{16}-\omega_{q}}
                 )\right)
                 n_{6}\beta_{24,6}^{q(-)}\right] - \\[4mm] -
                 \delta_{24}\sum\limits_{5}\Bigl[\left(
                 \fr{n_{1}}{\epsilon_{13}(\epsilon_{35}+\omega_{q})}+
                 \fr{1-n_{1}}{\epsilon_{15}+\omega_{q}}(\fr{1}{\epsilon_{13}}-\fr{1}
                                                                        {\omega+\epsilon_{25}+\omega_{q}}
                 )\right)
                 n_{5}\beta_{13,5}^{q(-)*} +
                 \\[4mm]
                  \left(
                \fr{1-n_{1}}{\epsilon_{31}(\epsilon_{53}+\omega_{q})}+
                 \fr{n_{1}}{\epsilon_{51}+\omega_{q}}(\fr{1}{\epsilon_{31}}+ \fr{1}
                                                                        {\omega-\epsilon_{52}-\omega_{q}}
                 )\right)
                 (1-n_{5})\beta_{13,5}^{q(+)*}\Bigr] +  \\[4mm]
               (-1)^{J+J_{q}}
\left\{
\begin{array}{ccc}
J&j_1&j_2\\
J_q&j_4&j_3
\end{array}
\right\}
\left[(-1)^{j_{3}-j_{2}}g_{13}^{q}g_{24}^{q*}n_{1}\times \right.\\[4mm]
                 \left(\fr{(1-n_{3})}{(\epsilon_{31}+\omega_{q})(\omega-\epsilon_{32}-\omega_{q})}
             -\fr{n_{3}}{(\epsilon_{42}+\omega_{q})(\omega + \epsilon_{41}+\omega_{q})}\right)+\\[4mm]
                    (-1)^{j_{1}-j_{4}}g_{31}^{q*}g_{42}^{q}(1-n_{1})
                 \left. \left.\left(\fr{(1-n_{3})}{(\epsilon_{24}+\omega_{q})(\omega-\epsilon_{14}-\omega_{q})}
                 -\fr{n_{3}}{(\epsilon_{13}+\omega_{q})(\omega + \epsilon_{23}+\omega_{q})}\right)                           \right]\right\},
                    \\[8mm]
                Q^{-+}_{12,34} = Q^{+-*}_{34,12}
  \end{array}\end{equation}

\begin{equation}\begin{array}{c}
P^{++}_{12,34}(\omega) =
\left[(1-n_{1})(1-n_{2})+n_{1}n_{2}\right]\left[(1-n_{3})(1-n_{4})+n_{3}n_{4}\right]
                        \times \\[4mm]
                       \sum\limits_{q}\Bigl\{
                        \delta_{13}
                        \sum\limits_{6}\fr{(n_{1}-n_{6})\beta_{24,6}^{q}}
                        {
           \left[\epsilon_{26}+(1-2n_{1})\omega_{q}\right]
           \left[\epsilon_{46}+(1-2n_{1})\omega_{q}\right]
           \left[\omega - \epsilon_{16}+(1-2n_{1})\omega_{q}\right]
                       }-    \\[4mm]
                         \delta_{24}\sum\limits_{5}\fr{(n_{1}-n_{5})\beta_{13,5}^{q*}}
                        {\left[\epsilon_{15}+(1-2n_{1})\omega_{q}\right]
                         \left[\epsilon_{35}+(1-2n_{1})\omega_{q}\right]
                         \left[\omega + \epsilon_{25}+(1-2n_{1})\omega_{q}\right]} +   \\[4mm]

                        (-1)^{J+J_{q}}
\left\{
\begin{array}{ccc}
J&j_1&j_2\\
J_q&j_4&j_3
\end{array}
\right\} \Bigl[
                       (-1)^{j_{3}-j_{2}}g_{13}^{q}g_{24}^{q*}\times \\[4mm]
                                                    \fr{n_{1}(1-n_{3})}
                                                       {(\epsilon_{31}+\omega_{q})(\epsilon_{42}+\omega_{q})}
                         \left(\fr{1}{\omega+\epsilon_{41} +\omega_{q}}-\fr{1}{\omega-\epsilon_{32}-\omega_{q}}
                          \right) + \\[4mm]
                          (-1)^{j_{1}-j_{4}}g_{31}^{q*}g_{42}^{q}
                                                    \fr{n_{3}(1-n_{1})}
                                                       {(\epsilon_{13}+\omega_{q})(\epsilon_{24}+\omega_{q})}
                         \left(\fr{1}{\omega+\epsilon_{23} +\omega_{q}}-\fr{1}{\omega-\epsilon_{14}-\omega_{q}}
                          \right)\Bigr]\Bigr\} .
\end{array}\end{equation}

\newpage
Here the following notations have been used:

\begin{equation}\begin{array}{c}
\beta_{12,3}^{q(+)} =
\fr{\delta_{j_{1}j_{2}}\delta_{l_{1}l_{2}}}{2j_{1}+1}
g_{13}^{q*}g_{23}^{q}, \beta_{12,3}^{q(-)} =
\fr{\delta_{j_{1}j_{2}}\delta_{l_{1}l_{2}}}{2j_{1}+1}
g_{31}^{q}g_{32}^{q*},\\[4mm]
\beta_{12,3}^{q}  = (1-n_{3})\beta_{12,3}^{q(+)} +
n_{3}\beta_{12,3}^{q(-)}.
\end{array}\end{equation}

Into all these formulae the reduced matrix elements $g^q_{12}\equiv
<1||g^q||2> (q\equiv{JLS})$ of the phonon creation amplitude at J
= L.
\begin{equation}
g({\bf r},{\bf {\bfsigma}})=\sum\limits_{JLSM} g_{JLS}(r)
T_{JLS}^M ({\bf n},{\bf {\bfsigma}})
\end{equation}
enter, where ${\bf n}={\bf r}/r$ and
\begin{equation}
T_{JLS}^M({\bf n},{\bf {\bfsigma}})=\sum\limits_{\mu,\nu}(L\mu
S\nu |JM)Y_{L\mu} ({\bf n})\sigma_{S\nu}
\end{equation}
with $\nu = \pm 1,0, \sigma_{0\nu}=1$ and
$\sigma_{1\nu}=\sigma_{\nu}, \sigma_{\nu}$  being cyclic
components of spin matrices. In the spherical single-particle
basis they are determined by
\begin{eqnarray}
<m_1l_1j_1n_1|g_{JLS}T_{JLS}^M|m_2l_2j_2n_2> =
(-1)^{-j_2+J+m_1}\times
\nonumber \\
\times \left(
\begin{array}{ccc}
j_1&J&j_2\\
-m_1&M&m_2
\end{array}
\right) <1||g^{JLS}||2>,
\end{eqnarray}
where
\begin{equation}
<1||g^{JLS}||2> =
<l_1j_1||T_{JLS}||l_2j_2>
\int R_{l_1j_1n_1}g_{JLS}(r)R_{l_2j_2n_2} r^2 dr.
\end{equation}

The reduced matrix element of the spherical tensor operator
$T_{JLS}$ has the form
\begin{equation}\begin{array}{c}
<l_1j_1||T_{JLS}||l_2j_2> = \nonumber \\ [4mm]
 = \frac{1}{2}[1 + (-1)^{L+l_{1}+l_{2}}]
(-1)^{S+j_{2}-\frac{1}{2}}\sqrt{\frac{(2J+1)(2L+1)(2j_{1}+1)(2j_{2}+1)}{4\pi}}
\times \nonumber  \\ [4mm] \Bigl[ \left(
\begin{array}{ccc}
J&L&S\\
0&0&0
\end{array}
\right) +(x_1 + (-1)^{J+L+S}x_2)\sqrt{\frac{S(S+1)}{J(J+1)}}
\left(
\begin{array}{ccc}
J&L&S\\
1&0&-1
\end{array}
\right) \Bigr] \left(
\begin{array}{ccc}
j_1&j_2&J\\
1/2&-1/2&0
\end{array}
\right),
\end{array}\end{equation}
where $x_{k} = (l_{k}-j_{k})(2j_{k}+1)$.


\section*{2.The refinement procedure}

Eq. (3.18) for the refined single-particle energies
$\tilde{\epsilon}_{\lambda}$
has the form
\begin{eqnarray}
\tilde{\epsilon}_{1} = \epsilon_{1} -
     \sum\limits_{3,q}\frac{\beta^{q}_{12,3}}{\epsilon_{1}-
      \tilde{\epsilon}_{3}-(1-2n_{3})\omega_{q}}
\end{eqnarray}
where index 1 stands for the set of the single-particle
quantum numbers  and  $\epsilon_{1}$ are
 the phenonomenological single-particle
energies.


\end{document}